\documentclass[10pt,a4paper]{article}
\usepackage{hyperref}
\usepackage[english]{babel}
\usepackage[utf8]{inputenc}
\usepackage[T1]{fontenc}
\usepackage{graphicx}
\usepackage{geometry}
\usepackage{tabularx}
\usepackage{psfrag}
\usepackage{fancyhdr}
\usepackage{xcolor}
\usepackage{sistyle}
\usepackage{amsfonts}
\usepackage{amsmath}
\usepackage{amssymb}
\usepackage{booktabs}
\usepackage{multirow}
\usepackage{amsthm}
\usepackage{subcaption}
\usepackage{sidecap}
\usepackage{pifont}
\usepackage[ruled]{algorithm2e}
\usepackage{enumerate}
\usepackage{enumitem}
\usepackage{bbold}
\usepackage[bottom]{footmisc}
\usepackage{tikz}
	\usetikzlibrary{backgrounds,automata,shapes}
	\usetikzlibrary{decorations.pathreplacing, arrows}
\usepackage{hyperref}
\usepackage{mathrsfs}
\usepackage{xr}

\theoremstyle{theorem}
\newtheorem{proposition}{Proposition}
\newtheorem{corollary}{Corollary}

\newtheorem{lemma}{Lemma}

\theoremstyle{definition}
\newtheorem{definition}{Definition}

\theoremstyle{remark}
\newtheorem{remark}{Remark}

\usepackage[numbers, comma, sort&compress]{natbib}
\usepackage{subcaption}
\usepackage{enumitem}
\usepackage{bbm}
\usepackage{footmisc}
\definecolor{mypink}{rgb}{0.56, 0.004, 0.32}
\definecolor{sapin}{RGB}{9,82,40}
\definecolor{carmin}{rgb}{0.7294,0.0392,0.0392}
\definecolor{sea}{rgb}{0.33, 0.45, 0.67}

\def\i{\mathrm{i}}

\def\1{\mathbbm{1}}
\def\wn{\xi}
\def\atom{f}
\def\snr{\mathsf{snr}}

\graphicspath{{figures/}}

\begin{document}

\title{Point Processes and spatial statistics in time-frequency analysis
\thanks{The authors acknowledge support from ERC grant Blackjack (ERC-2019-STG-851866) and ANR AI chair Baccarat (ANR-20-CHIA-0002), and thank Patrick Flandrin, Julien Flamant and Pierre Chainais for insightful discussions and courteously sharing some illustrations.}}

\author{Barbara Pascal, Rémi Bardenet.
\thanks{B. Pascal is with Nantes Université, École Centrale Nantes, CNRS, LS2N, UMR 6004, F-44000 Nantes, France, (e-mail: barbara.pascal@cnrs.fr, \textit{corresponding author}); 
R. Bardenet is with Univ. Lille, CNRS, Centrale Lille, UMR 9189 CRIStAL, F-59000 Lille, France,  (e-mail: remi.bardenet@cnrs.com).}}

\maketitle

\abstract{
A \emph{finite-energy} \emph{signal} is represented by a square-integrable, complex-valued function $t\mapsto s(t)$ of a real variable $t$, interpreted as time. 
Similarly, a noisy signal is represented by a random process.
\emph{Time-frequency analysis}, a subfield of signal processing, amounts to describing the temporal evolution of the frequency content of a signal.
Loosely speaking, if $s$ is the audio recording of a musical piece, time-frequency analysis somehow consists in writing the \emph{musical score} of the piece.
Mathematically, the operation is performed through a transform $\mathcal{V}$, mapping $s \in L^2(\mathbb{R})$ onto a complex-valued function $\mathcal{V}s \in L^2(\mathbb{R}^2)$ of time $t$ and angular frequency $\omega$.
The squared modulus $(t, \omega) \mapsto \vert\mathcal{V}s(t,\omega)\vert^2$ of the time-frequency representation is known as the \emph{spectrogram} of $s$; in the musical score analogy, a peaked spectrogram at $(t_0,\omega_0)$ corresponds to a musical note at angular frequency $\omega_0$ localized at time $t_0$.
More generally, the intuition is that upper level sets of the spectrogram contain relevant information about the original signal.
Hence, many signal processing algorithms revolve around identifying maxima of the spectrogram.
In contrast, \emph{zeros} of the spectrogram indicate perfect silence, that is, a time at which a particular frequency is absent. 
Assimilating $\mathbb{R}^2$ to $\mathbb{C}$ through $z = \omega + \mathrm{i}t$, this chapter focuses on time-frequency transforms $\mathcal{V}$ that map signals to analytic functions. 
The zeros of the spectrogram of a noisy signal are then the zeros of a random analytic function, hence forming a \emph{Point Process} in $\mathbb{C}$.
This chapter is devoted to the study of these Point Processes, to their links with zeros of Gaussian Analytic Functions, and to designing signal detection and denoising algorithms using spatial statistics, by identifying perturbations in the point pattern of silence.
}

\section{Introduction and chronological perspective}

\begin{figure}
\centering
\begin{subfigure}{0.32\linewidth}
\centering
\includegraphics[width = 0.9\linewidth]{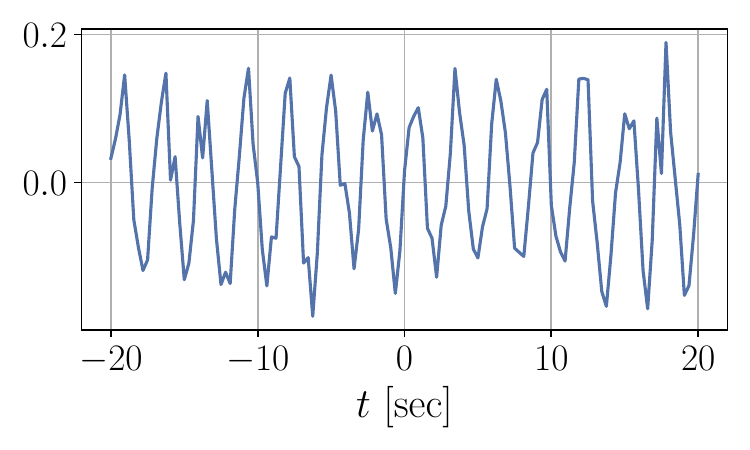}
\subcaption{\label{sfig:sine}Pure sine.}
\end{subfigure}
\begin{subfigure}{0.32\linewidth}
\centering
\includegraphics[width = 0.9\linewidth]{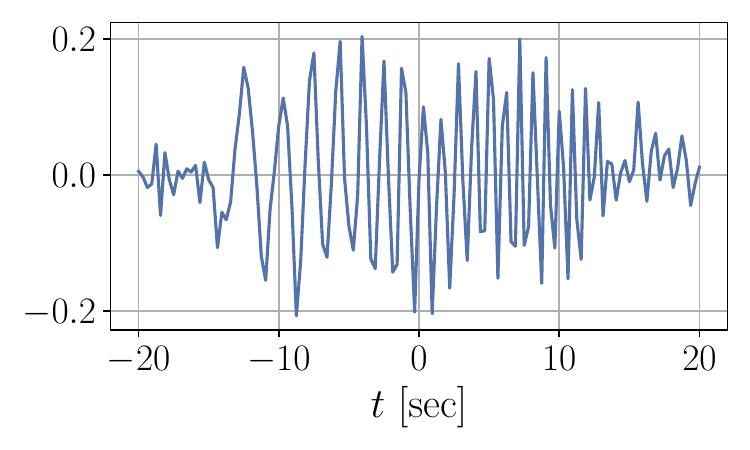}
\subcaption{\label{sfig:chirp}Modulated chirp.}
\end{subfigure}
\begin{subfigure}{0.32\linewidth}
\centering
\includegraphics[width = 0.9\linewidth]{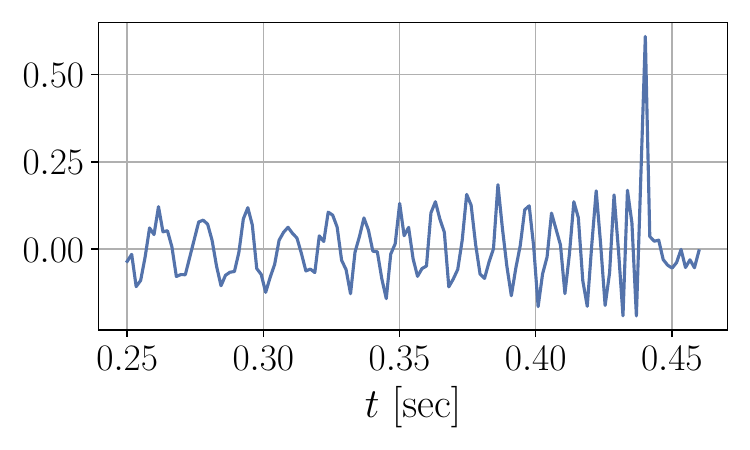}
\subcaption{\label{sfig:wave}Gravitational wave.}
\end{subfigure}
\caption{\label{fig:ex_sig}Examples of realistic elementary signals.}
\end{figure}

In the broadest sense, a \emph{signal} is a collection of data which carries information about a phenomenon of interest. 
This definition encompasses data of very diverse types, from physical measurements to epidemiological indicators, or even man-made data.
The present work focuses on \emph{temporal} signals, such as audio recordings,  physical quantities, e.g.,  light intensity, pressure or voltage,  measured by a sensor over time. Mathematically, these temporal signals $t\mapsto s(t)$ are represented as real or complex functions of a real variable $t$, referred to as \emph{time}. 
Three realistic signals are provided in Figure~\ref{fig:ex_sig}: a pure sine wave with constant amplitude and frequency in~\ref{sfig:sine}, a \emph{chirp} with linearly increasing frequency and amplitude following a smooth envelope in~\ref{sfig:chirp},  and a gravitational wave with exploding amplitude and frequency in~\ref{sfig:wave}.\\

\begin{figure}
\centering
\begin{subfigure}{0.32\linewidth}
\centering
\includegraphics[width = \linewidth]{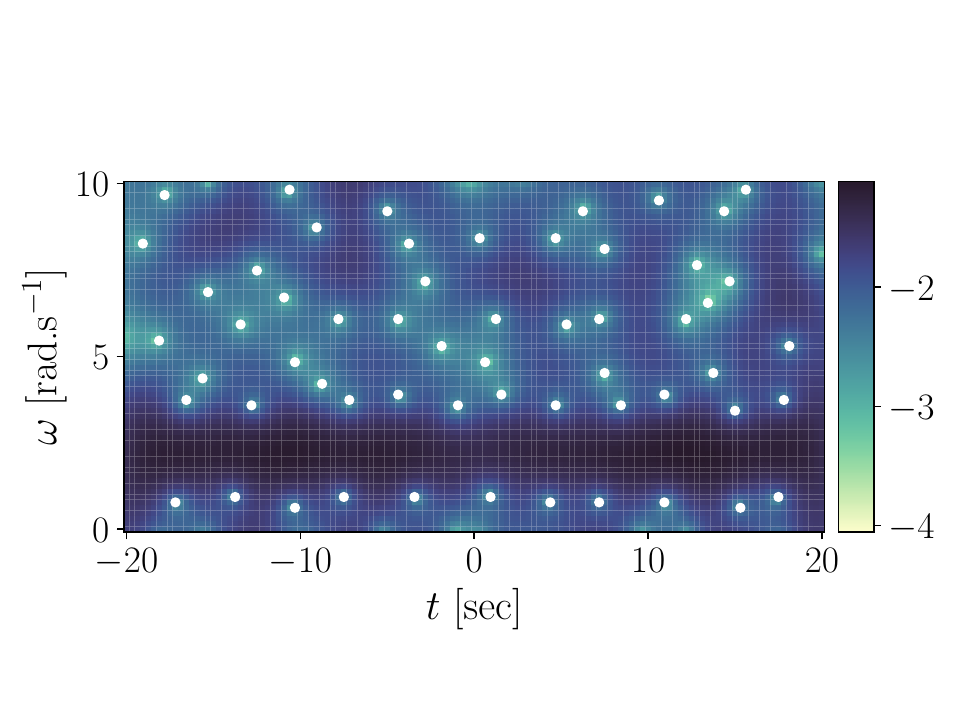}
\subcaption{\label{sfig:sine_spec}Pure sine.}
\end{subfigure}
\begin{subfigure}{0.32\linewidth}
\centering
\includegraphics[width = \linewidth]{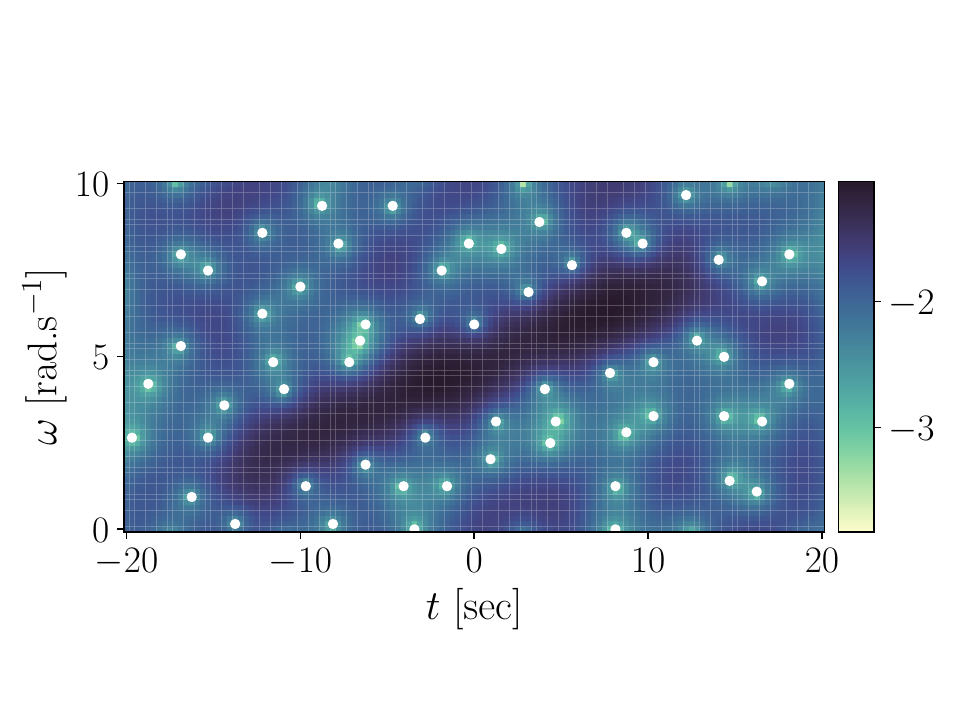}
\subcaption{\label{sfig:chirp_spec}Modulated chirp.}
\end{subfigure}
\begin{subfigure}{0.32\linewidth}
\centering
\includegraphics[width = \linewidth]{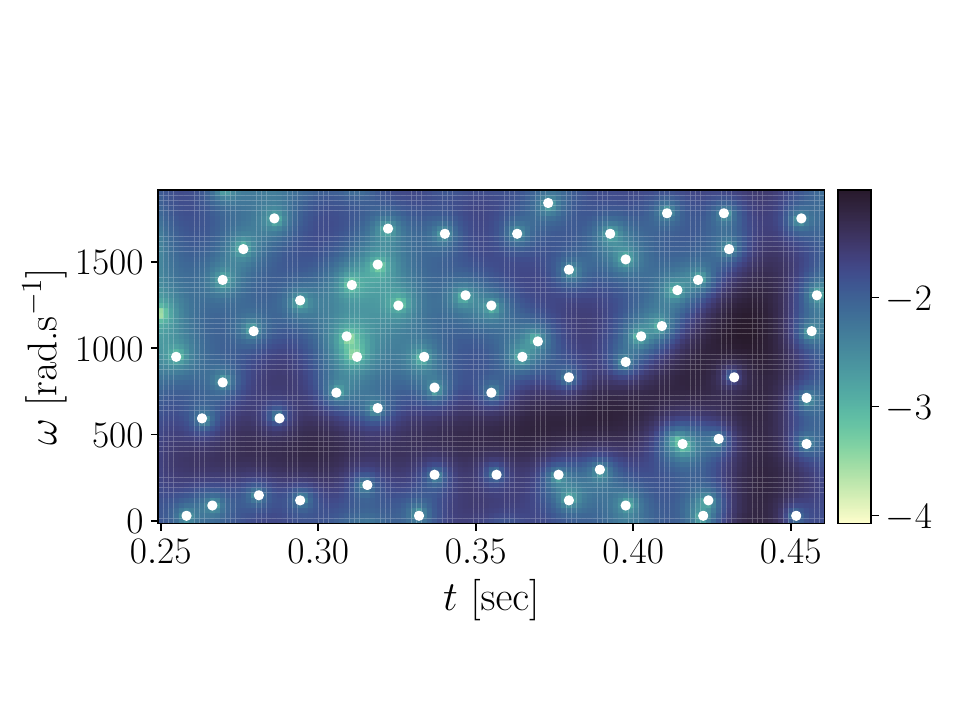}
\subcaption{\label{sfig:wave_spec}Gravitational wave.}
\end{subfigure}
\caption{\label{fig:ex_spec}Spectrograms of elementary signals displayed in log-scale colormaps. Maximum values are in dark blue, zeros are represented as white dots.}
\end{figure}

As in Figure~\ref{fig:ex_sig}, most signals are acquired in the time domain, and correspond to the measurement of some physical quantity in which time events are particularly visible, e.g., an earthquake and its aftershocks.
Yet, in many situations, the frequency content provides complementary information that is crucial to get insight into the studied phenomenon, e.g., the types of rocks through which the seismic wave propagated, enabling precise localization of the epicenter.
The most standard tool to probe the frequency content of a signal is the \emph{Continuous Fourier Transform}\footnote{
    See Section~\ref{sssec:CFT} for a formal definition and main properties.
} $\omega\mapsto\mathcal{F}s(\omega)$, which yields an overall \emph{spectral} density, depending on a real variable $\omega$ called the angular frequency. 
However, most of the time, one is interested in a joint time-frequency representation, in order to visualize the frequency properties of events localized in time.
To that aim, the Short-Time Fourier Transform\footnote{
    See Section~\ref{sssec:STFT} for a formal definition and properties.
} $(t, \omega)\mapsto\mathcal{V}_h s(t,\omega)$ has been introduced.
It amounts to computing \emph{local} Fourier transforms by sliding a localized window $h$ over the signal $s$, thus recording the local frequency content at each time $t$.
This results in the \emph{spectrogram}, defined as the squared modulus of $\mathcal{V}_h s(t,\omega)$.
By analogy with physics, the spectrogram of a signal is usually described as the energy distribution of that signal.
The spectrograms of the three signals from Figure~\ref{fig:ex_sig} are displayed in Figure~\ref{fig:ex_spec}: high values of the spectrogram, in dark blue, indicate high-energy regions, suggesting that important information is encapsulated in the spectrogram at this location of the time-frequency plane.
For instance, for the sine signal, Figure~\ref{sfig:sine_spec} shows a horizontal line of maxima reflecting the constant frequency.
For the chirp, the maxima of Figure~\ref{sfig:chirp_spec} form an increasing line, reflecting the linearly increasing frequency.
Finally, one can observe on Figure~\ref{sfig:wave_spec} the exploding frequency typical of a gravitational wave.

Both the signals of Figure~\ref{fig:ex_sig} and the spectrograms of Figure~\ref{fig:ex_spec} are \emph{easy} to read, in the sense that a clear structure can be identified through direct observation. 
Nevertheless, the reader might have remarked that the signals, and hence the spectrograms, contain higher-frequency oscillations typical of \emph{noise},\footnote{
    The definition of \emph{noise} will be thoroughly discussed in Section~\ref{ssec:white_noise}.
}
i.e., a unstructured perturbation superimposed on the pure signal.
    
Noise is omnipresent in signal processing, as any physical measurement is corrupted both by intrinsic and by environmental perturbations. Its presence translates into irregular fluctuations in the signals of Figure~\ref{fig:ex_sig}, both in time and in amplitude. 
Similarly, noise creates stochastic low-energy patterns in the spectrograms of Figure~\ref{fig:ex_spec}, appearing in medium blue, and popping up outside of the upper level set in dark blue.
The (numerical) zeros of the spectrograms are represented as white dots in Figure~\ref{fig:ex_spec}.
A striking observation is that, whatever the shape of the underlying signal, the spectrogram zeros spread very uniformly outside of the regions with large energy.
An unorthodox path has emerged in time-frequency analysis from this observation, shifting the focus from the maxima to the zeros of spectrograms.

\begin{figure}
\centering
 \begin{subfigure}[t]{0.32\linewidth}
\centering
\raisebox{2mm}{\includegraphics[width = 0.9\linewidth]{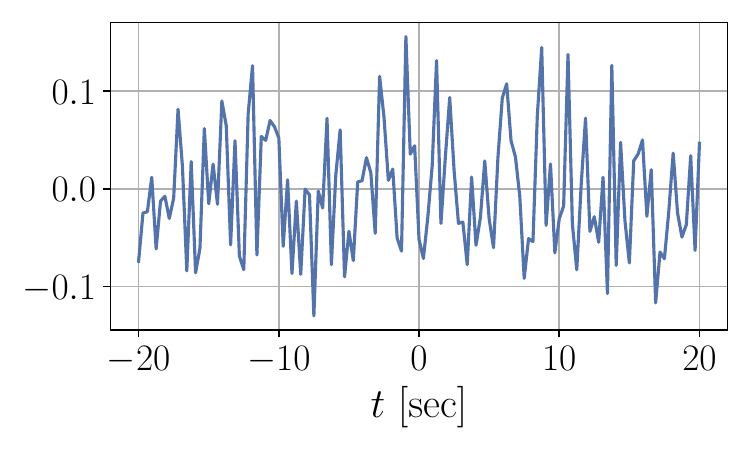}}
\vspace{-4mm}
\subcaption{\label{sfig:wnoise}White noise $\wn(t)$.}
\end{subfigure} \hspace{0.11\linewidth}
\begin{subfigure}[t]{0.32\linewidth}
\centering
\includegraphics[width = \linewidth]{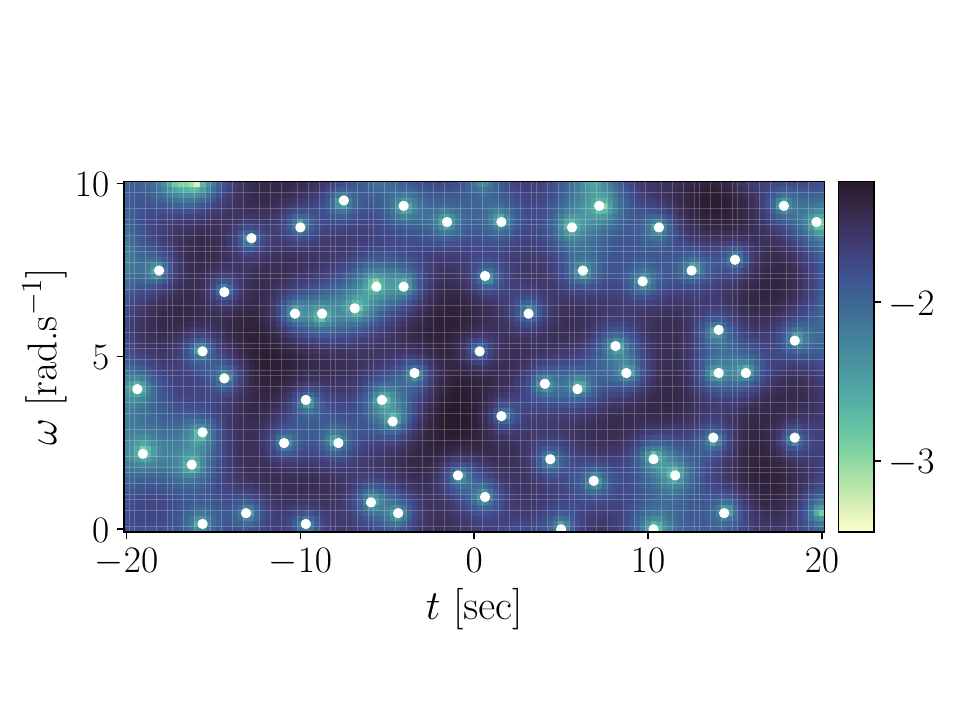}
\vspace{-7.5mm}
\subcaption{\label{sfig:wnoise_spec}White noise spectrogram.}
\end{subfigure} 
\caption{\label{fig:wnoise}Time-frequency analysis of pure noise.}
\end{figure}

The first interest in the zeros of spectrogram appeared in~\cite{gardner2006sparse} in which the authors emphasized the quite rigid distribution of zeros of the \textit{Gaussian spectrogram}\footnote{The Gaussian spectrogram is defined as $\left\lvert \mathcal{V}_g s(t,\omega)\right\rvert^2$, for $\mathcal{V}_g$ the Short-Time Fourier Transform using the circular Gaussian window $g(t) = \pi^{-1/2}\exp(-t^2/2)$. See Section~\ref{sssec:STFT}.} of \emph{white noise}\footnote{
    White noise consists in a random process $\wn(t)$ of zero mean and no correlations at different times. See Section~\ref{ssec:white_noise} for a formal definition.
} in the time-frequency plane. 
A sample of white noise and its spectrogram are displayed in Figures~\ref{sfig:wnoise}~and~\ref{sfig:wnoise_spec} respectively.
More precisely, ~\cite{gardner2006sparse} remarked that the zeros of the Gaussian spectrogram are uniformly spread, as can be seen in Figure~\ref{sfig:wnoise_spec}, each zero roughly occupying a unit time-frequency area in the time-frequency plane.
Elaborating on this observation, \cite{flandrin2015time} first showed that the zeros of the spectrogram almost fully characterize the analyzed signal. 
Furthermore, \cite{flandrin2015time} took advantage of the fact that the presence of a signal creates holes in the pattern of zeros to design zeros-based algorithms that denoise signals, and separate the holes into elementary signal components.
This pioneering work opened the way to a new paradigm in time-frequency analysis, revolving around zeros instead of maxima.
Continuing this approach, \cite{bardenet2020zeros} established that the zeros of the Gaussian spectrogram of white noise form a \emph{Point Process} --~a random configuration of points in the time-frequency plane~-- and that its law corresponds to the zeros of the \textit{planar Gaussian Analytic Function}.\footnote{
    See Section~\ref{sec:GAF}, Equation~\eqref{eq:planarGAF}.
}
Extending~\cite{bardenet2020zeros}, similar connections have been proven between standard representations in signal processing and canonical Gaussian Analytic Functions, such as the Daubechies-Paul scalogram, whose zeros are those of the \emph{hyperbolic Gaussian Analytic Function}~\cite{abreu2018filtering,bardenet2021time}, or the newly introduced Kravchuk transform, this time linked to the \emph{spherical Gaussian Analytic Function}~\cite{pascal2022covariant,pascal2022famille}.
These connections were used to design both signal detection procedures~\cite{bardenet2020zeros,pascal2022covariant}, denoising strategies~\cite{flandrin2015time,abreu2018filtering,MACLM24}, and \emph{unmixing}, consisting in recovering the components of a composite signal made of a linear superposition of elementary waveforms~\cite{flandrin2015time}.
Recently, the geometric analysis of spectrograms has extended from zeros to more general \emph{level sets}~\cite{ghosh2021estimation}. 
Leveraging the geometrical and statistical properties of level sets of the spectrogram of white noise and of noisy Hermite functions, \cite{ghosh2021estimation} designed detection and denoising strategies, accompanied with theoretical guarantees.

\noindent \textbf{Outline.} Section~\ref{sec:tf_basics} is devoted to a self-contained review of the main concepts of signal processing, with an emphasis on time-frequency analysis and a rigorous definition of white noise.
Then, Gaussian Analytic Functions are introduced in~\ref{ssec:def_GAF}, and some key properties of their zeros are presented. 
The connection between Gaussian Analytic Functions and time-frequency analysis is established in Section~\ref{ssec:zerosGAF_TF}.
Section~\ref{sec:sig_spat_stat} presents principles and tools from spatial statistics, and their application to the design of signal processing procedures.
Finally, Section~\ref{sec:extensions} highlights recent extensions and promising research directions.

\subsection{Notations}

$L^2(\mathbb{R})$ (resp. $L^2(\mathbb{C})$) denotes the complex-valued functions of a real (resp. complex) variable, which are square-integrable w.r.t. the Lebesgue measure over $\mathbb{R}$ (resp. $\mathbb{C}$).
Elements of $L^2(\mathbb{R})$ (resp. $L^2(\mathbb{C})$) are denoted in lower (resp. upper) case. 
Operators, linear or not, acting on $L^2(\mathbb{R})$ are written with upper-case calligraphic letters.
The set of integers is denoted by $\mathbb{Z}$, while the set of non-negative (resp. positive) integers is denoted by $\mathbb{N}$ (resp. $\mathbb{N}^*$).
The complex conjugate of $z\in \mathbb{C}$ is denoted by $\overline{z}$.
Bold lower- (resp. upper-) case refers to complex vectors (resp. matrices).
For $n,p \in\mathbb{N}^*$, a vector $\boldsymbol{s}\in \mathbb{C}^n$ (resp a matrix $\textbf{B} \in \mathbb{C}^{n\times p}$) has components $s_1, \hdots, s_n$ (resp. $\lbrace \mathrm{B}_{k,\ell}, \, 1 \leq k \leq n, 1\leq \ell \leq p \rbrace$). 
Independent identically distributed random variables are referred to as ``i.i.d.".
$\mathbb{E}[\cdot]$ denotes the \emph{expectation} of an random variable and $\mathbb{P}[\cdot]$ the \emph{probability} of an event. 

\section{Signal processing and time-frequency analysis}
\label{sec:tf_basics}

This section provides,  in Section~\ref{ssec:SP}, a short introduction to the main concepts of signal processing, followed by a brief, self-contained presentation of time-frequency analysis in Section~\ref{ssec:TF_analysis}, with a focus on the \emph{Short-Time Fourier Transform}, which is the main signal processing tool at stake in this chapter.
The classical material is adapted from the reference books \cite{flandrin1998time,grochenig2001foundations}, to which the interested reader can add the more recent \citep{flandrin2018explorations}, and the harmonic analysis viewpoint from \cite{folland2016harmonic}. 
The rigorous definition of \emph{white noise} in ~\ref{ssec:white_noise} is that of Gross in \cite{gross1967abstract}, as presented in \cite{bardenet2021time}.
It will have a prominent importance in connecting Gaussian Analytic Functions with time-frequency representations in Section~\ref{ssec:zerosGAF_TF}.
Finally, Section~\ref{ssec:num_impl} discusses numerical implementation in the straight line of~\cite{oppenheim1999discrete}.

\subsection{Basic concepts of signal processing}
\label{ssec:SP}

Deterministic signals are data consisting in a real or complex-valued, square-integrable function $t\mapsto y(t)$ of a real variable $t$, assimilated to time.
Whether to account for variability under repeated measurements, or for physical processes of lesser importance that have been neglected, measured data are often further modeled as a random process. 
The most common model is the signal-plus-noise model.

\subsubsection{The signal-plus-noise model}
\label{sssec:signalnoise}
\begin{figure}
\centering
\begin{subfigure}{0.32\linewidth}
\centering
\includegraphics[width =0.9 \linewidth]{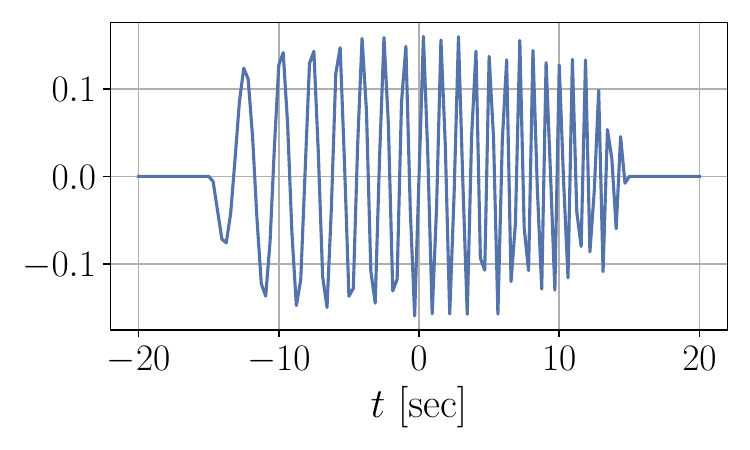}
\subcaption{\label{sfig:chirp_inf}$\snr = \infty$.}
\end{subfigure}
\begin{subfigure}{0.32\linewidth}
\centering
\includegraphics[width =0.9 \linewidth]{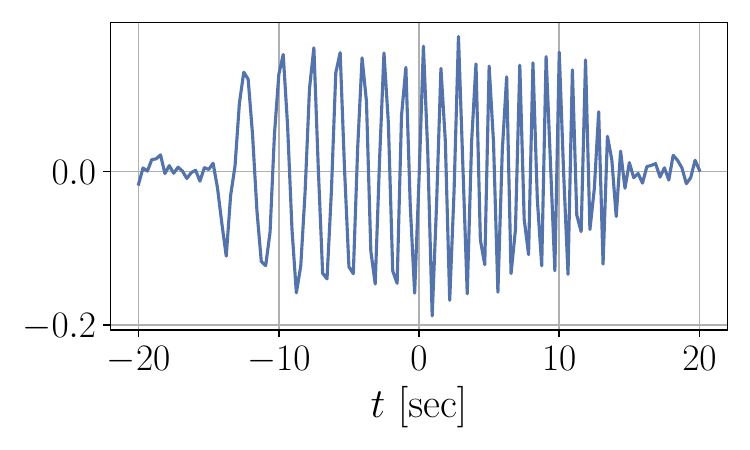}
\subcaption{\label{sfig:chirp_5}$\snr = 5$.}
\end{subfigure}
\begin{subfigure}{0.32\linewidth}
\centering
\includegraphics[width =0.9 \linewidth]{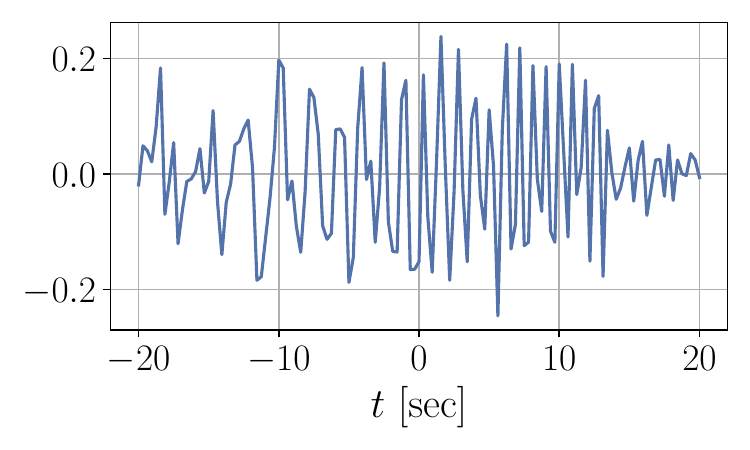}
\subcaption{\label{sfig:chirp_1}$\snr = 1$.}
\end{subfigure}
\caption{\label{fig:snr}Realizations of the signal-plus-noise model~\eqref{eq:signalnoise} with increasing noise levels.}
\end{figure}

In the \emph{signal-plus-noise} model, the observed signal decomposes as
\begin{align}
\label{eq:signalnoise}
    y = \snr \times s + \wn,
\end{align}
with $s \in L^2(\mathbb{R})$ the deterministic signal of interest, $\wn$ a random process accounting for the presence of noise, 
and $\snr \geq 0$ the so-called \emph{signal-to-noise} ratio, quantifying the corruption of the data $y$.
Figure~\ref{fig:snr} exemplifies the signal-plus-noise model on a template signal, namely a  \emph{chirp}, whose exact expression is given below at Equation~\eqref{eq:chirp}.
The value of $\snr$ decreases, from no noise at all in Figure~\ref{sfig:chirp_inf}, to a situation in which the signal is almost drowned in the noise in Figure~\ref{sfig:chirp_1}.

\subsubsection{A few template signals}

A common class of signals comes from modeling the propagation of \emph{waves}.
A wavelike signal is characterized by its time-varying amplitude $t\mapsto A(t)$ and phase $t\mapsto\varphi(t)$, 
\begin{align}
    s(t) = A(t) \sin(\varphi(t)),  \quad  \omega(t) = \dfrac{\mathrm{d} \varphi}{\mathrm{d}t}(t), \quad t\in\mathbb{R},
\end{align}
where $\omega(t)$ defines the instantaneous angular frequency.
In particular, a \emph{pure sine wave}, such as in Figure~\ref{sfig:sine}, corresponds to
\begin{align}
    s(t) = A \sin(\omega t)  \tag{$\mathsf{sine}$},
\end{align}
with both the amplitude $A > 0$ and the angular frequency $\omega$ constant. 
Other ubiquitous wavelike signals are \emph{linear chirps}, exemplified in Figure~\ref{sfig:chirp}, and described, for some duration $T>0$, as the transient waveform
\begin{align}
\label{eq:chirp}
    s(t) = A(t) \sin(\omega(t) t), \quad \omega(t) = \omega_1 + \frac{\omega_2-\omega_1}{2T}(t+T), \tag{$\mathsf{chirp}$}
\end{align}
with $A$ a smooth envelope whose support is contained in $[-T, T]$, and $\omega_1, \omega_2 > 0$ two fixed angular frequencies.
\eqref{eq:chirp} models a waveform modulated in amplitude, and with frequency evolving linearly from $\omega_1$ at time $-T$ to $\omega_2$ at time $T$.
More intricate wavelike signals are often encountered in modern physics to describe highly non-linear phenomena, such as gravitational waves, for which general relativity predicts an exploding chirp-like form,
\begin{align}
\label{eq:sing-chirp}
s(t) = C(t_0 - t)^{-1/4} \cos(2\pi d(t_0-t)^{5/8}+\varphi)\boldsymbol{1}_{(-\infty; t_0[}(t),  \tag{$\mathsf{wave}$}
\end{align}
where $C > 0$ is the amplitude, $d$ is a constant that encodes physical information about the cosmological event that produced the wave, $\varphi \in [0, 2\pi[$ is a pure phase, and  $t_0 \in \mathbb{R}$ is the collapsing time at which both the amplitude $A(t) = C(t_0 - t)^{-1/4}$ and the instantaneous angular frequency $\omega(t) = 10\pi d/8  (t_0-t)^{-3/8}$ are diverging. 
An example of such singular wave is provided in Figure~\ref{sfig:wave}.

\subsubsection{Aims and means of signal processing}
\label{sssec:aims}

The general goal of signal processing is to extract as much information as possible about the underlying signal $s$ from noisy observations $y$. 
Extracting information can mean different things depending on the context.
This chapter restricts to two representative tasks of signal processing,
\begin{itemize}
\item \emph{detection}: from noisy measurements $y$, decide whether there is an underlying signal $s$, that is, whether $\snr>0$ in Equation~\eqref{eq:signalnoise};
\item \emph{reconstruction}:\footnote{Also referred to as \emph{denoising} in the signal processing literature.} assuming $\snr > 0$, recover as accurately as possible the signal of interest $t\mapsto s(t)$.
\end{itemize}

\begin{figure}
\centering
\begin{tikzpicture}[scale = 2.35]
\draw[black!35!white, dashed] (0,0) -- ({cos(30)},{-sin(30)} );
\draw[black!35!white, dashed] (0,0) -- ({-cos(30)},{-sin(30)} );
\fill[white] (0,0) circle(0.5);
\draw[black!35!white, dashed] (0,0.25) -- (0,1);
\draw (0,1.15) node{physics};
\draw (0,1) node{$\bullet$};
\draw (0,1) -- ({-cos(30)},{-sin(30)} );
\draw ({-cos(30)},{-sin(30)} ) node{$\bullet$};
\draw ({-cos(30)-0.15},{-sin(30)-0.175} ) node{mathematics};
\draw (0,1) -- ({cos(30)},{-sin(30)} );
\draw ({cos(30)},{-sin(30)} ) node{$\bullet$};
\draw ({cos(30)+0.15},{-sin(30)-0.195} ) node{computer science};
\draw ({-cos(30)},{-sin(30)} ) -- ({cos(30)},{-sin(30)} );
\draw (0,0.15) node {\textbf{\color{sea} signal}};
\draw (0,-0.15) node {\textbf{\color{sea} processing}};
\end{tikzpicture}
\caption{\label{fig:gold_triangle} The cross-disciplinarity of signal processing, pictured as the so-called \emph{golden triangle of signal processing} of Patrick Flandrin, see for example~\cite[Chapter 1: Introduction]{flandrin2018explorations}.}
\end{figure}
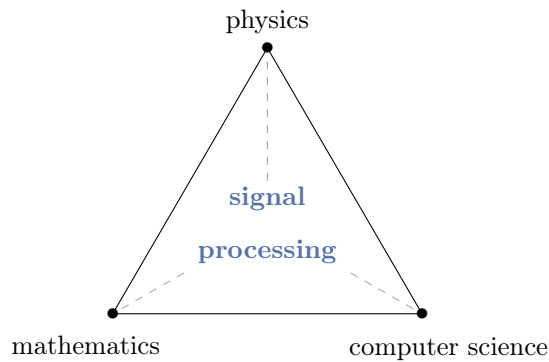

To meet these challenges, signal processing calls on the theory and tools of different disciplines, as illustrated in Figure~\ref{fig:gold_triangle}.
Indeed, signal processing lies at the intersection of \emph{physics}, for modeling phenomena in biology, mechanics, acoustics, etc; \emph{mathematics}, to formalize the problem and design performance criteria, and \emph{computer science} to provide efficient algorithms and implementations.
This complementarity is key in many results presented in this chapter.

\subsection{From time or frequency to joint time and frequency}
\label{ssec:TF_analysis}

As explained in the introduction, signals are often acquired in the time domain.
The standard representation of a finite-energy signal is then a square-integrable function of time,  the $L^2$ norm of the function being thought of as the energy of the signal.
This representation is convenient for precisely localizing an event in time, or for analyzing the dynamics of a phenomenon that involves a change in the signal over time, but it is impractical to visualize or detect mechanisms that create a superposition of oscillations.
The latter are more readily seen on a \emph{frequency} representation, such as the Fourier spectrum of the signal.
This section is devoted to a concise presentation of the signal processing tools used to probe the frequency content of a signal, with a focus on Fourier-based representations.

\subsubsection{Fourier analysis}
\label{sssec:CFT}

The \emph{Continuous Fourier Transform} of a signal can be interpreted as a comprehensive description of the elementary oscillations composing the signal.

\begin{definition} 
    Let $s \in L^1(\mathbb{R})\cap L^2(\mathbb{R})$, its \emph{Continuous Fourier Transform} $\mathcal{F}s$ is the function of the angular frequency $\omega\in \mathbb{R}$ defined by
    \begin{align*}
        \mathcal{F}s(\omega) = \int_{\mathbb{R}} s(t) \exp(-\mathrm{i}\omega t) \, \mathrm{d}t.
    \end{align*}
    Since $L^1(\mathbb{R})\cap L^2(\mathbb{R})$ is dense in $L^2(\mathbb{R})$, the Continuous Fourier Transform continuously extends to all finite-energy signals, i.e. to all $s \in L^2(\mathbb{R})$.
    \label{def:Fourier_transform} 
\end{definition}

An instructive example is the \emph{Gaussian window},  ubiquitous in signal processing.  

\begin{proposition}
Let  the unit-energy Gaussian window of width $\sigma > 0$ be defined as 
\begin{align*}
g_{\sigma}(t) = \frac{1}{(\pi \sigma^2)^{1/4}} \exp\left( - \frac{t^2}{2\sigma^2}\right).
\end{align*}
Then, $g_{\sigma} \in L^1(\mathbb{R})\cap L^2(\mathbb{R})$ and its Fourier transform is also a Gaussian window
\begin{align*}
\mathcal{F}g_{\sigma}(\omega) = (\pi \sigma^2)^{1/4} \exp\left( - \frac{\sigma^2\omega^2}{2}\right), \quad \omega\in\mathbb{R},
\end{align*}
whose standard deviation is the \emph{inverse} of the original window's standard deviation.
\label{prop:fourier-window}
\end{proposition}

Proposition~\ref{prop:fourier-window} reflects a general principle of Fourier analysis, which can be enunciated as: \emph{the narrower the signal, the wider its Fourier transform and vice versa}.

\begin{proposition}
    The Continuous Fourier Transform $\mathcal{F} : L^2(\mathbb{R}) \rightarrow  L^2(\mathbb{R})$ is linear, continuous and invertible.
    Furthermore, if $s \in L^1(\mathbb{R})$ and $\mathcal{F}s \in L^1(\mathbb{R})$, then the reconstruction formula 
    \begin{align*}
    s(t) = \frac{1}{2\pi} \int_{\mathbb{R}} \mathcal{F}s(\omega) \exp(\mathrm{i}\omega t) \, \mathrm{d}\omega,
    \end{align*}
    holds for almost every $t\in \mathbb{R}$ with respect to Lebesgue measure on $\mathbb{R}$.
\end{proposition}

\begin{proposition}
    Let $s \in  L^2(\mathbb{R})$, then $\mathcal{F}s \in L^2(\mathbb{R})$, and
    \begin{align*}
    \lVert \mathcal{F}s \rVert_2^2 = 2\pi \lVert s \rVert_2^2,
    \end{align*}
    where $\forall h \in L^2(\mathbb{R})$, $\lVert h \rVert_2^2 =\int_{\mathbb{R}} \lvert h(u)\rvert^2 \, \mathrm{d}u$.
    From a signal processing point of view, the Fourier transform is said to preserve the \emph{energy} of the signal.
\end{proposition}

\begin{figure}
\centering
\begin{subfigure}[t]{0.32\linewidth}
\centering
\raisebox{2mm}{\includegraphics[width = 0.9\linewidth]{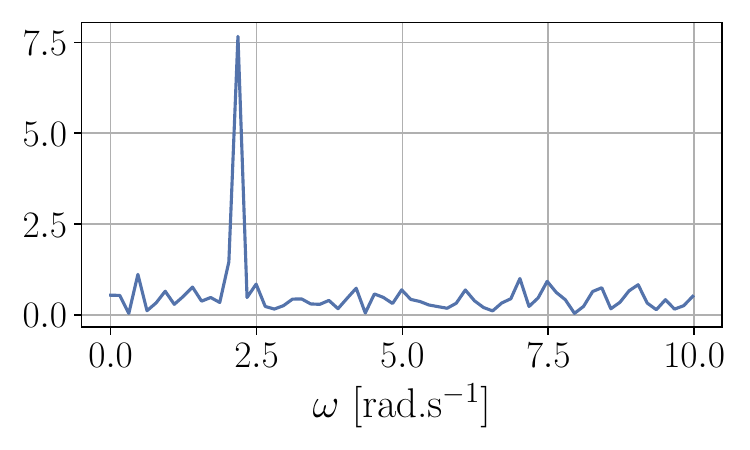}}
\vspace{-4mm}
\subcaption{\label{sfig:sine_fft}Pure sine.}
\end{subfigure}
\begin{subfigure}[t]{0.32\linewidth}
\centering
\raisebox{2mm}{\includegraphics[width = 0.9\linewidth]{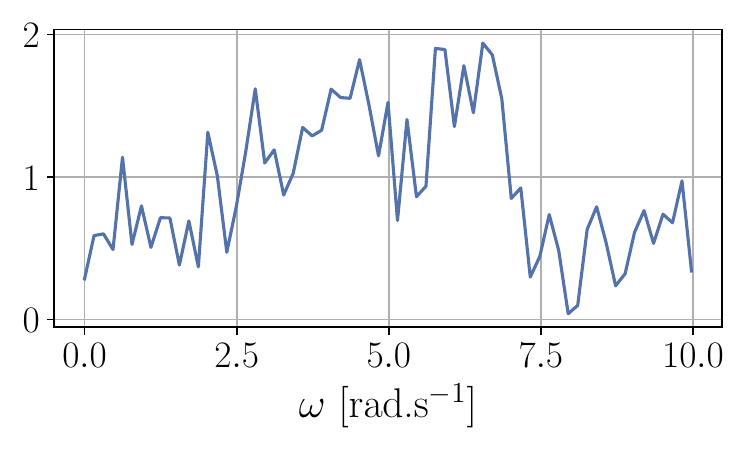}}
\vspace{-4mm}
\subcaption{\label{sfig:chirp_fft}Modulated chirp.}
\end{subfigure}
\begin{subfigure}[t]{0.32\linewidth}
\centering
\raisebox{2mm}{\includegraphics[width = 0.9\linewidth]{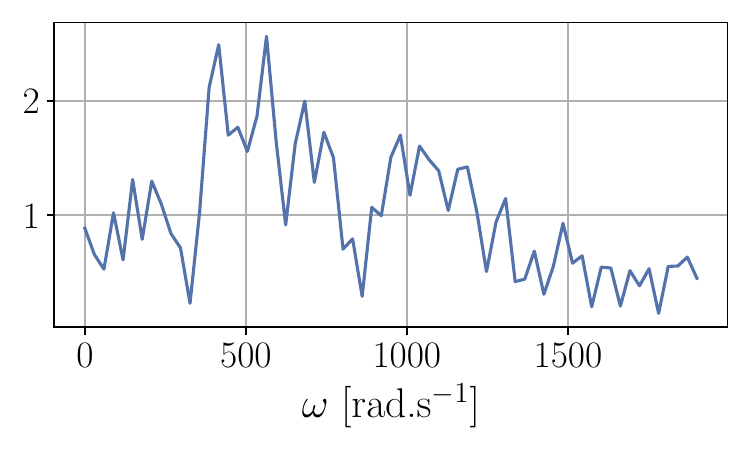}}
\vspace{-4mm}
\subcaption{\label{sfig:wave_fft}Gravitational wave.}
\end{subfigure}
\caption{\label{fig:fft}Fourier spectra  $\lvert \mathcal{F} s (\omega)\rvert$ of the template signals of Figure~\ref{fig:ex_sig}.}
\end{figure}

The \emph{Fourier spectrum} of a signal is defined as the modulus of its Continuous Fourier Transform: it quantifies the power of a frequency in the signal. Figure~\ref{fig:fft} yields the Fourier spectrum of each of the template signals displayed in Figure~\ref{fig:ex_sig}.
For the pure sine, characterized by a unique frequency, the Fourier spectrum provided in Figure~\ref{sfig:sine_fft} is, up to small fluctuations due to the additive noise, a Dirac mass at this frequency.
For the chirp in Figure~\ref{sfig:chirp_fft}, as for the gravitational wave in Figure~\ref{sfig:wave_fft}, the Fourier spectrum spreads over a wide frequency range, accounting for all frequencies that emerge over time.
This shows a limitation of the Fourier spectrum as a signal representation: the Fourier spectrum only indicates whether a frequency is \emph{globally} present in the signal, but it does not give any visual hint at \emph{when} it appears. 
The time localization of each frequency event is lost.
Finally, a key property of the Fourier transform of the \emph{pointwise product} of two signals will appear useful in the next section to understand the duality between time and frequency localization.
\begin{proposition}
Let $s_1,s_2 \in L^1(\mathbb{R})$ be such that $s_1s_2, \mathcal{F}s_1, \mathcal{F}s_2 \in L^1(\mathbb{R})$, then
\begin{align}
\label{eq:prod-conv}
2\pi \, \mathcal{F}s_1s_2 = \mathcal{F}s_1 \ast \mathcal{F}s_2
\end{align}
where $\ast$ denotes the \emph{convolution product}\footnote{Let $f,g \in L^1(\mathbb{R})$,  their convolution product is defined as $f \ast g (t) = \int_{-\infty}^\infty f(u) g(t-u) \, \mathrm{d}u$.} and the equality holds in $L^1(\mathbb{R})$.
\label{prop:prod-conv}
\end{proposition}

\subsubsection{The Short-Time Fourier Transform}
\label{sssec:STFT}

The aforementioned limitation of Fourier analysis is addressed by \emph{time-frequency analysis}, which aims at representing simultaneously the temporal dynamics and the frequency content of a signal.
Recalling the analogy with music transcription mentioned in the abstract, time-frequency representations yield an overview of which frequencies, seen as musical notes, are active, or played, at a given time.
This leads to an energy map on the time-frequency plane that can be read in a similar way as a musical score,  the temporal dynamics being analogous to the rhythm, and frequency to the pitch~\cite[Section 2.1]{grochenig2001foundations}.

\begin{figure}
\centering
\begin{subfigure}{0.32\linewidth}
\centering
\includegraphics[width = 0.9\linewidth]{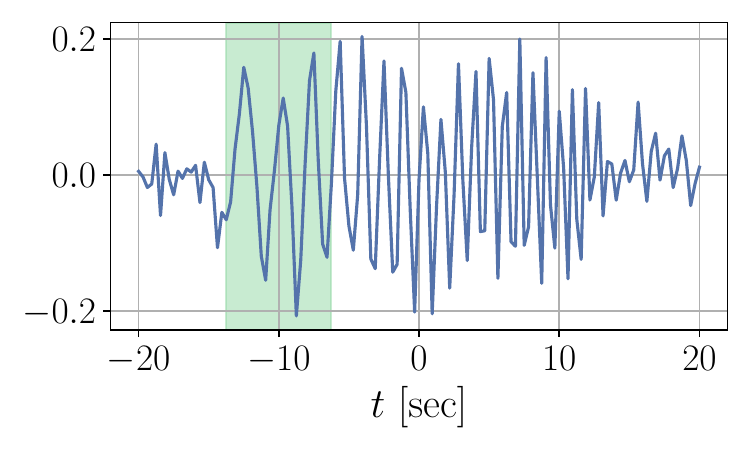}
\subcaption{\label{sfig:win1}$t = -10$ sec.}
\end{subfigure}
\begin{subfigure}{0.32\linewidth}
\centering
\includegraphics[width = 0.9\linewidth]{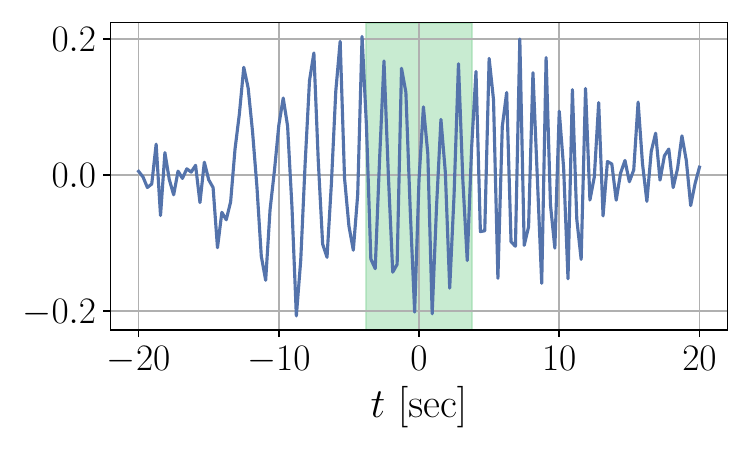}
\subcaption{\label{sfig:win2}$t=0$ sec.}
\end{subfigure}
\begin{subfigure}{0.32\linewidth}
\centering
\includegraphics[width = 0.9\linewidth]{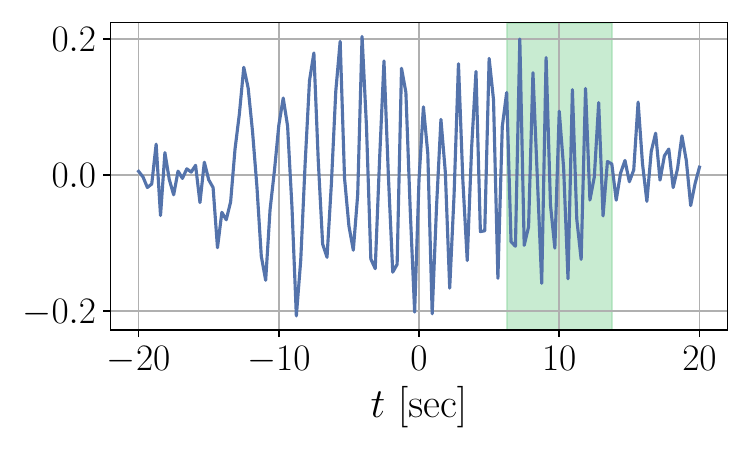}
\subcaption{\label{sfig:win3}$t = 10$ sec.}
\end{subfigure}
\caption{\label{fig:stft-moving}The Short-Time Fourier Transform seen as a moving Fourier transform. 
The window $h$ is centered at $t \in \lbrace -10, 0, 10\rbrace$. The green area is the support of $h(\cdot - t)$.}
\end{figure}

The \emph{Short-Time Fourier Transform} amounts to probe the \emph{local} frequency content of a signal $s$, where $\emph{local}$ means duing a \emph{short-time} window $h$ centered at time $t$. 
Basically, it consists in computing the Fourier transform of a small portion of the signal: in the case of the chirp, illustrated in Figure~\ref{fig:stft-moving} when the window is centered at $t = -10$ seconds, small frequencies are detected, while when analyzing the vicinity of $t=10$ seconds, one sees larger frequencies.
The narrower the window $h$, the better the temporal resolution. 
However, it is not possible to capture correctly the frequency content of a signal, in particular low frequencies, if the window is not broad enough. 
The width and shape of $h$ hence need to be chosen according to the context and goals, as discussed in Section~\ref{ssec:zerosGAF_TF}.
For now, a \emph{short-time} window is an element of $L^2(\mathbb{R})$, which is expected to be somehow \emph{localized in time}.

\begin{definition}
Let $h \in L^2(\mathbb{R})$ and $s \in L^2(\mathbb{R})$.
The Short-Time Fourier Transform of $s$ is the joint function of time and angular frequency defined by
\begin{align}
\label{eq:STFT}
\mathcal{V}_h s(t, \omega) = \int_{-\infty}^{\infty} s(u) \overline{h(u - t)} \exp( -\mathrm{i}  \omega u ) \, \mathrm{d}u, \quad  (t,\omega) \in \mathbb{R}^2.
\end{align}
The \emph{spectrogram} of a finite-energy signal $s$ is defined as the squared modulus of its Short-Time Fourier Transform $\lvert \mathcal{V}_h s(t,\omega) \rvert^2$.
\label{def:STFT}
\end{definition}

The spectrograms of the three elementary signals of Figure~\ref{fig:ex_sig} are provided in~Figure~\ref{fig:ex_spec}: for the transient wave forms, namely the chirp, Figure~\ref{sfig:chirp_spec}, and the gravitational wave, Figure~\ref{sfig:wave_spec}, the increasing and exploding instantaneous angular frequencies are very clearly visible on this joint time-frequency representation, contrary to the frequency representations provided by the Fourier spectrum of Figures~\ref{sfig:chirp_fft} and~\ref{sfig:wave_fft}.

\begin{remark}
Equation~\eqref{eq:STFT} provides a mathematical ground to the assertion that the width of the analysis window controls the \emph{temporal} resolution of the Short-Time Fourier Transform: if the support of $h$ is small, the frequency content is extracted from a small portion of the signal around $t$.
To devise how the window's width impacts the \emph{frequency} resolution,  note, first, that the right-hand side of Equation~\eqref{eq:STFT} is the Fourier transform of the \emph{pointwise} product of the signal $s$ and the conjugated and translated window $\overline{h(\cdot - t)}$. 
Then, by Proposition~\ref{prop:prod-conv}, the Short-Time Fourier Transform rewrites as the \emph{convolution} product of the Fourier transforms of the signal and of the conjugated translated window:
\begin{align*}
2\pi \, \mathcal{V}_h s(t, \omega) = \mathcal{F}_{s} \ast \mathcal{F}_{\overline{h(\cdot-t)}},
\end{align*}
where, from direct computations, $\forall \omega \in \mathbb{R}, \, \mathcal{F}_{\overline{h(\cdot-t)}}(w) = \exp(-\mathrm{i}\omega t) \overline{\mathcal{F}_{h}(-\omega)}$.
Altogether, this shows that the \emph{frequency} resolution of the Short-Time Fourier Transform is controlled by the width of the Fourier Transform of the analysis window.
But, as exemplified on the Gaussian window in Proposition~\ref{prop:fourier-window}, if $h$ has a small support, then its Fourier transform is widely spread, and vice versa. Thus, choosing the width of the analysis window involves a trade-off between the \emph{time} and \emph{frequency} resolutions, which cannot be simultaneously set to arbitrarily small values. This trade-off, colloquially known as the \emph{Heisenberg uncertainty principle}, is a core element of time-frequency analysis, explained in details in~\cite[Chapter 2]{grochenig2001foundations} and \cite[Chapters 5 and 7]{flandrin2018explorations}.
It is worth noting that the results presented in Section~\ref{sec:GAF}, and the ensuing signal processing procedures designed in Section~\ref{sec:sig_spat_stat},  require the window to be Gaussian with \emph{unit} variance, which corresponds to equal time and frequency resolutions.
\end{remark}

The Short-Time Fourier Transform benefits from particularly useful properties, part of them inherited from Fourier analysis.

\begin{proposition}
For any two windows $h, h' \in L^2(\mathbb{R})$ of unit energy, $\lVert h \rVert_2 =\lVert h' \rVert_2=1$, and any two signals $s,s' \in L^2(\mathbb{R})$, the following holds.
\begin{enumerate}
\item The Short-Time Fourier Transform maps the inner product of $L^2(\mathbb{R})$ onto the one on $L^2(\mathbb{R}^2)$, and hence preserves the energy:
\begin{align}
\int_{\mathbb{R}} \int_{\mathbb{R}} \mathcal{V}_h s(t,\omega) \overline{\mathcal{V}_h s'(t,\omega)} \, \mathrm{d}t \frac{\mathrm{d}\omega}{2\pi} =  \int_{\mathbb{R}} s(t) \overline{s'(t)} \, \mathrm{d}t.
\end{align}
\item It induces a Reproducing Kernel Hilbert Space, meaning that
\begin{align}
\label{eq:RKHS}
    \forall (t,\omega) \in \mathbb{R}^2, \quad \mathcal{V}_h s(t, \omega) = \int_{\mathbb{R}} \int_{\mathbb{R}} \mathsf{K}(t,\omega;t', \omega') \mathcal{V}_h s(t',\omega') \, \mathrm{d}t' \frac{\mathrm{d}\omega'}{2\pi},
\end{align}
where the kernel $\mathsf{K}$ is defined by $\mathsf{K}(t,\omega ; t', \omega') = \mathcal{V}_h h (t-t', \omega - \omega')$.
\item If $\langle h , h'\rangle \neq 0$, then the following reconstruction formula is satisfied:
\begin{align}
s(t') =\frac{1}{\langle h, h'\rangle} \int_{\mathbb{R}} \int_{\mathbb{R}} \mathcal{V}_{h} s (t, \omega)  h'(t' - t) \exp(\mathrm{i}\omega t) \, \mathrm{d}t  \frac{\mathrm{d\omega}}{2\pi} \quad \text{a.e.},
\end{align}
where $a.e.$ means for Lebesgue-almost all $t'\in\mathbb{R}$.
\end{enumerate}
\label{prop:STFT}
\end{proposition}

\begin{remark}
The Reproducing Kernel Hilbert Space property \eqref{eq:RKHS} shows that an arbitrary element of $L^2(\mathbb{R}^2)$ is not necessarily the Short-Time Fourier Transform of a finite-energy signal.
On the contrary, $\mathcal{V}_h(t,\omega)$ can be reconstructed from the values of $\mathcal{V}_h(t',\omega')$ for $(t',\omega')$ in a neighborhood of $(t,\omega)$, of diameter roughly the width of the analyzing window. 
From a geometrical point of view, this means that the landscape of a spectrogram is highly constrained, and in a certain sense, smooth.
\end{remark}

A careful reader might have noticed that, although the Short-Time Fourier Transform introduced in Equation~\eqref{eq:STFT} is defined for angular frequencies $\omega \in \mathbb{R}$, the spectrograms of Figures~\ref{fig:ex_spec} and~\ref{sfig:wnoise_spec} are displayed only for non-negative angular frequencies. This is motivated by the following symmetry result.

\begin{proposition}
Let $h, s\in L^2(\mathbb{R})$ be both \emph{real-valued}.
Then,
\begin{align}
\forall (t,\omega) \in \mathbb{R}^2, \quad \overline{\mathcal{V}_h s(t, \omega)} = \mathcal{V}_h s(t, -\omega).
\end{align}
In particular, the spectrogram  is symmetric with respect to the time axis.
\end{proposition}

In Figure~\ref{fig:ex_spec}, both the Gaussian window used in the Short-Time Fourier Transform and the signals of interest are real-valued.
Hence the corresponding spectrograms are symmetric under frequency flip.
In particular, the curve of maxima that is the signature of the time-varying angular frequency is exactly mirrored in the region $\omega<0$, justifying to focus on $\omega\geq 0$ to avoid redundancy.
Spectrograms covering positive and negative frequencies are exemplified later in the case of Hermite functions in~Figures~\ref{sfig:hk_spec} and~\ref{sfig:thk_spec}.

\begin{remark}
    It is worth noting that the zeros on Figures~\ref{sfig:hk_spec} and~\ref{sfig:thk_spec} correspond to a complex-valued signal, since the noise is complex-valued in the signal-plus-noise model of Section~\ref{sssec:signalnoise}.
    This explains why the zeros on these two figures are not symmetric with respect to the time axis.
    Still, as will be shown in Section~\ref{ssec:zerosGAF_TF}, the \emph{distribution} of the zeros of the spectrogram of white noise is \emph{stationary}, that is, it is invariant under translations of the time-frequency plane. 
\end{remark}

\subsubsection{The family of Hermite functions}
\label{sssec:hermite}
The family of \emph{Hermite functions} will be of utmost importance to construct the continuous white noise $\wn$ in Section~\ref{ssec:white_noise}, and to link its Short-Time Fourier Transform to a Gaussian Ananlytic Function in Section~\ref{ssec:zerosGAF_TF}.
Two major properties of Hermite functions  will be used: first, they form a Hilbert basis of the space of finite-energy signals, second their Gaussian spectrograms have very simple forms.

\begin{definition}
    Let $k \in \mathbb{N}$, the Hermite function of order $k$ is
    \begin{align}
    \label{eq:def-hermite} h_k(t) = \frac{(-1)^k}{\sqrt{\sqrt{\pi}2^k k!}} \, \exp(t^2/2) \dfrac{\mathrm{d^k}}{\mathrm{d}t^k}  \mathrm{e}^{-t^2}, \quad t\in\mathbb{R}.
    \end{align}
    \label{def:hermite}
\end{definition}

\begin{proposition}
The family of Hermite functions $(h_k)_{k\in\mathbb{N}}$ is a \emph{Hilbert basis} of $L^2(\mathbb{R})$.
In particular, Hermite functions form an orthogonal family, that is
\begin{align}
    \forall k,\ell \in \mathbb{N}, \quad \int_{\mathbb{R}} h_k(t) h_{\ell} (t) \, \mathrm{d}t = \delta_{k,\ell},
\end{align}
where $\delta$ denotes the \emph{Kronecker delta}, satisfying $\delta_{k,\ell} = 0$ if $k\neq \ell$ and $\delta_{k,k} = 1$.
\end{proposition}

\begin{figure}
\centering
\includegraphics[width = 7.5cm]{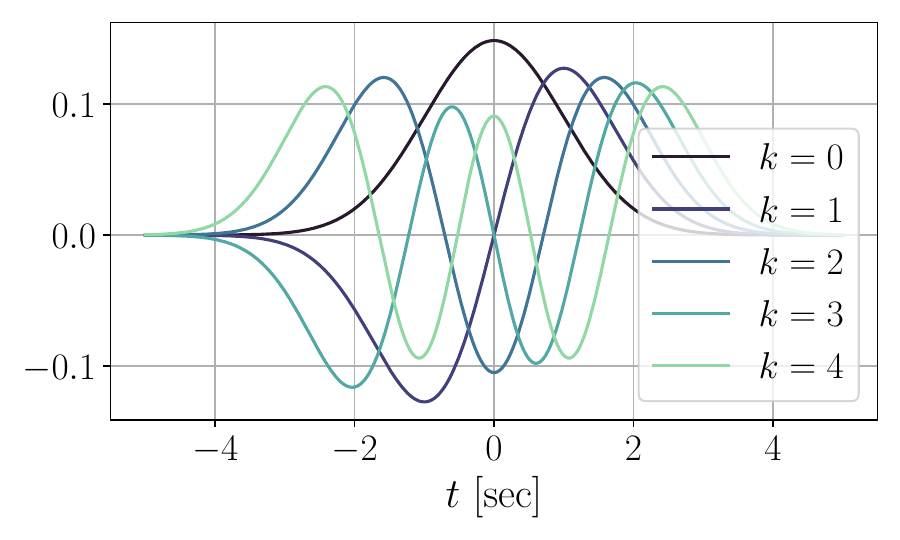}
\caption{\label{fig:hermite_first}Hermite functions of order $k=0$ to $k=4$.}
\end{figure}

To illustrate Definition~\ref{def:hermite}, the first five Hermite functions are displayed in Figure~\ref{fig:hermite_first}.
When $k$ is even (resp. odd), so is the Hermite function. It can further be shown that the Hermite function of order $k$ has exactly $k$ distinct zeros.
The larger $k$, the more $h_k$ oscillates.  
As an example, the Hermite function of order $k=16$ with a small amount of additional noise is displayed in the top plot of Figure~\ref{sfig:hk-thk}. 

\begin{proposition}
Let $g(t) = \pi^{-1/4}\exp(-t^2/2)$ be the so-called circular Gaussian window and $k \in \mathbb{N}$.\footnote{The terminology \emph{circular} comes from the \emph{ambiguity function} of $g$ which writes
\begin{align*}
\mathcal{A}g(\xi,\tau) = \int_{-\infty}^\infty g\left(s + \frac{\tau}{2}\right) \overline{g\left(s - \frac{\tau}{2}\right)} \exp(\mathrm{i} \xi s ) \, \mathrm{d}s = \exp \left( -\frac{1}{4}(\xi^2 + \tau^2)\right),
\end{align*}
and whose level set are circles in the time-frequency plane $(\xi,\tau)$.} 
The Short-Time Fourier Transform with analyzing window $g$ of the Hermite function of order $k$ is given by
\begin{align}
    \forall (t,\omega)\in \mathbb{R}^2, \quad\mathcal{V}_g h_k(t,\omega) =  \frac{(\omega + \mathrm{i}t)^k}{\sqrt{2^k k!}} \exp\left( -\frac{(t^2+\omega^2)}{4} - \frac{\mathrm{i}\omega t}{2}\right).
\end{align}
Hence, as stated in~\cite[Equation (15)]{flandrin2013note}, the Gaussian spectrogram of $h_k$ writes
\begin{align}
\label{eq:hk-spec}
\forall (t,\omega)\in \mathbb{R}^2, \quad\left\lvert \mathcal{V}_g h_k(t,\omega) \right\rvert^2= \frac{1}{2^k k!} \left( t^2 + \omega^2\right)^k \exp\left(-\frac{1}{2}(t^2+\omega^2)\right).
\end{align}
\label{prop:hermite-ring}
\end{proposition}

Maximizing w.r.t. $(t,\omega)$ the explicit expression~\eqref{eq:hk-spec} shows that the spectrogram of the Hermite function of order $k \in \mathbb{N}^*$ admits a \emph{ring} of maxima, centered at the origin of the time-frequency plane and of radius $\sqrt{2k}$.
This is illustrated in Figure~\ref{sfig:hk_spec}, which presents the Gaussian spectrogram of the Hermite function of order $k=16$, with a small amount of additive noise according to the signal-plus-noise model~\eqref{eq:signalnoise}.

\subsubsection{Covariance of the Short-Time Fourier Transform}
\label{sssec:covariance}

\begin{figure}
    \centering
    \begin{subfigure}{0.32\linewidth}
    \centering
    \raisebox{6.5mm}{\includegraphics[width = \linewidth]{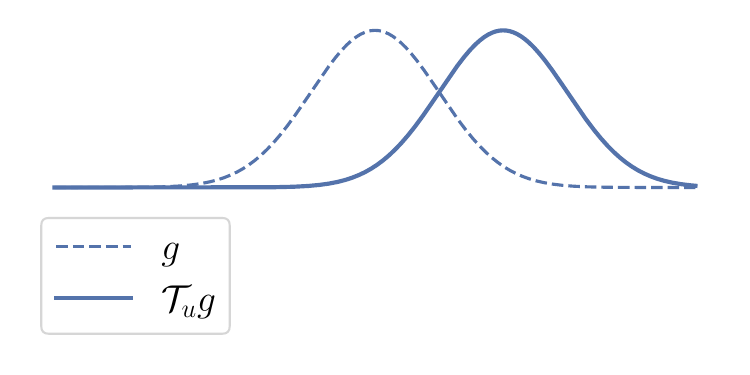}}
    \end{subfigure} \hspace{0.11\linewidth}
    \begin{subfigure}{0.32\linewidth}
    \centering
    \raisebox{6.5mm}{\includegraphics[width = \linewidth]{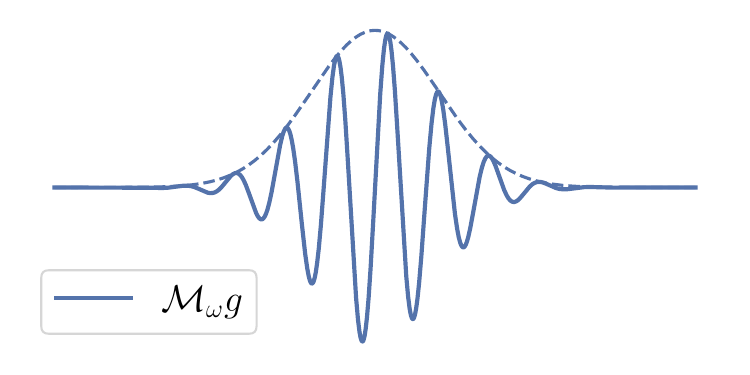}}
    \end{subfigure}
    \caption{\label{fig:MvTu}Time translation and frequency modulations of the circular Gaussian window $g(t) = \pi^{-1/2}\exp(-t^2/2)$.}
\end{figure}

\begin{figure}
    \centering
    \begin{subfigure}[t]{0.32\linewidth}
    \centering
    \includegraphics[width = 0.9\linewidth]{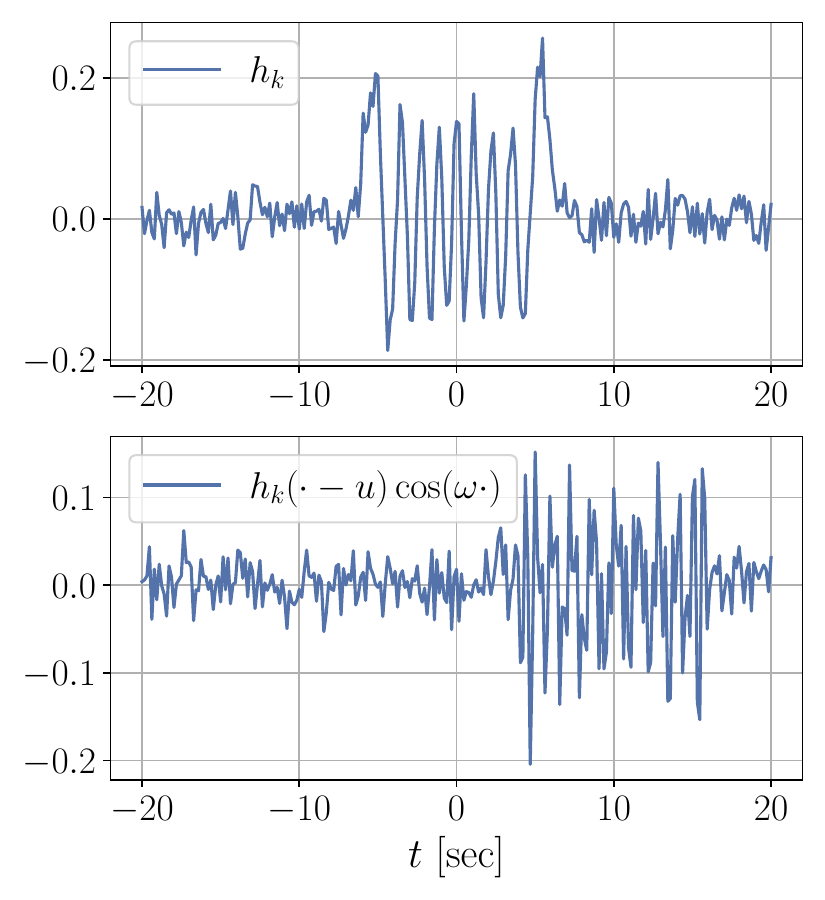} 
    \subcaption{\label{sfig:hk-thk}$h_k$ and $\mathcal{M}_{\omega}\mathcal{T}_uh_k$.}
    \end{subfigure}
    \begin{subfigure}[t]{0.32\linewidth}
    \centering
    \raisebox{5mm}{\includegraphics[width = \linewidth]{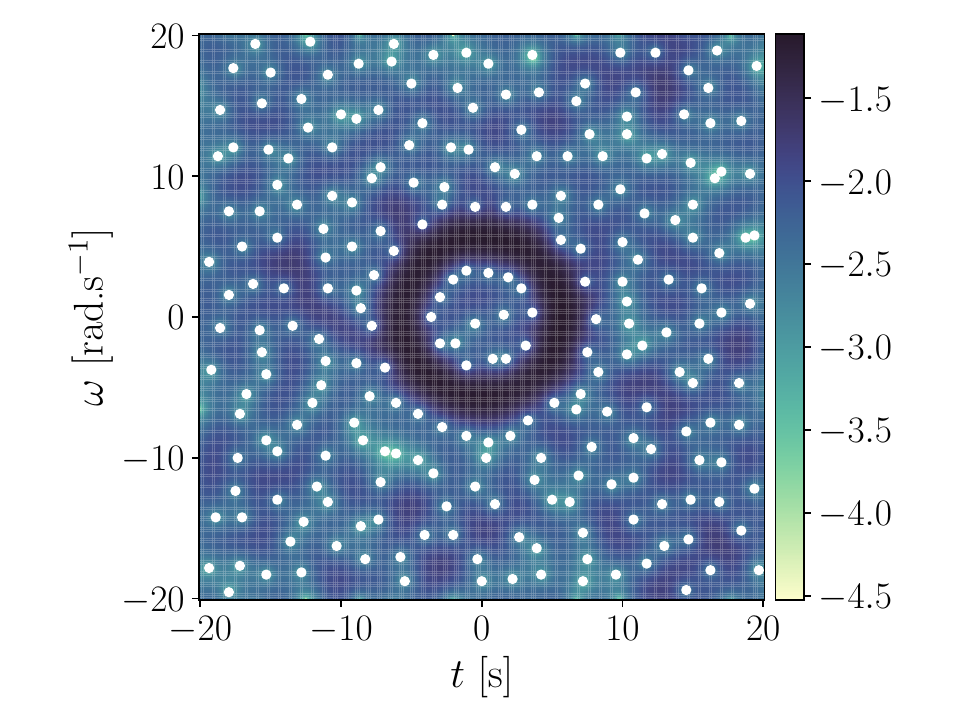}}
    \subcaption{\label{sfig:hk_spec}Spectrogram of $h_k$.}
    \end{subfigure}
    \begin{subfigure}[t]{0.32\linewidth}
    \centering
    \raisebox{5mm}{\includegraphics[width = \linewidth]{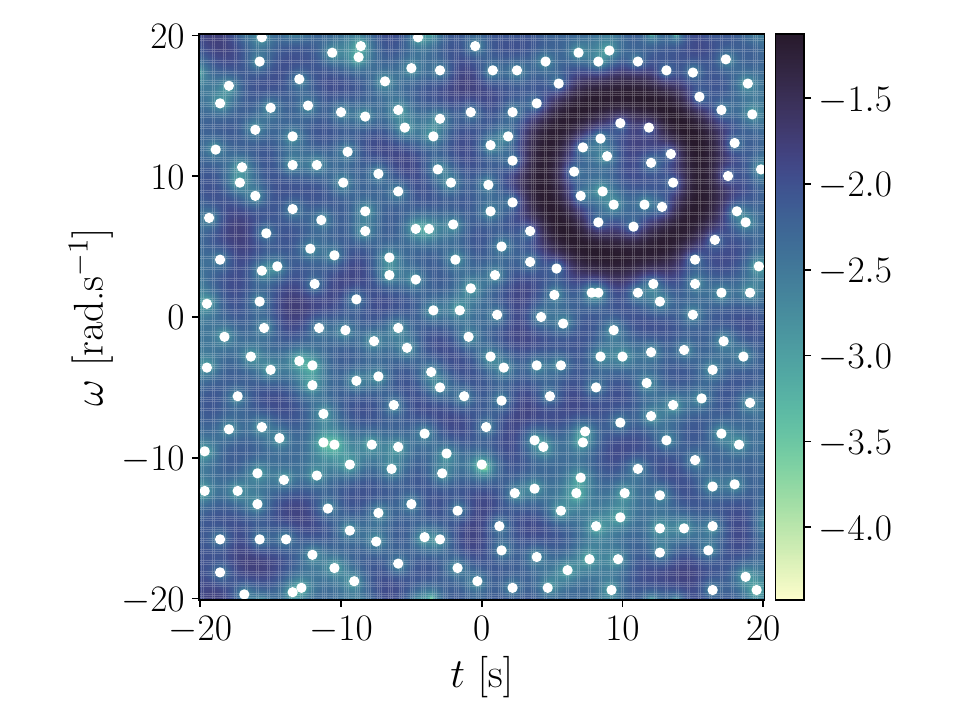}}
    \subcaption{\label{sfig:thk_spec}Spectrogram of $\mathcal{M}_{\omega}\mathcal{T}_uh_k$.}
    \end{subfigure}
    \caption{\label{fig:hermite}Covariance of the spectrogram exemplified for the Hermite function $h_k$ with $k = 16$, displayed in top plot of (a) and associated spectrogram in (b), and the modulated and translated Hermite function $\mathcal{M}_{\omega}\mathcal{T}_uh_k$ with $u = 10$ [sec] and $\omega = 10$ [rad.s$^{-1}$], displayed in bottom plot of (a) and associated spectrogram in (c).}
\end{figure}

The Short-Time Fourier Transform of Definition~\ref{def:STFT} can be reinterpreted as the decomposition of a finite-energy signal on a family of \emph{atoms}, obtained by applying elementary operations on the analyzing window.
Namely, these operations consist in time translations and frequency modulations, as displayed in Figure~\ref{fig:MvTu} in the case when the analyzing window is the circular Gaussian window.

\begin{proposition}
    Let $(u,\omega) \in \mathbb{R}^2$, and consider the \emph{translation} and \emph{modulation} endomorphisms $\mathcal{T}_u : L^2(\mathbb{R}) \rightarrow L^2(\mathbb{R})$ and $\mathcal{M}_{\omega}: L^2(\mathbb{R}) \rightarrow L^2(\mathbb{R})$, defined by
    \begin{align}
    \forall h \in L^2(\mathbb{R}),\, \forall t \in \mathbb{R}, \quad \left( \mathcal{T}_u h \right)(t) = h(t-u), \quad \left( \mathcal{M}_{\omega} h \right)(t) = h(t) \mathrm{e}^{-\mathrm{i} \omega t }.
    \end{align}
    Then the Short-Time Fourier Transform of $s \in L^2(\mathbb{R})$ rewrites
    \begin{align}
    \forall (t,\omega) \in \mathbb{R}^2, \quad \mathcal{V}_h s (t,\omega) = \langle s, \mathcal{M}_{\omega} \mathcal{T}_t h\rangle
    \end{align}
\end{proposition}

The covariance of the spectrogram now means that if one translates and modulates the analyzed signal, its spectrogram is a translated version of the original one.

\begin{proposition}
Let $h \in L^2(\mathbb{R})$ and $s \in L^2(\mathbb{R})$. 
For any time-shift $u\in \mathbb{R}$ and modulation frequency $\omega' \in \mathbb{R}$,
\begin{align}
    \forall (t,\omega) \in \mathbb{R}^2, \quad \mathcal{V}_h [\mathcal{M}_{\omega'} \mathcal{T}_u s](t, \omega) = \mathrm{e}^{-\mathrm{i}(\omega-\omega')u} \mathcal{V}_h s(t-u,\omega-\omega'),
\end{align}
Hence, the spectrogram is \emph{covariant} under time-shifts and frequency modulations:
\begin{align}
\forall (t,\omega) \in \mathbb{R}^2, \quad \left\lvert \mathcal{V}_h [\mathcal{M}_{\omega'} \mathcal{T}_u s](t, \omega)\right\rvert^2 = \left\lvert  \mathcal{V}_h s(t-u,\omega-\omega')\right\rvert^2.
\end{align}
\end{proposition}

This property is illustrated in Figure~\ref{fig:hermite}, where the spectrogram of the original noisy Hermite function, displayed in Figure~\ref{sfig:hk_spec}, presents a ring of maxima centered at the origin, while the spectrogram of the translated and modulated noisy Hermite function, in Figure~\ref{sfig:thk_spec}, presents the same overall geometry,  though shifted: the ring of maxima is translated of $u$ seconds in time and $\omega$ rad.s$^{-1}$ in frequency.

\subsection{Numerical implementation}
\label{ssec:num_impl}

While it is handy in theory to identify signals to functions of a real variable, in practice, one never has access to the full observed function $y:t \in \mathbb{R}\mapsto y(t) \in \mathbb{\mathbb{C}}$, but only to a finite number of measurements $\boldsymbol{y} = (y_1, \hdots, y_n) \in \mathbb{C}^n$.
It is often implicitly assumed that these measurements are regular evaluations of an underlying function during a finite time interval, at angular frequency $\omega_s = 2\pi/\delta t$ for some time step $\delta t > 0$, i.e., $y_k = y(k\cdot \delta t)$, $k = 1, \hdots, n$.
The Continuous Fourier Transform is usually replaced in practice by the discrete Fourier transform (DFT).

\begin{definition}
    For $n\in\mathbb{N}^*$, let $\boldsymbol{s} = (s_1, \hdots, s_n) \in \mathbb{C}^n$ be a \emph{discrete} signal.
    The \emph{Discrete Fourier Transform} of $\boldsymbol{s}$ is a vector $\widehat{\boldsymbol{s}} \in \mathbb{C}^n$,  defined by
    \begin{align}
        \label{eq:dft}
        \forall k \in \lbrace 0, \hdots, n-1\rbrace, \quad  \widehat{s}_{k+1} = \sum_{j = 1}^{n} s_j \exp\left(-2\mathrm{i}\pi k \cdot \frac{j - 1}{n}\right).
    \end{align}
    Similarly, the \emph{Discrete Fourier spectrum} of a discrete signal is defined as the componentwise squared modulus of its Discrete Fourier Transform, the $k$-th component quantifying the global energy of the signal at angular frequency $2 \pi k/n \cdot \delta t^{-1}$.
    \label{def:dft}
\end{definition}

One major advantage of \eqref{eq:dft} is that there exists a fast algorithm, namely the Fast Fourier Transform algorithm, that computes $\widehat{\boldsymbol{s}}\in\mathbb{C}^n$ in $O(n\ln(n))$ operations.
In comparison, the naive approach that successively computes~\eqref{eq:dft} for each $k = 0, \hdots, n-1$ requires $O(n^2)$ operations.

\begin{definition}
    For $n\in\mathbb{N}^*$, let $\boldsymbol{s} = (s_1, \hdots, s_n) \in \mathbb{C}^n$ a \emph{discrete} signal.
    For $m\in\mathbb{N}^*$, the    \emph{Discrete Short-Time Fourier Transform} of $\boldsymbol{s}$ is a matrix $\mathrm{V} \boldsymbol{s} \in \mathbb{C}^{n\times m}$, whose elements write, for $ k \in \lbrace 0, \hdots, n-1\rbrace, \, \ell \in \lbrace 0, \hdots, m-1\rbrace$,
    \begin{align}
    \label{eq:dstft}
    \left( \mathrm{V}\boldsymbol{s}\right)_{k,\ell}= \sum_{j = 1}^{n} s_j \overline{h\left(k\cdot \delta t - j \cdot \delta t\right)}\exp\left(-\mathrm{i} ( \ell \cdot \delta \omega )( (j-1)\cdot \delta t )\right) \, \delta t.
    \end{align}
    The \emph{discrete spectrogram} is defined as the squared modulus of the discrete Short-Time Fourier Transform, and its $(k,\ell)$-element describes the energy density at time $k\cdot \delta t$ and angular frequency $ \ell  \cdot \delta \omega$~\cite{oppenheim1999discrete}.
    \label{def:dstft}
\end{definition}

Provided that the time and frequency discretizations are compatible, i.e., that $\delta t \times \delta \omega = 2\pi n$ for some $n \in \mathbb{N}^*$,
the \emph{Discrete} Short-Time Fourier Transform can be implemented efficiently in $\mathcal{O}(n^2\log n)$ operations by leveraging the Fast Fourier Transform algorithm into a sliding algorithm~\cite{jacobsen2003sliding}.

\subsection{From random processes to a rigorous definition of white noise}
\label{ssec:white_noise}

Going further in the analysis of noisy signals $t\mapsto y(t)$ in the signal-plus-noise model~\eqref{eq:signalnoise} requires to provide a rigorous mathematical description of \emph{noise}.
The noise $t\mapsto \wn(t)$, exemplified in Figure~\ref{sfig:wnoise}, is a random process with no temporal structure, unlike the underlying signals of interest $s(t)$ such as pure sines, linear chirps or gravitational waves of Figure~\ref{fig:ex_sig}.
The ideal noise, which is ubiquitous in signal processing, should have three major properties: \textit{i)}~for each $t$, $\wn(t)$ has zero expectation, also known as the \emph{unbiased measurements hypothesis}; \textit{ii)}~at two different times $t$ and $t'$, the random variables $\wn(t)$ and $\wn(t')$ are uncorrelated, the so-called \emph{independent measurements hypothesis}; \textit{iii)}~$\wn(t)$ is a Gaussian variable, whose variance does not depend on $t$, a property known as \emph{homoscedasticity}.
Such a random process $\wn$ is called \emph{white noise}. 
The purpose of this section is to discuss the existence and the construction of white noise in both the discrete, finite-dimensional, setting and in the continuous, infinite-dimensional, setting, and the links between the two.

The signals considered in this chapter are \emph{complex}-valued, and hence this section describes the construction of complex white noise.
Not only the complex framework is more general, but also, as will be thoroughly discussed in Section~\ref{sec:GAF}, it yields a wealth of symmetries, which can then be leveraged efficiently to design signal processing procedures presented in Section~\ref{sec:sig_spat_stat}.
Though, it is worth noting that it is possible to consider \emph{real}-valued signals contaminated by real-valued white noise, as discussed in~\cite[Section~3]{bardenet2020zeros}.

\subsubsection{Discrete white noise as a Gaussian vector}
\label{sssec:gauss-vect}

First of all, a \emph{complex} analogue of the real standard Gaussian distribution is needed. 
If $x,y  \sim \mathcal{N}(0,1)$ are independent real standard Gaussian random variables, then 
\begin{align}
\label{eq:NC_01}
z = \frac{x + \mathrm{i}y}{\sqrt{2}} \sim \mathcal{N}_{\mathbb{C}}(0,1) 
\end{align}
defines a standard complex Gaussian random variable,  whose distribution is denoted $\mathcal{N}_{\mathbb{C}}(0,1)$.
Equivalently, the standard complex Gaussian distribution is characterized by its density with respect to Lebesgue on $\mathbb{C}$, given by $\pi^{-1} \mathrm{e}^{-\lvert z \rvert^2}$.
The factor $1/\sqrt{2}$ in Equation~\eqref{eq:NC_01} ensures that the variance of the complex variable $z$, defined as $\mathbb{V}[z] = \mathbb{E}\left[\lvert z\rvert^2\right] - \left\lvert \mathbb{E}[z]\right\rvert^2$, is equal to one.
It can further be checked that, if $z\sim\mathcal{N}_\mathbb{C}(0,1)$, then $\mathbb{E}[z] = \mathbb{E}[zz] = \mathbb{E}[\overline{z}\overline{z}] = 0$, and $\mathbb{E}[z\overline{z}] = 1$.

Once the standard complex Gaussian is defined, it is possible to extend it to obtain complex Gaussian vectors,  generalizing real Gaussian vectors.
Let $n \in \mathbb{N}^*$ and $\wn_1, \dots, \wn_n$ denote i.i.d. standard \emph{complex} Gaussian random variables, stacked into a vector $\boldsymbol{\wn}\in \mathbb{C}^n$, and consider $\boldsymbol{s}\in\mathbb{C}^n$, $\textbf{B}\in\mathbb{C}^{n \times n}$ and $\textbf{C} = \textbf{BB}^*$.
The distribution of the random vector
\begin{align}
\label{eq:gauss-vect}
\boldsymbol{y} = \boldsymbol{s}+\textbf{B} \boldsymbol{\wn}
\end{align}
is the complex multivariate Gaussian distribution of mean $\boldsymbol{s}$ and covariance $\textbf{C}$, shortened as $\boldsymbol{y}\sim\mathcal{N}_{\mathbb{C}}(\boldsymbol{s}, \textbf{C})$.
Complex Gaussian vectors behave as their real counterparts in all basic respects \cite[Section 2.1]{hough2009zeros}.
Note that they are sometimes alternatively called \emph{circular Gaussian vectors}, e.g., in~\cite{haimi2020zeros}, in order to emphasize the stability of the ensemble of Gaussian vectors under unitary transforms of $\mathbb{C}^n$.

\begin{definition}
Let $n\in \mathbb{N}^*$,  $\boldsymbol{0}_n$ the zero-vector of length $n$ and $\textbf{I}_{n\times n}$ the identity matrix of size $n$, the \emph{finite-dimensional complex white noise} is the Gaussian vector of distribution~$\mathcal{N}_{\mathbb{C}}(\boldsymbol{0}_n, \textbf{I}_{n\times n})$.
Equivalently, given an orthonormal basis $(\boldsymbol{e}_k)_{k = 1}^n$ of $\mathbb{C}^n$ and $\wn_1,\hdots,\wn_n$ i.i.d. standard complex Gaussian random variables,
\begin{align}
    \label{eq:wn-finite-sum}
    \boldsymbol{\wn} = \sum_{k=1}^n \wn_k \boldsymbol{e}_k.
\end{align}
\label{def:finite-dim-wn}
\end{definition}

\begin{remark}
The distribution $\mathcal{N}_{\mathbb{C}}(\boldsymbol{0}_n, \textbf{I}_{n\times n})$ is invariant under unitary transforms of $\mathbb{C}^n$.
In particular, the law of the complex Gaussian vector in Equation~\eqref{eq:wn-finite-sum} is independent of the choice of the orthonormal basis.
\end{remark}

\begin{figure}
\centering
\begin{subfigure}[t]{0.32\linewidth}
\centering
\raisebox{2mm}{\includegraphics[width = 0.9\linewidth]{wnoise.pdf}}
\vspace{-4mm}
\subcaption{\label{sfig:wnoise2}Real part of white noise.}
\end{subfigure}
\begin{subfigure}[t]{0.32\linewidth}
\centering
\raisebox{2mm}{\includegraphics[width = 0.9\linewidth]{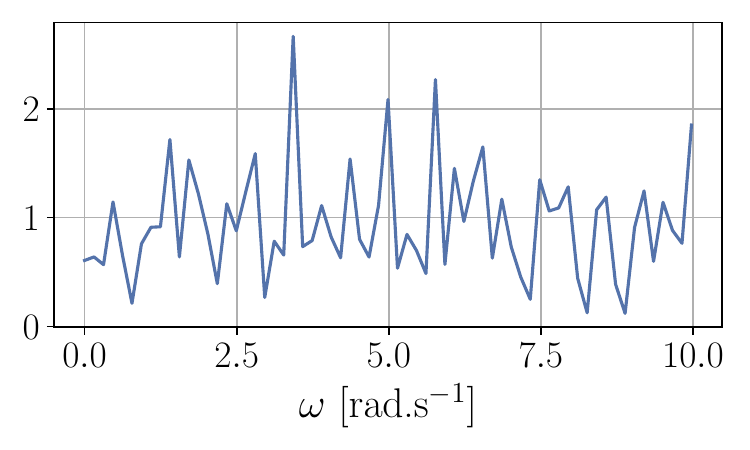}}
\vspace{-4mm}
\subcaption{\label{sfig:wnoise_fft2}Fourier spectrum.}
\end{subfigure}
\begin{subfigure}[t]{0.32\linewidth}
\centering
\includegraphics[width = \linewidth]{wnoise_spec.pdf}
\vspace{-7.5mm}
\subcaption{\label{sfig:wnoise_spec2}Gaussian spectrogram.}
\end{subfigure}
\caption{\label{fig:wnoise2}Time and frequency vs. time-frequency analysis of white noise.}
\end{figure}

As an illustration, the real part of a realization of complex white noise in dimension $n=128$ is displayed in Figure~\ref{sfig:wnoise2}. From a signal processing point of view it models the measurement noise, completely uncorrelated from one sampling time to another. 
As the Discrete Fourier Transform, introduced in Definition~\ref{def:dft}, consists in a change of orthonormal basis, the Discrete Fourier transform of the complex white noise is the complex white noise itself.\footnote{
    Up to the normalizing constant $1/\sqrt{n}$ which ensures unit variance when added to~\eqref{eq:dft}. 
}
Thus, its expected Fourier spectrum, the second moment of each component, is constant: all frequencies are equally represented. 
Figure~\ref{sfig:wnoise_fft2} presents the discrete Fourier spectrum of the white noise realization of Figure~\ref{sfig:wnoise2}:
it randomly oscillates around a constant value.
More informally for the moment, this absence of deterministic structure is also visible in the Gaussian spectrogram of Figure~\ref{sfig:wnoise_spec2}: the Short-Time Fourier Transform of white noise is a smooth Gaussian random field with constant mean, and its zeros are very evenly distributed in the time-frequency plane.

Finally, to prepare the ground for the infinite-dimensional white noise,  note that Definition~\ref{def:finite-dim-wn} is equivalent to specifying a Gaussian characteristic function for the random vector $\boldsymbol{\wn}$.
\begin{proposition}
    The random vector $\boldsymbol{\wn}\in\mathbb{C}^n$ is a finite-dimensional Gaussian white noise if and only if
    \begin{align}
        \label{eq:FT_wn}
        \forall \boldsymbol{s} \in \mathbb{C}^n, \quad  \mathbb{E}_{\boldsymbol{\wn}} [\exp(\mathrm{i}\langle \boldsymbol{\wn}, \boldsymbol{s}\rangle )] = \exp(-\lVert \boldsymbol{s}\rVert^2/2),
    \end{align}
    where $\langle \boldsymbol{\cdot}, \boldsymbol{\cdot} \rangle$ (resp. $\lVert \boldsymbol{\cdot} \rVert$) is the Euclidean inner product (resp. norm) of $\mathbb{C}^n$.
\end{proposition}

\subsubsection{White noise in an infinite-dimensional space}
\label{sssec:inf-wn}
Going further in the analysis of continuous \emph{noisy} signals $t\mapsto \snr\times s(t) +\wn(t)$, described in Section~\ref{sssec:signalnoise}, requires a rigorous definition of the noise process $\wn(t)$, and hence to extend Definition~\ref{def:finite-dim-wn} to the continuous setting, to get an \emph{infinite-dimensional} white noise.
Although considering the Hilbert space $L^2(\mathbb{R})$ of continuous finite-energy signals would be enough for this chapter, the construction of infinite-dimensional white noise will be presented in the general case of a separable Hilbert space of signals, as it does not imply further technicality and will allow easier connections to more recent work in Section~\ref{sec:extensions}.
A thorough introduction to Gaussian measures on Banach spaces can be found in~\cite[Chapter 8]{stroock2010probability}.

Let $\mathfrak{H}$ be a complex separable Hilbert space, with norm $\lVert \cdot \rVert_{\mathfrak{H}}$, and $(\atom_k)_{k\in\mathbb{N}}$ a Hilbert basis of $\mathfrak{H}$.  
Mimicking the above definition of discrete white noise, it would be tempting to define infinite-dimensional white noise as an element $\wn$ of the Hilbert space satisfying, $\forall k \in \mathbb{N},  \langle \wn, \atom_k \rangle = \wn_k$ with $(\wn_k)_{k\in \mathbb{N}}$ an infinite sequence of i.i.d. standard complex Gaussian random variables, or equivalently as the series
\begin{align}
\label{eq:wn-series}
\wn = \sum_{k\in \mathbb{N}} \wn_k \atom_k.
\end{align}
However, as stated in~\cite[Section 3.2.2]{bardenet2021time}, with probability one, the series~\eqref{eq:wn-series} diverges in~$\mathfrak{H}$.
The solution proposed by~\cite{gross1967abstract}, and used in~\cite[Section~3.2]{bardenet2021time}, consists in considering a weaker norm $\lVert \cdot \rVert_{\mathfrak{G}}$ on $\mathfrak{H}$, defined as
\begin{equation}
\label{eq:weak-norm}
\forall s \in \mathfrak{H}, \quad \lVert s \rVert_{\mathfrak{G}}^2:=\sum_{k\in\mathbb{N}}\frac{\left\lvert\langle s,\atom_k\rangle\right\rvert^2}{1+k^2}
\end{equation}
enabling to define an extension $\mathfrak{G}$ of $\mathfrak{H}$ in which the convergence of~\eqref{eq:wn-series} is ensured.

\begin{proposition}
Let $\lVert \cdot \rVert_{\mathfrak{G}}$ be the norm on $\mathfrak{H}$ defined in \eqref{eq:weak-norm}, $\mathfrak{G}$ the completion of $\mathfrak{H}$ induced by $\lVert \cdot \rVert_{\mathfrak{G}}$,  and $(\atom_k)_{k\in \mathbb{N}}$ a \emph{fixed} Hilbert basis of $\mathfrak{H}$.
Then, for $(\wn_k)_{k\in\mathbb{N}}$ a sequence of i.i.d. standard complex Gaussian random variables, the series of Equation~\eqref{eq:wn-series} converges almost surely in $\mathfrak{G}$.
The limit $\xi$ of the series is called \emph{white noise}, with distribution $\mu$.
Moreover, 
\begin{equation}
\label{FT}
\forall \chi\in\mathfrak{G}^*, \quad \mathbb{E}_{\wn\sim\mu} \left[\exp (\i \langle \wn, \chi \rangle )\right]=\exp\left(-\lVert h_\chi\rVert_{\mathfrak{H}}^2\big/2\right),
\end{equation}
where $\mathfrak{G}^*$ is the topological dual of $\mathfrak{G}$, $\langle \cdot, \cdot \rangle$ is the natural pairing between $\mathfrak{G}$ and its dual,  and $h_\chi\in \mathfrak{H}$ is associated to $\chi\in \mathfrak{G}^*\subset \mathfrak{H}^*$ by Riesz's representation theorem.\footnote{
    By construction, $\mathfrak{H} \subset \mathfrak{G}$, thus $\mathfrak{G}^* \subset \mathfrak{H}^*$, justifying the existence of a representing function in $\mathfrak{H}$.
}
\label{prop:white-noise}
\end{proposition}

Several comments are in order. 
First, Equation~\eqref{FT} is an extension of the finite-dimensional characteristic function \eqref{eq:FT_wn}.
Second, the convergence of the series is a consequence of the definition of the norm $\lVert \cdot \rVert_{\mathfrak{G}}$ in Equation~\eqref{eq:weak-norm} and the fact that $(\wn_k)_{k\in\mathbb{N}}$ is a sequence of i.i.d. standard complex Gaussian random variables; indeed, 
\begin{align}
    \label{Err}
    \left\lVert \sum_{k\in\mathbb{N}} \wn_k \atom_k\right \rVert_{\mathfrak{G}}^2= \sum_{k\in\mathbb{N}}\frac{|\wn_k|^2}{1+k^2}
\end{align}
is almost surely finite, using e.g. Kolmogorov's two-series theorem.
Third, note that the definition of the space $\mathfrak{G}$ depends on the choice of the Hilbert basis $(\atom_k)_{k\in \mathbb{N}}$. Hence, unlike the finite-dimensional case of Section~\ref{sssec:gauss-vect}, the definition of white noise here is \emph{basis-dependent}.
Finally, the definition implies that for any $n\in \mathbb{N}^*$, and any \emph{finite} orthonormal family $(\atom_k)_{k = 1}^n$  of $\mathfrak{H}$, the vector of $\mathbb{C}^n$ defined as
\begin{align}
    \left( \langle \wn, \atom_k\rangle \right)_{k = 1}^n \sim \mathcal{N}_{\mathbb{C}}(\boldsymbol{0}_n, \textbf{I}_{n\times n})
\end{align}
is an $n$-dimensional white noise. 
Hence, in signal processing where $\mathfrak{H} = L^2(\mathbb{R})$, the definition of white noise in Proposition~\ref{prop:white-noise} is consistent with the intuition of an uncorrelated continuous random process with Gaussian marginals; see also \cite[Definition~2.1.7]{cohen2013fractional}.

\subsubsection{From discrete to continuous:  approximation results}

The infinite-dimensional white noise is described in Proposition~\ref{prop:white-noise} as an infinite sum, though, in numerical experiments, only \emph{finite} sums can be computed.
One can thus wonder, given $(\atom_k)_{k\in \mathbb{N}}$ a Hilbert basis of $L^2(\mathbb{R})$,  $(\wn_k)_{k\in \mathbb{N}}$ a sequence of i.i.d. standard complex Gaussian variables, and $n \in \mathbb{N}^*$, to what extent a truncated sum of the form
\begin{align}
\label{eq:truncated-wn}
\wn^{(n)} = \sum_{k = 0}^n \wn_k \atom_k \sim \mu^{(n)}
\end{align}
is a good approximation of the white noise $\wn$ constructed in Proposition~\ref{prop:white-noise}; or, in other words,  whether the sequence of distributions $\left(\mu^{(n)}\right)_{n \in \mathbb{N^*}}$ converges to the distribution $\mu$ of the infinite-dimensional white noise.

\begin{proposition}
Let $\mu$ be the distribution of the infinite-dimensional white noise constructed in Proposition~\ref{prop:white-noise}, and $\mu^{(n)}$ the distribution of the truncated white noise of Equation~\eqref{eq:truncated-wn}. 
Then, \cite{bardenet2021time} showed that for any $n \in \mathbb{N}^*$, 
\begin{align}
\mathfrak{W}_2\left(\mu^{(n)},\mu\right) \leq \frac{1}{\sqrt{n}},
\end{align}
where $\mathfrak{W}_2$ denotes the $2$-Wasserstein-Kantorovich distance.\footnote{
    \label{ft:Wasserstein}Let $p\geq 1$, then the $p$-Wasserstein-Kantorovich distance $\mathfrak{W}_p(\nu_1,\nu_2)$ between two probability measures $\nu_1, \nu_2$ on $\mathfrak{G}$ is defined by $\mathfrak{W}_p(\nu_1,\nu_2)^p = \inf \iint_{\mathfrak{G} \times \mathfrak{G}} \lVert h_1-h_2 \rVert^p_{\mathfrak{G}} \, \mathrm{d}\kappa(h_1,h_2)$, where the infimum is taken over probability measures $\kappa$ on $\mathfrak{G} \times \mathfrak{G}$ with marginals $\nu_1$ and $\nu_2$.
}.
\label{prop:W2_wn}
\end{proposition}

Remarking that, for any $\nu_1,\nu_2$ probability measures on $\mathfrak{G}$, the Cauchy-Schwarz inequality\footref{ft:Wasserstein} ensures $\mathfrak{W}_1\left(\nu_1,\nu_2\right) \leq \mathfrak{W}_2\left(\nu_1,\nu_2\right)$, and that $\mathfrak{W}_1$ admits the Kantorovich-Rubinstein dual representation
\begin{align}
    \mathfrak{W}_1(\nu_1,\nu_2) =\underset{\mathcal{L}}{\sup} \left\vert \int_{\mathfrak{G}} \mathcal{L} \, \mathrm{d}\nu_1 - \int_{\mathfrak{G}} \mathcal{L} \, \mathrm{d}\nu_2\right\vert,
\end{align}
where the supremum is taken over all Lipschitz-continuous functions $\mathcal{L}:\mathfrak{G} \rightarrow\mathbb{R}$ whose Lipschitz constant $\lVert \mathcal{L}\rVert_{\text{Lip}}$ is smaller or equal to one, the authors of~\cite{bardenet2021time} used Proposition~\ref{prop:W2_wn} to demonstrate the following corollary.
This will prove useful when studying time-frequency transforms of white noise in Section~\ref{ssec:zerosGAF_TF}.

\begin{corollary}
    Let $\mathcal{L}:\mathfrak{G} \rightarrow\mathbb{R}$,  $\wn$ the infinite-dimensional white noise and $\mu$ its distribution introduced in Propositions~\ref{prop:white-noise} and \ref{prop:W2_wn}, $n\in \mathbb{N}^*$, $\wn^{(n)}$ defined in~\eqref{eq:truncated-wn}, and $\mu^{(n)}$ in \eqref{eq:truncated-wn}.
    If $\mathcal{L}$ is Lipschitz, then
    \begin{align}
    \left\lvert \mathbb{E} [\mathcal{L}(\wn^{(n)})]  - \mathbb{E} \left[\mathcal{L}(\wn)\right] \right\rvert \leq \frac{\lVert \mathcal{L}\rVert_{\text{Lip}}}{\sqrt{n}}.
    \end{align}
    \label{cor:dist-L-wn}
\end{corollary}

\section{Gaussian Analytic Functions in time-frequency analysis}
\label{sec:GAF}

The continuous white noise has been defined in Section~\ref{sssec:inf-wn} as an infinite series,  in which each term is the product of a standard Gaussian random variable and a normalized elementary function, the Gaussian variables being independent.
Hence, applying to the infinite-dimensional white noise a \emph{linear} time-frequency transform, such as the Gaussian Short-Time Fourier Transform $\mathcal{V}_h$, associated to a window $h$ of Section~\ref{sssec:STFT}, should informally yield a series of the form
\begin{align}
    \label{eq:Vgw}
    \mathcal{V}_h  \wn = \sum_{k \in \mathbb{N}} \wn_k  \mathcal{V}_h  \atom_k 
\end{align}
where $(\atom_k)_{k\in\mathbb{N}}$ is the Hilbert basis of $L^2(\mathbb{R})$ used in the definition of $\wn$~\eqref{eq:wn-series}.
As will be shown in Section~\ref{ssec:zerosGAF_TF}, assimilating the time-frequency plane to the complex plane and appropriately choosing the analysis window $h$ and the Hilbert basis $(\atom_k)_{k\in\mathbb{N}}$, 
\begin{align}
    \label{eq:Vh-atom}
    \mathcal{V}_h  \atom_k(t,\omega)  = \psi(z) \mathsf{F}_k(z)
\end{align}
expresses as the product of an analytic function $\mathsf{F}_k$ of $z = \omega + \mathrm{i}t \in \mathbb{C}$, and a non-vanishing function $\psi$, common to all atoms $f_k$.
Then, it will follow from~\eqref{eq:Vgw} that, up to multiplication by a non-vanishing function, $\mathcal{V}_h  \wn $ is an infinite series of analytic functions with i.i.d. standard complex Gaussian weights, known as a \emph{Gaussian Analytic Function} in probability theory and denoted
\begin{align}
\label{eq:def-GAF}
\mathsf{GAF} = \sum_{k\in\mathbb{N}} \wn_k \mathsf{F}_k.
\end{align}
Section~\ref{ssec:def_GAF} provides a self-contained introduction to the theory of Gaussian Analytic Functions, with a focus on the geometry of their level sets and in particular of their zeros.
It will closely follow the classical reference \cite{hough2009zeros}, so widely recommended that it is commonly referred to as \emph{The GAF book}.
Then, in Section~\ref{ssec:zerosGAF_TF}, closely following \cite{bardenet2021time}, a rigorous mathematical meaning will be given to the time-frequency transform of white noise $\mathcal{V}_h \wn$ informally defined as the right-hand side of~\eqref{eq:Vgw}.

\subsection{A short introduction to Gaussian Analytic Functions}
\label{ssec:def_GAF}

Let $\mathfrak{L} \subset \mathbb{C}$ be an open subset of the complex plane, and $A(\mathfrak{L})$ be the space of analytic functions on $\mathfrak{L}$ endowed with the topology of uniform convergence on compact sets. 
Then $A(\mathfrak{L})$ is a complete separable metric space, which allows to rigorously define random variables taking values in the space of analytic functions on $\mathfrak{L}$.

\subsubsection{Complex Gaussian processes}

Studying the Short-Time Fourier Transform of complex white noise requires to give a rigorous mathematical definition to the $A(\mathfrak{L})$-valued \emph{Gaussian} variables introduced in~\eqref{eq:def-GAF}.
A way to build \emph{Gaussian} random functions is to constrain all marginals to be complex Gaussian vectors, as defined in Section~\ref{sssec:gauss-vect}.

\begin{definition}
    A random function $\mathsf{F}\in A(\mathfrak{L})$ is a Gaussian Analytic Function if and only if for all $p\geq 1$, $z_1, \dots, z_p\in\mathfrak{L}$, 
    the marginal $\left ( \mathsf{F}(z_1), \dots, \mathsf{F}(z_p) \right)$ is a zero-mean complex Gaussian vector $\mathcal{N}_{\mathbb{C}}(\boldsymbol{0}, \textbf{K})$. 
    In particular, 
    \begin{align}
    \label{eq:def-kernel}
    \mathsf{K}(z,z') = \mathbb{E}\left[ \mathsf{F}(z)\overline{\mathsf{F}(z')}\right], \quad z, z'\in \mathbb{C}
    \end{align}
    is called the \emph{covariance kernel} of $\mathsf{F}$.
    \label{def:GAF}
\end{definition}

From Definition~\ref{def:GAF} it is not \emph{a priori} obvious whether non-trivial random analytic functions satisfying such marginal conditions exist. 
If it exists, the law of a Gaussian Analytic Function is characterized by its kernel. 
The proof comes from a standard, though technical, argument.\footnote{
    See, for instance, the extension result in \citep[Proposition 3.2]{kallenberg2001foundations}, which can be adapted to the topology of compact convergence by noting that analytic functions are in particular continuous.
} 
As for the existence, rather than resorting to Kolmogorov extension and continuity theorems, Proposition~\ref{prop:constructive_lemma} below follows the exposition in \cite{hough2009zeros}, providing an explicit construction of particular Gaussian Analytic Functions as entire series with random coefficients of the form~\eqref{eq:def-GAF}.
This construction will be key to rigorously establishing the link between Gaussian Analytic Functions and the spectrogram of white noise in Section~\ref{ssec:zerosGAF_TF}.

\begin{proposition}[Lemma 2.2.3 in~\cite{hough2009zeros}]
    Let $\mathsf{F}_1, \mathsf{F}_2, \dots$ be analytic functions on $\mathfrak{L}$, and $(\wn_k)_{k\in \mathbb{N}}$ be an infinite sequence of i.i.d. zero-mean unit-variance complex random variables.
    Assume that
    \begin{align}
    \forall \mathfrak{K}\subset \mathfrak{L} \text{ compact}, \quad \sup_{z\in \mathfrak{K}}\,\sum_{k\in\mathbb{N}} |\mathsf{F}_k(z)|^2<\infty.  \tag{$\mathsf{uniform}-\mathfrak{K}$}
    \label{eq:uniform-K}
    \end{align}
    Then, almost surely, $\sum_{k \in \mathbb{N}} \wn_k \mathsf{F}_k$ converges uniformly on compact subsets of $\mathfrak{L}$.
    The limit thus defines a random analytic function.

    Additionally, if the coefficients $\wn_k$ are standard complex Gaussian variables, then the limit is the Gaussian Analytic Function with covariance kernel
    $$
    \mathsf{K}(z,z') = \sum_{k = 0}^{\infty} \mathsf{F}_k(z) \overline{\mathsf{F}_k(z')}.
    $$
    \label{prop:constructive_lemma}
\end{proposition}

Independently of its use to build Gaussian Analytic Functions, Proposition~\ref{prop:constructive_lemma} is also of interest for its proof, which shows a beautiful interplay of probability and analysis.
This is a good example of how the theory of random analytic functions combines these two domains.
The proof here is from \cite[Lemma 2.2.3]{hough2009zeros}, with a few of their voluntary gaps filled for the sake of completeness.

\begin{proof}
From the analytic functions $\mathsf{F}_1, \mathsf{F}_2, \hdots$ and the sequence of i.i.d. random variables $(\wn_k)_{k\in \mathbb{N}}$ in Proposition~\ref{prop:constructive_lemma}, consider the random variables $ \mathsf{X}_1,  \mathsf{X}_2, \hdots$ defined as the partial sums
\begin{align}
    \forall n \in \mathbb{N}, \, z\in \mathfrak{L}, \quad  \mathsf{X}_n(z) =  \sum_{\ell = 0}^n \wn_\ell \mathsf{F}_\ell(z).
\end{align}
Clearly, $\mathsf X_1, \mathsf X_2, \dots$ are all in $A(\mathfrak{L})$.
The proof proceeds in three steps: 1) demonstrating that, for any compact $\mathfrak{K} \subset \mathfrak{L}$, $\left( \mathsf{X}_n \right)_{n\in \mathbb{N}}$ is a Cauchy sequence in $L^2(\mathfrak{K})$; then 2) showing that $\left(\mathsf{X}_n\right)_{n \in \mathbb{N}}$ converges uniformly on compact sets and hence that $\mathsf{X}\in A(\mathfrak{L})$; and 3) concluding about the Gaussianity of $\mathsf{X}$ and the expression of its covariance kernel.

\emph{Step 1.} Let $\lVert \cdot \rVert_{\mathfrak{K}}$ denote the norm in $L^2(\mathfrak{K})$, and $\langle \cdot, \cdot\rangle _{\mathfrak{K}}$ the inner product.
Let $k \leq n$.
It can be shown that 
\begin{equation}
    \label{e:kolmo}
    \mathbb{E}\left[ \lVert \mathsf{X}_n\rVert_{\mathfrak{K}}^2 \lvert \wn_1, \hdots, \wn_k\right] = \lVert \mathsf{X}_k \rVert_{\mathfrak{K}}^2 + \sum_{\ell = k+1}^n \lVert \mathsf{F}_\ell \rVert_{\mathfrak{K}}^2.
\end{equation}
Indeed, for any Borel set $\mathfrak{B} $ of $\mathbb{C}^k$,
\begin{align*}
\mathbb{E} \left[ \lVert \mathsf{X}_n \rVert_{\mathfrak{K}}^2 \1_{(\wn_1, \hdots, \wn_k) \in \mathfrak{B}} \right] &= \mathbb{E}\left[ \lVert \mathsf{X}_k\rVert_{\mathfrak{K}}^2  \1_{(\wn_1, \hdots, \wn_k) \in \mathfrak{B}} \right]\\
&+2 \mathsf{Re} \left(\mathbb{E}\left[ \left( \sum_{ 1 \leq i \leq k} \sum_{j >k} \wn_i \overline{\wn_j} \left\langle \mathsf{F}_i, \mathsf{F}_j \right\rangle_{\mathfrak{K}} \right)  \1_{(\wn_1, \hdots, \wn_k) \in \mathfrak{B}}\right]\right)\\
&+ \mathbb{E}\left[ \left(\sum_{i,j >k} \wn_i \overline{\wn_j} \left\langle \mathsf{F}_i, \mathsf{F}_j \right\rangle_{\mathfrak{K}} \right)  \1_{(\wn_1, \hdots, \wn_k) \in \mathfrak{B}}\right]
\end{align*}
where $\mathsf{Re}(\cdot)$ stands for the real part of a complex number.
Because $\mathbb{E}\left[\wn_i \overline{\wn_j}\right] = \delta_{i,j}$, the second term vanishes, while the third term reduces to 
$$ 
    \mathbb{E}\left[ \left(\sum_{\ell =k+1}^n  \lVert \mathsf{F}_\ell\rVert_{\mathfrak{K}}^2 \right)  \1_{(\wn_1, \hdots, \wn_k) \in \mathfrak{B}}\right].
$$ 
This proves \eqref{e:kolmo}.

Then, for $\varepsilon > 0$, define the stopping time  $\tau = \inf \lbrace k, \,  \lVert \mathsf{X}_k \rVert_{\mathfrak{K}} \geq \varepsilon\rbrace$. 
Using \eqref{e:kolmo}, it comes
\begin{align*}
\mathbb{E} \left[ \lVert \mathsf{X}_n \rVert_{\mathfrak{K}}^2\right] &\geq \sum_{k = 1}^n \mathbb{E} \left[ \lVert \mathsf{X}_n \rVert_{\mathfrak{K}}^2 \1_{\tau = k}\right]\\ &= \sum_{k = 1}^n \mathbb{E}\left[ \1_{\tau = k} \mathbb{E}\left[ \lVert \mathsf{X}_n\rVert_{\mathfrak{K}}^2 \mid \wn_j, j \leq k \right] \right]\\
& \geq \sum_{k = 1}^n \mathbb{E} \left[  \1_{\tau = k} \lVert \mathsf{X}_k \rVert_{\mathfrak{K}}^2\right]\\
& \geq \varepsilon^2 \mathbb{P}\left[ \tau \leq n  \right],
\end{align*}
It follows that 
\begin{align}
    \label{eq:kolmo}
    \mathbb{P} \left[ \underset{j \leq n}{\sup} \, \lVert \mathsf{X}_j \rVert_{\mathfrak{K}}^2  \geq \varepsilon \right] = \mathbb{P}\left[ \tau \leq n \right] \leq \frac{1}{\varepsilon^2}\mathbb{E}\left[ \lVert \mathsf{X}_n \rVert_{\mathfrak{K}}^2\right] = \frac{1}{\varepsilon^2} \sum_{\ell=1}^n \lVert \mathsf{F}_\ell\rVert_{\mathfrak{K}}^2.
\end{align}
In particular, the monotone convergence theorem implies that 
\begin{align}
    \label{eq:kolmo}
    \mathbb{P} \left[ \underset{j}{\sup} \, \lVert \mathsf{X}_j \rVert_{\mathfrak{K}}^2  \geq \varepsilon \right] \leq \frac{1}{\varepsilon^2} \sum_{\ell=1}^\infty \lVert \mathsf{F}_\ell\rVert_{\mathfrak{K}}^2,
\end{align}
where the right-hand side is finite by \eqref{eq:uniform-K}.

For any fixed $p\in \mathbb{N}$, a similar computation applied to $\left( \mathsf{X}_{p+n} - \mathsf{X}_p \right)_{n\in \mathbb{N}}$ yields
\begin{align}
    \label{eq:pre-cauchy-bound}
    \mathbb{P} \left[  \underset{n\geq 1}{\sup} \, \lVert \mathsf{X}_{p+n} - \mathsf{X}_{p} \rVert_{\mathfrak{K}}^2 \geq  \varepsilon \right] \leq \frac{1}{\varepsilon^2} \sum_{j = p+1}^\infty \lVert \mathsf{F}_j \rVert_{\mathfrak{K}}^2,
\end{align}
which, by Hypothesis~\eqref{eq:uniform-K}, converges to zero as $p \rightarrow\infty$.
On the other hand, by dominated convergence,
$$
    \mathbb{E} \left[  \lim_{p\rightarrow\infty} \1_{\left\{{\sup_{n\geq 1}} \, \lVert \mathsf{X}_{p+n} - \mathsf{X}_{p} \rVert_{\mathfrak{K}}^2 \geq \varepsilon\right\}} \right] = \lim_{p\rightarrow\infty} \mathbb{E} \left[ \1_{\left\{{\sup_{n\geq 1}} \, \lVert \mathsf{X}_{p+n} - \mathsf{X}_{p} \rVert_{\mathfrak{K}}^2 \geq \varepsilon\right\}} \right], 
$$
which is zero by \eqref{eq:pre-cauchy-bound}.
Consequently, 
\begin{align}
\mathbb{P}\left[ \exists p\in\mathbb{N} \text{ such that } \forall n, \, \lVert X_{p+n} - X_p \rVert \leq \varepsilon\right] = 1.
\end{align}
In other words, almost surely, $\left( \mathsf{X}_n\right)_{n\in\mathbb{N}}$ is a Cauchy sequence in the Hilbert space $L^2(\mathfrak{K})$, and thus converges in $L^2(\mathfrak{K})$ to some $\mathsf{X} \in L^2(\mathfrak{K})$.

\emph{Step 2.}
The second step consists in proving that $\mathsf{X}_n$ converges uniformly on the compact subsets of $\mathfrak{L}$, so that the limit is analytic.
Let $z_0 \in \mathfrak{L}$ and take $R>0$ such that the closed disk $\mathsf{D}(z_0,4R)$ centered at $z_0$ and of radius $4R$ is included in $\mathfrak{L}$. 
As a sum of analytic functions, $\mathsf{X}_n$ is analytic, thus it satisfies Cauchy's formula: for any $r< 4R$,
$$
    \forall z \in \mathfrak{L}, \, \lvert z - z_0\rvert \leq  r, \quad \mathsf{X}_n(z) = \frac{1}{2\pi \mathrm{i}} \int_{\mathsf{C}_r} \frac{\mathsf{X}_n(\zeta)}{\zeta - z} \,  \mathrm{Leb}(\mathrm{d}\zeta),
$$
where the integral is over the circle $\mathsf{C}_r = \{z_0 + r\mathrm{e}^{\mathrm{i}t}$, $0\leq t \leq 2 \pi\}$ and $\mathrm{Leb}$ denotes the Lebesgue measure on $\mathbb{C}$. 
Let $z \in \mathsf{D}(z_0, R)$, and average Cauchy's formula over $r\in (2R, 3R)$, to obtain
\begin{align}
    \mathsf{X}_n(z) 
    &= \frac{1}{2\pi \mathrm{i}} \frac{1}{R} \int_{2R}^{3R} \int_0^{2\pi} \frac{\mathsf{X}_n(z_0 + r\mathrm{e}^{\mathrm{i}t})}{z_0 + r\mathrm{e}^{\mathrm{i}t} - z} \, \mathrm{i}r\mathrm{e}^{\mathrm{i}t} \mathrm{d}t \mathrm{d}r \nonumber\\
    &=  \frac{1}{2\pi}\int_\mathsf{A} \mathsf{X}_n(\zeta) \mathsf{G}(\zeta ;z) \, \mathrm{Leb}(\mathrm{d}\zeta), \quad \mathsf{G}(\zeta ;z) = \frac{\zeta }{ \lvert \zeta\rvert R (\zeta - z)},
\label{eq:cauchy-av}
\end{align}
where $\mathsf{A}$ is the annulus $\lbrace \zeta \in \mathfrak{L}, \, \,  2R \leq \lvert \zeta - z_0 \rvert \leq 3R\rbrace$.
Then, on the one hand, 
\begin{align}
\label{eq:unif-bound}
\forall z  \in \mathsf{D}(z_0,R), \zeta \in \mathsf{A}, \quad  \left\lvert \mathsf{G}(\zeta ;z) \right\rvert \leq 1 \times \frac{1}{R} \times \frac{1}{R} = \frac{1}{R^2}.
\end{align}
On the other hand, Step 1 implies that, on an event of probability one, there exists $\mathsf{X}$ in $L^2(\mathsf{D}(z_0,4R))$ such that $\mathsf X_n\rightarrow \mathsf X$ in $L^2(\mathsf{D}(z_0,4R))$.
Cauchy-Schwarz inequality applied to \eqref{eq:cauchy-av} then shows that, on the same event,
$$
    \mathsf{X}_n \rightarrow \frac{1}{2\pi}\int_\mathsf{A} \mathsf{X}(\zeta) \mathsf{G}(\zeta ;\cdot) \, \mathrm{Leb}(\mathrm{d}\zeta)
$$
uniformly over $D(z_0,R)$.
Since $z_0$ and $R$ were chosen arbitrarily, $\left( \mathsf{X}_n \right)_{n\in \mathbb{N}}$ converges uniformly on any compact subset $\mathfrak{K}$ of $\mathfrak{L}$ to some $\mathsf{X}^{\mathfrak{K}} \in A(\mathfrak{L})$. 
Uniform convergence implying simple convergence, for $z\in\mathfrak{L}$, the limit $\mathsf{X}^{\mathfrak{K}}(z)$ is independent of $\mathfrak{K}$, which defines a function $\mathsf{X}$ on $\mathfrak{L}$. 
Since $\mathsf{X}$ restricted to any compact set is analytic,  then $\mathsf{X} \in A(\mathfrak{L})$.

\emph{Step 3.} 
If the coefficients $\left( \wn_k \right)_{n\in\mathbb{N}}$ are i.i.d. complex Gaussian variables, then for any $n \in \mathbb{N}$, $\mathsf{X}_n$ is a Gaussian Analytic Function in the sense of Definition~\ref{def:GAF}, with covariance kernel 
\begin{align}
    \label{eq:partial_kernel}
    \mathsf{K}_n(z,z') = \sum_{k=0}^n \mathsf{F}_k(z) \overline{\mathsf{F}_k(z')}.
\end{align}
In particular, any marginal vector $\boldsymbol{x}_n = (\mathsf{X}_n(z_1), \hdots, \mathsf{X}_n(z_p))$ of evaluations of $\mathsf{X}_n$ at $z_1, \hdots,  z_p \in \mathbb{C}$ is a complex Gaussian vector, say with covariance matrix $\mathbf{K}_n$.
On the other hand, almost surely, $\mathsf X_n\rightarrow \mathsf X$ uniformly on compact subsets, so that in particular $\boldsymbol{x}_n$ converges in distribution towards the vector $\boldsymbol{x}$ of evaluations of $\mathsf X$ at the same points.
Weak limits of Gaussian vectors are Gaussian, and the generic term of the covariance matrix of $\boldsymbol{x}$ is necessarily the limit of the corresponding term in $\mathbf{K}_n$; see e.g. \citep[Exercise 2.1.4]{hough2009zeros}.
Hence, $\mathsf{X}$ is a Gaussian Analytic Function, and its covariance kernel is obtained by letting $n\rightarrow \infty$ in~\eqref{eq:partial_kernel}.
This concludes the proof.
\end{proof}

\subsubsection{A key example: the planar Gaussian Analytic Function}
\label{sssec:planarGAF}
The construction of Proposition~\ref{prop:constructive_lemma} yields a wealth of Gaussian Analytic Functions.
The simplest building blocks are monomials, that is, taking $\mathsf{F}_k(z) \propto z^k$ for all $k \in \mathbb{N}$: the resulting Gaussian Analytic Function then consists in a Taylor expansion with random coefficients.
A suitable choice of normalization in $\mathsf{F}_k$ leads to a first key example: the planar Gaussian Analytic Function.

\begin{definition}
Let $\mathfrak{L} = \mathbb{C}$, $ \gamma > 0$,  and $\left(\wn_k\right)_{k\in\mathbb{N}}$ be i.i.d. standard complex Gaussian variables.
The \emph{planar Gaussian Analytic Function} of parameter $\gamma$ is the entire function
\begin{align}
\label{eq:planarGAF}
\mathsf{GAF}_{\mathbb{C}}^{(\gamma)}(z) = \sum_{k=0}^\infty \wn_k \sqrt{\frac{\gamma^k}{k!}} z^k, \quad z\in\mathbb{C}.
\end{align}
In the following, when $\gamma = 1$, the superscript is omitted and the planar Gaussian Analytic Function is simply denoted $\mathsf{GAF}_{\mathbb{C}}$.
\label{def:planarGAF}
\end{definition}

The demonstration that the planar Gaussian Analytic Function is well defined, analytic on $\mathbb{C}$ and Gaussian, relies on the verification that, whatever $\gamma > 0$, Hypothesis~\eqref{eq:uniform-K} holds for $\mathfrak{L}=\mathbb{C}$ and 
\begin{align}
\mathsf{F}_k(z) = \sqrt{\frac{\gamma^k}{k!}}z^k.
\end{align}
This allows to invoke Proposition~\ref{prop:constructive_lemma}.
It follows that the covariance kernel of the planar Gaussian Analytic Function writes
\begin{align}
\label{eq:K-alpha}
\mathsf{K}^{(\gamma)}(z,w) = \sum_{k = 0}^\infty \sqrt{\frac{\gamma^k}{k!}} z^k \overline{\sqrt{\frac{\gamma^k}{k!}} w^k} = \sum_{k =0}^\infty \frac{(\gamma z\overline{w})^k}{ k!}= \exp\left(\gamma z\overline{w}\right).
\end{align}

\begin{figure}
\centering
\includegraphics[width = 6cm]{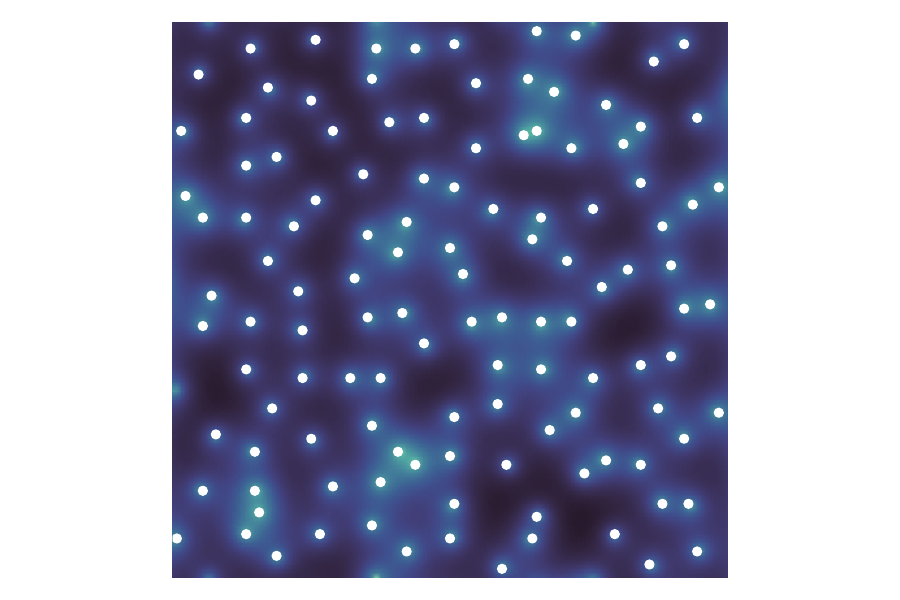}
\caption{Zeros $\mathfrak{Z}_{\mathbb{C}}$ of the planar Gaussian Analytic Function~$\mathsf{GAF}_{\mathbb{C}}$ of Definition~\ref{def:planarGAF}.
The distribution of $\mathfrak{Z}_{\mathbb{C}}$ is invariant under the isometries of the complex plane, hence the uniform spread. (Adapted from \cite{bardenet2021time})  }
\label{fig:planar}
\end{figure}

By construction, the planar Gaussian Analytic Function is a random function, and so are its level sets.
In particular, because $\mathsf{GAF}_{\mathbb{C}}^{(\gamma)}$ is analytic, its \emph{zeros}
\begin{align}
    \label{eq:Z-alpha}
    \mathfrak{Z}^{(\gamma)}_{\mathbb{C}} =\left( \mathsf{GAF}_{\mathbb{C}}^{(\gamma)}\right)^{-1}(\lbrace 0 \rbrace)
\end{align}
form a random set of \emph{isolated} points, that is, a \emph{Point Process}\footnote{See Section~\ref{ssec:SPat_stat} for a proper definition, properties and analysis tools.} on $\mathbb{C}$.
A sample of the zeros of the planar Gaussian Analytic Function of parameter $\gamma =1$, denoted\footnote{When the superscript is omitted, it is assumed that $\gamma = 1$.} by $\mathfrak{Z}_{\mathbb{C}}$, is shown in Figure~\ref{fig:planar}. 
Strikingly, the zeros of $\mathsf{GAF}_{\mathbb{C}}$ are very evenly spread in the plane, a property that the next sections shall make more precise.

\subsubsection{The random zeros of a Gaussian Analytic Function form a Point Process}
\label{ssec:SPat_stat}

Though Point Processes can be defined in more general spaces, this chapter restricts to a complete metric space $\mathfrak{S}$, which will be $\mathbb{C}$ most of the time.

\begin{definition}
A \emph{locally finite point configuration} in $\mathfrak{S}$ is a subset $\mathfrak{X}$ of $\mathfrak{S}$ such that, for any bounded $\mathfrak{B}\subset\mathfrak{S}$, the cardinality of $\mathfrak{X}\cap \mathfrak{B}$ is finite. 
A \emph{Point Process} on $\mathfrak{S}$ is a random variable taking values in the locally finite point configurations in $\mathfrak{S}$.
\label{def:PP}
\end{definition}

First, note that according to Definition~\ref{def:PP}, a Point Process is almost surely a set, so that points naturally come with multiplicity one. 
Some authors prefer to allow for points with arbitrary multiplicity, in which case Definition~\ref{def:PP} becomes that a of a \emph{simple} Point Process.
Second, the fact that zero sets of Gaussian Analytic Functions are Point Processes is a consequence of the analyticity of a Gaussian Analytic Function, which forbids accumulation points. 
Finally, it can be shown that the \emph{random} zeros of a Gaussian Analytic Function, i.e., those that are not shared by every sample, are all simple zeros with probability one~\cite[Lemma 2.4.1]{hough2009zeros}. 
Hence,  Definition~\ref{def:PP}, which does not take multiplicity into account, accurately describes the set of random zeros of a Gaussian Analytic Function.
To make things concrete, 
$$
    z\mapsto (z-z_0)^r \times \mathsf{GAF}_{\mathbb{C}}(z)
$$
for $z_0\in \mathbb{C}$ and $r \in \mathbb{N}^*$, is a Gaussian Analytic Function.
It has a deterministic zero in $z_0$ with multiplicity $r$, while its random zeros are those of $\mathsf{GAF}_{\mathbb{C}}$, which all have of multiplicity one, almost surely.

Of particular interest when studying the zero set of a Gaussian Analytic Functions zero sets are its \emph{symmetries}. 
Indeed, as observed in Figure~\ref{fig:planar}, the zeros of the planar Gaussian Analytic Function looks rather uniformly distributed, which turns out to reflect a deeper invariance property of the distribution of zeros.
Consider the isometries of $\mathbb{C}$,  described as the transformations
\begin{align}
\mathcal{I}_{\eta, \vartheta} : \left \lbrace
\begin{array}{ccc}
\mathbb{C} & \rightarrow & \mathbb{C} \\
z & \mapsto & \eta z +\vartheta
\end{array}
\right. ,
\end{align}
for $\eta,\vartheta \in \mathbb{C}$ with $\lvert \eta\rvert = 1$. 
They consist in compositions of rotations and translations.

\begin{proposition}{\cite[Proposition 2.3.4]{hough2009zeros}}
Let $\gamma > 0$, the zero set of $\mathsf{GAF}_{\mathbb{C}}^{(\gamma)}$ is invariant, in distribution, under the complex plane isometries $\lbrace \mathcal{I}_{\eta, \vartheta}, \, \eta, \vartheta \in \mathbb{C}, \lvert \eta \rvert =1\rbrace$.
\label{prop:inv-iso}
\end{proposition}

The proof below is taken verbatim from \citep{hough2009zeros}, and is here to show how basic properties of a Gaussian Analytic Function transfer to its zeros. 
\begin{proof}
For some $\gamma > 0$,  let $\eta, \vartheta \in \mathbb{C}$ with $\lvert \eta\rvert = 1$ and define 
$$
    \mathsf{G}(z) = \mathsf{GAF}_{\mathbb{C}}^{(\gamma)}(\eta z +\vartheta).
$$ 
$\mathsf{G}$ is a Gaussian Analytic Function and its kernel writes
\begin{align}
\mathsf{K}_{\mathsf{G}}(z, w) &= \mathbb{E}\left[\mathsf{G}(z)\overline{\mathsf{G}(w)}\right] \\
						&= \mathbb{E}\left[ \mathsf{GAF}_{\mathbb{C}}^{(\gamma)}(\eta z +\vartheta) \overline{\mathsf{GAF}_{\mathbb{C}}^{(\gamma)}(\eta w +\vartheta)} \right].
\end{align}
Then, using the expression of the kernel of $\mathsf{GAF}_{\mathbb{C}}^{(\gamma)}$ provided in Equation~\eqref{eq:K-alpha},
\begin{align}						
\mathsf{K}_{\mathsf{G}}(z, w) &= \exp \left( \gamma (\eta z +\vartheta)(\overline{\eta w +\vartheta})\right)\\
                           &= \exp \left( \gamma z \overline{w} + \gamma \eta \overline{\vartheta} z + \gamma \vartheta\overline{\eta} \overline{w} +\gamma \lvert \vartheta \rvert^2\right)\\
                           &=\exp(\gamma z \overline{w}) \exp (\gamma \eta \overline{\vartheta} z +  \lvert \vartheta\rvert^2/2) \overline{\exp(\gamma \eta \overline{\vartheta}w + \lvert \vartheta \rvert^2/2)}.
                           \label{eq:recognize_a_kernel}
\end{align}
Since $\exp(\gamma z \overline{w})$ is the covariance kernel of $\mathsf{GAF}_{\mathbb{C}}^{(\gamma)}$, one can recognize in \eqref{eq:recognize_a_kernel} the covariance kernel of 
$$
    \mathsf{H}: z\mapsto \exp (\gamma \eta \overline{\vartheta} z +  \lvert \vartheta \rvert^2/2) \times \mathsf{GAF}_{\mathbb{C}}^{(\gamma)}.
$$
As emphasized in Section~\ref{ssec:def_GAF}, a Gaussian Analytic Function is characterized by its kernel, so that $\mathsf{G}$ and $\mathsf{H}$ are equal in law, and thus the Point Processes of their zeros as well.
Further, $\mathsf{H}$ coincides with the planar Gaussian Analytic Function up to multiplication by a non-vanishing function, thus its zero set coincides with the zero set of $\mathsf{GAF}_{\mathbb{C}}^{(\gamma)}$, which concludes the proof.
\end{proof}

\begin{remark}
    \label{rem:canonical_gafs}
    Apart from the planar case, there exist two other \emph{canonical} Gaussian Analytic Functions, respectively referred to as \emph{spherical} and \emph{hyperbolic}, whose zero sets are respectively invariant under isometries of the sphere and of the hyperbolic plane~\cite[Proposition~2.3.4]{hough2009zeros}.
    All three canonical Gaussian Analytic Functions have been connected to (generalized) time-frequency transforms of white noise.
    While this chapter focuses on the planar Gaussian Analytic Function zeros in standard time-frequency analysis, though similar results and methodologies have been applied in both the hyperbolic and the spherical settings~\cite{abreu2018filtering,bardenet2020zeros,bardenet2021time,pascal2022covariant}, as discussed in Section~\ref{sec:extensions}.
\end{remark}

\subsubsection{Spatial statistics of zero sets of Gaussian Analytic Functions}
\label{sssec:spat-stat-Z}

To characterize second- (or higher-) order properties in a Point Process, such as the apparent absence of pairs of points too close to each other in Figure~\ref{fig:planar}, one can use the \textit{joint intensities}.

\begin{definition}
Let $\mathfrak{X}$ be a Point Process, defined on an open subset $\mathfrak{L} \subset \mathbb{C}$ and $n\in \mathbb{N}^*$. Following~\cite[Definition 1.2.2]{hough2009zeros}, the $n$-point \emph{joint intensity} $\rho_n$ of $\mathfrak{X}$ with respect to the Lebesgue measure $\mathrm{Leb}$ on $\mathbb{C}$, is defined, when it exists, by
\begin{align}
    \int_{\mathfrak{L}^n} \! \! \! \! \Psi(\zeta_1, \hdots,\zeta_n)  \rho_n(\zeta_1, \hdots, \zeta_n) \, \mathrm{Leb}(\mathrm{d}\zeta_1)\hdots \mathrm{Leb}(\mathrm{d}\zeta_n) = 
    \mathbb{E}\left[ \!  \sum_{\begin{array}{c} {\scriptstyle (z_1, \hdots, z_n) \in \mathfrak{X}^n} \\ {\scriptstyle z_1 \neq \hdots \neq z_n} \end{array}} \! \! \! \! \! \! \ \ \! \! \! \! \ \! \! \! \!\Psi(z_1, \hdots, z_n)\right],
\end{align}
for any bounded compactly supported measurable map $\Psi : \mathfrak{L}^n \rightarrow \mathbb{C}$.
\label{def:joint}
\end{definition}

If $\rho_n$ exists, Definition~\ref{def:joint} intuitively  means that given $n$ points $(\zeta_1, \hdots, \zeta_n) \in \mathbb{C}^n$ and infinitesimal radii $(\mathrm{d}\zeta_1, \hdots, \mathrm{d}\zeta_n) \in \mathbb{R}_+^n$,   the joint probability that there is at least one point of $\mathfrak{X}$ falling in each infinitesimal disk $\mathsf{D}(\zeta_k, \mathrm{d}\zeta_k)$, of radius $\mathrm{d}\zeta_k$ and centered at $\zeta_k$, is
\begin{align}
\rho_n(\zeta_1, \hdots, \zeta_n) \mathrm{Leb}(\mathrm{d}\zeta_1) \hdots \mathrm{Leb}(\mathrm{d}\zeta_n).
\end{align}

\begin{remark}
    If the Point Process $\mathfrak{X}$ has \emph{exponential tails}, that is, there exists a constant $c> 0$ such that for any compact $\mathfrak{K}\subset\mathfrak{L}$,  $\mathbb{P}\left[ \mathrm{card}(\mathfrak{X}\cap\mathfrak{K} )> n \right] \leq \exp(-cn)$, where $\mathrm{card}$ denotes the number of elements of a discrete ensemble, then if they exist, the collection of all joint intensities determines the distribution of $\mathfrak{X}$~\cite[Remark 1.2.4]{hough2009zeros}.
\end{remark}

For $k=1$, the first intensity function $\rho_1$ can be interpreted as the unnormalized \emph{density} of points of $\mathfrak{X}$ in $\mathfrak{L}$. 
Indeed, it follows from Definition~\ref{def:joint} that for any compact set $\mathfrak{K} \subset \mathfrak{L}$,
\begin{align}
    \mathbb{E}\left[ \mathrm{card}(\mathfrak{K}\cap \mathfrak{X})\right] = \int_{\mathfrak{K}} \rho_1(\zeta)\, \mathrm{Leb}(\mathrm{d}\zeta).
\end{align}
For Gaussian Analytic Functions, the first intensity exists and is given by the so-called Edelman-Kostlan formula.
\begin{proposition}
    Let $\mathfrak{Z}$ be the zero set of a Gaussian Analytic Function of covariance kernel $\mathsf{K}$, as defined in~\eqref{eq:def-kernel}.
    Up to multiplication by a deterministic function, one can assume $\mathfrak{Z}$ only consists of random zeros.
    Then the first intensity of $\mathfrak{Z}$ exists, and\footnote{One can show that $z\mapsto \mathsf{K}(z,z)$ is real analytic, so that differentiation is allowed. 
    By assumption, $\mathsf{K}$ does not vanish: if it did, the Gaussian Analytic Function would have a deterministic zero there; see \cite[Section 2.4.1]{hough2009zeros}.}
    \begin{align}
        \rho_1(z) = \frac{1}{4\pi} \Delta \log \mathsf{K}(z,z)
        \label{eq:edelman-kostlan}
    \end{align}
    where $\Delta$ is the Laplacian operator. \footnote{For a function $\mathsf{F}$ of the complex variable $z$, seen as a function of two real variables through $z = x+\mathrm{i}y$, the Laplacian of $\mathsf{F}$ is defined as $\Delta \mathsf{F} = \dfrac{\partial^2 \mathsf{F}}{\partial x^2} + \dfrac{\partial^2 \mathsf{F}}{\partial y^2}$.}
    \label{prop:edelman-kostlan}
\end{proposition}

Three different proofs of Proposition~\ref{prop:edelman-kostlan} are given in~\cite[Section 2.4]{hough2009zeros}, all beautifully illustrating the interplay of geometry, harmonic analysis and probability.

Applying the Edelman-Kostlan formula \eqref{eq:edelman-kostlan} to the covariance kernel of the planar Gaussian Analytic Function in \eqref{eq:K-alpha}, one obtains a constant first intensity 
\begin{align}
\label{eq:rho1_pGAF}
\rho_1 (z) = \frac{\gamma}{\pi}, \quad z \in \mathbb{C}
\end{align}
for the Point Process $\mathfrak{Z}^{(\gamma)}_{\mathbb{C}}$ of Equation~\eqref{eq:Z-alpha}. 
This is consistent with the invariance of the zeros under translations in Proposition~\ref{prop:inv-iso}, as well as with the visibly uniform distribution of zeros in Figure~\ref{fig:planar}. 

Gaussian Analytic Functions are suprisingly constrained objects. 
As an edifying example, a property called \emph{Calabi's rigidity}, proved by Sodin \citep{sodin2000rigidity} and discussed in \cite[Theorem 2.5.1]{hough2009zeros}, states that the first intensities of the zero sets of two Gaussian Analytic Functions defined on $\mathfrak{L}$, $\mathsf{F}$ and $\mathsf{G}$, are equal is and only if there exists a deterministic non-vanishing analytic function $\psi : \mathfrak{L} \rightarrow\mathbb{C}$ such that $\mathsf{G} = \psi \mathsf{F}$. 
In other words, the first intensity \emph{alone} essentially characterizes the zeros of a Gaussian Analytic Function. 
In particular, this implies that the planar Gaussian Analytic Function of Definition~\ref{def:planarGAF} is the only one, up to multiplication by a non-vanishing analytic function, whose zeros are invariant under isometries of $\mathbb{C}$.
The same is true for the other two canonical Gaussian Analytic Functions of Remark~\ref{rem:canonical_gafs}.

Among the numerous properties that make Gaussian Analytic Functions remarkable mathematical objects,  the existence of explicit expressions for their joint intensities~\cite[Corollary 3.4.1]{hough2009zeros} has important practical implications. 

\begin{proposition}
    Let $\mathsf{F}$ be the Gaussian Analytic Function defined on $\mathfrak{L}$ with covariance kernel $\mathsf{K}$, and denote its zero set by $\mathfrak{Z}$. 
    If 
    $$ 
        \zeta_1, \dots, \zeta_n \mapsto \mathrm{det}\left(\left(\mathsf{K}(\zeta_k,\zeta_\ell)\right)_{1 \leq k,\ell \leq n}\right)
    $$ does not vanish anywhere, then the $n$-point joint intensity function of $\mathfrak{Z}$ exists, and writes as the ratio of an $n\times n$ permament and an $n\times n$ determinant,
    \begin{align}
    \forall (\zeta_1, \hdots, \zeta_n) \in \mathfrak{L}^n, \quad \rho_n(\zeta_1, \hdots, \zeta_n) = \frac{\mathrm{per}(\mathbf{C}-\mathbf{B}\mathbf{A}^{-1}\mathbf{B})}{\mathrm{det}(\pi \mathbf{A})},
    \end{align}
    where $\mathbf{A}, \mathbf{B},  \mathbf{C} \in \mathbb{C}^{n\times n}$ are defined by, for $1  \leq k,\ell \leq n$,
    $$
        \mathrm{A}_{k,\ell} = \mathbb{E}\left[ \mathsf{F}(\zeta_k) \overline{\mathsf{F}(\zeta_\ell)}\right], \, 
        \mathrm{B}_{k,\ell} = \mathbb{E}\left[ \mathsf{F}'(\zeta_k) \overline{\mathsf{F}(\zeta_\ell)}\right], \,
        \mathrm{C}_{k,\ell} = \mathbb{E}\left[ \mathsf{F}'(\zeta_k) \overline{\mathsf{F}'(\zeta_\ell)}\right],
    $$
    and $\mathsf{F}'$ the derivative of $\mathsf{F}$. \footnote{A careful reader might have noted that, interestingly, $\mathrm{A}_{k,\ell} = \mathsf{K}(\zeta_k,\zeta_\ell)$.}
\label{prop:joint-intensity}
\end{proposition}

Proposition~\ref{prop:joint-intensity} has been leveraged to compute spatial statistics of the zeros of the Gaussian spectrogram of white noise~\cite{bardenet2020zeros}, as later presented in Section~\ref{sssec:pair-corr}.
Not only does this explicit formula enable to explain the joint behavior of the zeros observed in Figure~\ref{sfig:wnoise_spec2}, but it also can be used to design zero-based signal processing procedures~\cite{bardenet2020zeros}, as discussed in Section~\ref{sec:sig_spat_stat}.

\subsubsection{Hyperuniformity and rigidity}

To go deeper in the understanding of the geometry of zero set of Gaussian Analytic Functions, it is useful to have a reference Point Process in mind. 
This role is often filled by the intensively studied  \emph{Poisson Point Process}, described in great detail in~\cite[Chapter 2]{daley2003introduction},~\cite[Chapter 1]{hough2009zeros}. 
Unlike the zero sets of Gaussian Analytic Functions, the distribution of its zeros shows spatial independence by definition.

\begin{definition}
    Let $\nu$ be a measure on $\mathbb{C}$, the \emph{Poisson Point Process} $\mathfrak{P}$ of underlying measure $\nu$ is the unique Point Process such that,
    $\forall p \in \mathbb{N}^*$, $\forall \mathfrak{B}_1, \hdots, \mathfrak{B}_p$ \emph{disjoint} Borel subsets of $\mathbb{C}$,
    \begin{align}
    \mathbb{P}\left[ \mathrm{card}(\mathfrak{P} \cap \mathfrak{B}_n) = n_k, \, 1\leq k \leq p \right]=\prod_{k=1}^p \exp(-\nu(\mathfrak{B}_k)) \frac{\nu(\mathfrak{B}_k)^{n_k}}{n_k!}
    \end{align}
    where by convention the right-hand side vanishes if $\nu(\mathbb{B}_k) = \infty$ for at least one $k$.
    \label{def:PoissonPP}
\end{definition}

In particular, the numbers of points of a Poisson Point Process falling in disjoint subsets are \emph{independent} random variables. 
Hence, in contrast to the zeros of $\mathsf{GAF}_{\mathbb{C}}$ displayed in Figure~\ref{fig:planar}, Poisson Point Processes do not visually exhibit any sort of regular spacings, as can be observed in Figure~1 of~\cite[Chapter 1]{hough2009zeros}.

Many properties of Poisson Point Processes are documented.
As an illustrative example,  for $\mathfrak{P}$ the Poisson Point Process with respect to the measure $\lambda \mathrm{Leb}$, with $\mathrm{Leb}$ the Lebesgue measure on $\mathbb{C}$ and $\lambda > 0$ the so-called \emph{intensity} of $\mathfrak{P}$,  Definition~\ref{def:PoissonPP} ensures that for any $r>0$, $\mathrm{card}( \mathfrak{P} \cap \mathsf{D}(0,r) )$ is a Poisson random variable of parameter $\lambda \times \pi r^2 $, so that $\mathbb{V} \left[\mathrm{card}( \mathfrak{P} \cap \mathsf{D}(0,r) )\right] = \lambda  \times \pi r^2$. 
This reference behavior is used to define the \emph{hyperuniformity} property.

\begin{definition}
A Point Process $\mathfrak{X}$ on $\mathbb{C}$ is said to be \emph{hyperuniform} if 
\begin{align}
\mathbb{V} \left[\mathrm{card}( \mathfrak{X} \cap \mathsf{D}(0,r) )\right] \underset{r \rightarrow\infty}{=} o(r^2).
\end{align}
\end{definition}

This sub-Poissonian growth is interpreted as a sign of organization at large scales. A thorough review on hyperuniformity can be found in~\cite{Cos21}, spatial statistics aspects are treated in~\cite{HGBL23,klatt2022genuine}.
Remarkably, the Point Process $\mathfrak{Z}_{\mathbb{C}}$ of the zeros of the planar Gaussian Analytic Function is hyperuniform; more precisely,
\begin{align}
    \lim\limits_{r\rightarrow \infty} \frac{\mathbb{V} \left[\mathrm{card}( \mathfrak{Z}_{\mathbb{C}} \cap \mathsf{D}(0,r) )\right] }{r} = \frac{\zeta_{\mathrm{R}}(3/2)}{4\pi^{3/2}},
\end{align}
where $\zeta_{\mathrm{R}}$ denotes the Riemann zeta function~\cite[Equation (4.16)]{forrester1999exact}.

Another interesting property quantifying the regularity of a Point Process is \emph{rigidity}~\cite{ghosh2017fluctuations,Cos21}.
In words, ridigity is the ability to predict some moments of the configuration of points within a compact window, given the outside.

\begin{definition}
    Let $\mathfrak{X}$ be a Point Process in $\mathbb{C}$. 
    If for any bounded open subset $\mathfrak{B} \subset \mathbb{C}$, there exists an integer-valued measurable function $\mathsf{N}$, defined on the configurations of points in $\mathfrak{B}^\mathrm{c} =\mathbb{C}\setminus\mathfrak{B}$, such that almost surely
    \begin{align}
        \label{e:number_rigidity}
        \mathrm{card} (\mathfrak{X}\cap \mathfrak{B}) = \mathsf{N}(\mathfrak{X}\cap \mathfrak{B}^\mathrm{c}),
    \end{align}
    then $\mathfrak{X}$ is said to be \emph{number-rigid}.\footnote{\emph{Number-rigidity} is not to be confused with \emph{Calabi's rigidity} mentioned in Section~\ref{sssec:spat-stat-Z}.}

    In words, \emph{number-rigidity} means that given the configuration of $\mathfrak{X}$ outside of $\mathfrak{B}$, the number of points of $\mathfrak{X}$ falling inside $\mathfrak{B}$ is known.
    Similarly, a Point Process is said to be \emph{barycenter-rigid} if the same property holds for the \emph{barycenter} of the elements of $\mathfrak{X}\cap \mathfrak{B}$, that is replacing the cardinality with the barycenter in Equation~\eqref{e:number_rigidity}.
\end{definition}

The Point Process of the zeros of the planar Gaussian Analytic Function has been shown to be both number- and barycenter-rigid, as explained in the survey~\cite{GhPe17}. 
Yet, not all zero sets of Gaussian Analytic Functions are number-rigid.
For example the zeros of the hyperbolic Gaussian Analytic Function~\cite[Equation (2.3.6)]{hough2009zeros} are not number rigid~\cite{HoSo13}.
Furthermode, considering that number-rigidity is of order $0$, and barycenter-rigidity of order $1$, \cite{GhKr21} proposed a parametrized variant of the planar Gaussian Analytic Function which shows rigidity at an arbitrary order.

\subsection{Zeros of the time-frequency representation of white noise}
\label{ssec:zerosGAF_TF}

Heuristic Equations~\eqref{eq:Vgw}~and~\eqref{eq:Vh-atom} in the preamble of Section~\ref{sec:GAF} advocate for the existence of a bridge between Gaussian Analytic Functions and time-frequency representations of white noise, provided that, first, the time-frequency transforms of the Hilbert basis elements are, up to multiplication by a non-vanishing function, analytic functions defining a proper Gaussian Analytic Function and then, that the Short-Time Fourier Transform of white noise is given a rigorous mathematical sense.

A key element to prove these claims is the connection between time-frequency representations and complex analysis embodied by the \emph{Bargmann transform}~\cite{segal1963mathematical}, originating in mathematical physics, which associates a finite-energy signal to an \emph{analytic} function tightly connected to its Short-Time Fourier Transform.

\subsubsection{The Bargmann transform}

\begin{definition}
Let $s\in L^2(\mathbb{R})$, the Bargmann transform $\mathcal{B}s$ of the signal $s$ is
\begin{align}
\mathcal{B}s: z \mapsto \pi^{-1/4} \exp ( - z^2/2 ) \int_{\mathbb{R}} s(t) \exp \left( \sqrt{2}tz - t^2/2 \right) \, \mathrm{d}t, \quad z \in \mathbb{C}.
\end{align}
\end{definition}
Integrability of the integrand over $\mathbb{R}$ is ensured by $s\in L^2(\mathbb{R})$.
As stated in Proposition~\ref{prop:Barg-analytic} below, the image of $\mathcal{B}$ is a subset of the analytic functions of the complex plane~\cite[Chapter 12]{flandrin2018explorations}.
Furthermore, the Bargman transform preservers the \emph{energy}, i.e., the $L^2$-norm, of the signal.

\begin{proposition}{\cite[Proposition 3.4.1]{grochenig2001foundations}}
The Bargmann transform $\mathcal{B}$ isometrically maps $L^2(\mathbb{R})$ onto the \emph{Bargmann-Fock space} 
$$
    \mathfrak{F} = \left\{ \mathsf{F} \in A(\mathbb{C}), \,  \lVert \mathsf{F} \rVert^2_{\mathfrak{F} } = \int_{\mathbb{C}} \lvert \mathsf{F}(z)\rvert^2 \exp(-\lvert z \rvert^2) \, \mathrm{Leb}(\mathrm{d}z) < \infty \right\}.
$$
\label{prop:Barg-analytic}
\end{proposition}

Assimilating the time-frequency plane to the complex plane, the Bargmann transform connects with the Gaussian Short-Time Fourier Transform~\cite[Section 12.1]{flandrin2018explorations}.

\begin{proposition}
    Let $g(t) = \pi^{-1/2}\exp(-t^2/2)$ be the circular Gaussian window, and $s\in L^2(\mathbb{R})$.
    The Short-Time Fourier Transform with window $g$ of the signal $s$ coincides, up to multiplication by a non-vanishing function, with its Bargmann transform:
    \begin{align}
    \forall (t,\omega) \in \mathbb{R}^2, \quad \mathcal{V}_g s(t,\omega) = \exp( - \lvert z\rvert^2/4-\mathrm{i}\omega t /2) \times \mathcal{B}s\left(\frac{z}{\sqrt{2}}\right),
    \label{eq:STFT_Barg}
    \end{align}
    where $z = \omega + \mathrm{i}t$.
    \label{prop:STFT_Barg}
\end{proposition}

One major interest of the \emph{Bargmann transform} is that it maps the basis formed by the Hermite functions, introduced in Section~\ref{sssec:hermite}, onto \emph{monomials}.

\begin{lemma}{\cite[Proposition 3.4.4]{grochenig2001foundations}}
    For any $k \in \mathbb{N}$, let $h_k$ be the Hermite function of order $k$, introduced in Definition~\ref{def:finite-dim-wn}. 
    Then, the Bargmann transform of $h_k$ reads
    \begin{align}
    \mathcal{B}h_k(z) = \frac{z^k}{\sqrt{k!}}, \quad z\in\mathbb{C}.
    \label{eq:Barg-hk}
    \end{align}
\end{lemma}
The family of monomials $\left(\mathcal{B}h_k\right)_{k\in\mathbb{N}}$ satisfies Hypothesis~\eqref{eq:uniform-K}, and thus lead to a Gaussian Analytic Function in Proposition~\ref{prop:constructive_lemma}, namely the planar Gaussian Analytic Function presented in detail in Section~\ref{sssec:planarGAF}. 

\begin{remark}
    Combining the link between the Short-Time Fourier Transform and the Bargmann transform~\eqref{eq:STFT_Barg}, and the closed-form expression of the Bargmann transform of the Hermite function $h_k$~\eqref{eq:Barg-hk}, one recovers the Gaussian spectrogram of $h_k$~\eqref{eq:hk-spec}.
\end{remark}

\subsubsection{The Short-Time Fourier Transform of white noise}

By Proposition~\ref{prop:STFT_Barg}, to define the Short-Time Fourier transform of white noise, it is enough to define its Bargmann transform. 
The definition of white Gaussian noise in Section~\ref{ssec:white_noise} gives a central role to a given basis of $L^2(\mathbb{R})$, which from now on will be chosen as the Hermite basis. 

\begin{proposition}{\cite[Section 3.5]{bardenet2021time}}
    Let $\mathfrak{G}$ be the completion of $L^2(\mathbb{R})$ defined in Section~\ref{sssec:inf-wn}, using the Hermite basis. 
    The Bargmann transform extends continuously to $\mathfrak{G}$.
    \label{prop:extend-Barg}    
\end{proposition}

\begin{proof}
    First, remark that the involved monomials satisfy a stronger assumption than \eqref{eq:uniform-K}. Indeed, for any compact $\mathfrak{K}\subset\mathbb{C}$,
    \begin{equation}
        C_{\mathfrak{K}}:=\sup_{z\in \mathfrak{K}}\,\sum_{k\in\mathbb{N}} (1+k^2)\left\lvert \frac{z^k}{\sqrt{k!}}\right\rvert^2<\infty.
    \end{equation}
    Letting  $s \in L^2(\mathbb{R})$, it follows from the expression~\eqref{eq:weak-norm} of $\lVert \cdot \rVert_{\mathfrak{G}}$ and the Cauchy-Schwarz inequality that
    \begin{align}
        \forall z \in \mathfrak{K}, \quad \lvert \mathcal{B}s(z) \rvert \leq \sqrt{C_{\mathfrak{K}}} \lVert s \rVert_{\mathfrak{G}}.
        \label{eq:Bargman-Lip}
    \end{align}
    Since $\mathfrak{G}$ is the \emph{completion} of $L^2(\mathbb{R})$, for any $y\in \mathfrak{G}$, there exists a sequence $(s_n)_{n\in \mathbb{N}}$ in $L^2(\mathbb{R})$ which converges to $y$ in $\mathfrak{G}$.
    By \eqref{eq:Bargman-Lip}, the sequence $\left( \mathcal{B}s_n \right)_{n\in\mathbb{N}}$ has the Cauchy property in $A(\mathbb{C})$, which is complete.
    Consequently, $\left( \mathcal{B}s_n \right)_{n\in\mathbb{N}}$ converges to a unique $\mathcal{B}y \in A(\mathbb{C})$, independently of the choice of the sequence.
    This defines a unique continuous extension $\mathcal{B} : \mathfrak{G} \rightarrow A(\mathbb{C})$.
\end{proof}

In particular, for $\xi$ the Gaussian white noise as built in Section~\ref{ssec:white_noise}, Proposition~\ref{prop:extend-Barg} defines a random variable $\mathcal{B}\xi$.
To see that the latter expresses as a random series, let $\wn^{(n)}$ be the truncation of $\xi$ introduced in \eqref{eq:truncated-wn}.
By linearity,
\begin{align}
    \mathcal{B}\wn^{(n)} = \sum_{k=0}^n \wn_k \frac{z^k}{\sqrt{k!}},
\label{eq:Bwn-trunc}
\end{align}
where $\left(\wn_k\right)_{k\in \mathbb{N}}$ are i.i.d. standard complex Gaussian variables. 
Since $\wn^{(n)} \rightarrow \wn$ almost surely in $\mathfrak{G}$, and by continuity of $\mathcal{B}$, taking the limit in both sides of~\eqref{eq:Bwn-trunc} yields
\begin{align}
    \mathcal{B}\wn(z) = \sum_{k=0}^\infty \wn_k \frac{z^k}{\sqrt{k!}} = \mathsf{GAF}_{\mathbb{C}}(z)
\end{align}
where convergence is almost sure in $A(\mathbb{C})$. 
In particular, the probability distribution of the Bargmann transform of white noise coincides with that of the planar Gaussian Analytic Function.

\begin{remark}
    The Bargmann transform being linear, Equation~\eqref{eq:Bargman-Lip} implies that, for any fixed compact set $\mathfrak{K}\subset \mathbb{C}$, $\mathcal{B} : L^2(\mathbb{R}) \rightarrow A(\mathfrak{K})$ is $\sqrt{C_\mathfrak{K}}$-Lipschitz.
    Hence Corollary~\ref{cor:dist-L-wn} applies and provides an upper bound of the rate at which $\mathcal{B}\wn^{(n)}$ converges toward $\mathcal{B}\wn$.
    This is the starting point of~\cite[Section 5.3]{bardenet2021time}, culminating in~\cite[Theorem 5.5]{bardenet2021time} which gives control on the number of zeros of $\mathcal{B}\wn$ based only on the truncation $\mathcal{B}\wn^{(n)}$ for $n$ large enough.
\end{remark}

Equation~\eqref{eq:STFT_Barg}, which links the Gaussian Short-Time Fourier Transform and the Bargmann transform, allows us to conclude as to the meaning of the Short-Time Fourier transform of white noise.

\begin{proposition}
    Let $\wn$ be the infinite-dimensional white noise associated to the Hilbert basis of Hermite functions. 
    The Short-Time Fourier Transform with circular Gaussian window of $\xi$, 
    $$
        \omega+\mathrm{i}t\mapsto \mathcal{V}_g\xi(t, \omega),
    $$ 
    has the same distribution as 
    \begin{align}
        z\mapsto \psi(z) \mathsf{GAF}_{\mathbb{C}}^{(1/2)}(z),
        \label{eq:Vgwn}
    \end{align}
    where $\psi:z\mapsto\exp( - \lvert z\rvert^2/4-\mathrm{i}\omega t /2)$ is a non-vanishing function.
\label{prop:Vgwn-GAF}
\end{proposition}

\begin{remark}
    Proposition~\ref{prop:Vgwn-GAF} has interesting implications in applied probability. 
    Indeed, sampling the zeros of random polynomials or analytic functions is known to be a very difficult task:  Gaussian Analytic Functions and their truncations can behave very wildly as soon as $\lvert z \rvert > 1$, rapidly exceeding numerical encoding capacities.
    This is not the case of time-frequency representations, which can be evaluated for discrete signals of large length in a very stable way, through the Fast Fourier Transform algorithm as explained in Section~\ref{ssec:num_impl}.
    The numerical stability is ensured by the non-vanishing prefactor $\psi$ in Equation~\eqref{eq:Vgwn}, which naturally performs a \emph{regularization} at infinity without impacting the distribution of the zeros.
\end{remark}

\subsubsection{Spatial statistics for the zeros of the Gaussian spectrogram of white noise}
\label{sssec:pair-corr}

From Proposition~\ref{prop:Vgwn-GAF}, since $\psi$ is non-vanishing, the zeros of the Gaussian spectrogram of white noise are exactly the zeros of the planar Gaussian Analytic Function $\mathsf{GAF}_{\mathbb{C}}^{(1/2)}$, which have been thoroughly presented in Section~\ref{ssec:def_GAF}.
In particular, Proposition~\ref{prop:inv-iso} implies that the zeros of the Gaussian spectrogram of white noise form a stationary and isotropic Point Process, and Proposition~\ref{prop:joint-intensity} yields explicit expressions for all its $k$-point statistics.

A major tool in spatial statistics is the two-point joint intensity introduced in Definition~\ref{def:joint}, which, roughly speaking, encapsulates the statistics of \emph{distances} between points.
In practice, its is often normalized by the first intensity, to define the \textit{pair correlation function} 
\begin{align}
\label{eq:pair-cor-fun}
    \mathsf{g}(z,z') = \frac{\rho_2(z,z')}{\rho_1(z) \rho_1(z')}, \quad z, z' \in \mathbb{R}^2.
\end{align}
For \emph{stationary} Point Processes, two-point statistics, such as $\mathsf{g}$, only depend on $z-z'$. 
If additionally, the Point Process is \emph{isotropic}, then they only depend on $\lvert z - z'\rvert$.

Expanding the intricate formula provided in Proposition~\ref{prop:joint-intensity}, a closed-form expression of the pair correlation function of the zeros of the Gaussian spectrogram of white noise has been derived\footnote{
    Proposition~\ref{prop:g0-wn} is slightly different from \cite[Proposition 5]{bardenet2020zeros} due to different conventions in the definition of the Short-Time Fourier Transform. The formula provided in~\eqref{eq:g0-wn} can be obtained by the change of variable $z \rightarrow z /\sqrt{2\pi}$ in~\cite[Equation (15)]{bardenet2020zeros}.}
in~\cite[Proposition 5]{bardenet2020zeros}.

\begin{proposition}{\cite[Proposition 5]{bardenet2020zeros}}
Let $\mathsf{g}_\mathbb{C}$ be the pair correlation function of the zeros of the Gaussian spectrogram of white noise. 
Then $\mathsf{g}_{\mathbb{C}}$ is well-defined, only depends on $r = \lvert z -z' \rvert$, and writes
\begin{align}
\mathsf{g}_\mathbb{C} (r) = \frac{\left[ \sinh^2\left(\frac{r^2}{4}\right) + \frac{r^4}{16} \right]\cosh\left(\frac{r^2}{4}\right) -  \frac{r^2}{2} \sinh\left( \frac{r^2}{4}\right)}{\sinh^3\left(\frac{r^2}{4}\right)}.
\label{eq:g0-wn}
\end{align}
\label{prop:g0-wn}
\end{proposition}
    
\begin{figure}
\centering
\includegraphics[width = 7.5cm]{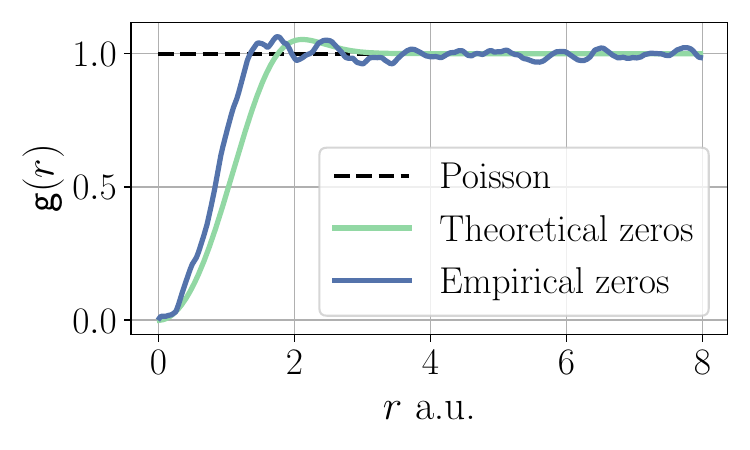}
\caption{\label{fig:g0}Pair correlation function of the zeros of the Gaussian spectrogram of white noise compared to the constant pair correlation function of the Poisson Process. The theoretical $\mathsf{g}_{\mathbb{C}}$ is given in~\eqref{eq:g0-wn} and the estimation is made with the \textit{R} package \texttt{spatstat}~\cite{baddeley2015spatial}.}
\end{figure}

The pair correlation function of the zeros of the Gaussian spectrogram of white noise is displayed in Figure~\ref{fig:g0}. It is to be compared with the pair correlation function of the Poisson Point Process which is constant equal to one.
The fact that $\mathsf{g}_{\mathbb{C}}$ vanishes at small scales indicates a short-range repulsion between zeros, which contrasts with the independence characterizing a Poisson Point Process.
Furthermore, the zeros of the Gaussian spectrogram of white noise have a small ring of attraction around $r =2$,  corresponding to $\mathsf{g}_{\mathbb{C}}(r) > 1$, which prevents them to constitute a \emph{Determinantal Point Process} with Hermitian kernel.\footnote{See~\cite[Chapter 4]{hough2009zeros} for a thorough presentation of Determinantal Point Processes.}

The so-called \emph{hole probability} is another common tool to study the dispersion of points in a stationary Point Process.

\begin{definition}
    For a stationary Point Process  $\mathfrak{X}$ on $\mathbb{C}$, and $r>0$, the \emph{hole probability} $p_r$ is the probability that no point of $\mathfrak{X}$ falls into the disk of radius $r$ centered at $0$,
    \begin{align}
        p_r = \mathbb{P}\left[ \mathfrak{X} \cap \mathsf{D}(0,r) = \emptyset\right].
    \label{eq:hole-proba}
    \end{align}
\end{definition}

Note that by stationarity of the Point Process, the choice of $0$ as the center in \eqref{eq:hole-proba} is arbitrary.
The asymptotic behavior of the hole probability of the zeros of the planar Gaussian Analytic Function $\mathsf{GAF}_{\mathbb{C}}$,\footnote{Note that this corresponds to $\gamma =1$ in Definition~\ref{def:planarGAF}.} and hence of the zeros of the Gaussian spectrogram of white noise, has been studied in~\cite{nishry2010asymptotics}, who showed that 
\begin{align}
r^{-4} \log p_r \underset{r \rightarrow \infty}{\longrightarrow} \frac{-3\mathrm{e}^2}{4}.
\end{align}
This decay of the hole probability as $\exp(-r^4)$ is to be compared to the slower decay of the hole probability of any stationary \emph{Poisson} Point Process, which goes as $\exp(-r^2)$~\cite[Proposition 7.2.1]{hough2009zeros}.
This comparison supports the observation that the zeros of the Gaussian spectrogram of white noise are more evenly spread in the time-frequency plane than a Poisson Point Process.

\section{Signal processing based on spatial statistics of zeros}
\label{sec:sig_spat_stat}

The comprehensive knowledge about the distribution of the zeros of the Gaussian spectrogram of white noise has been leveraged to design \emph{zero-based} signal processing procedures targeting signal detection and reconstruction, the two major signal processing tasks introduced in Section~\ref{sssec:aims}.
While the seminal procedure proposed in~\cite{flandrin2015time}, and the recent generalization to zeros of wavelet transforms~\cite{koliander2019filtering}, rely on the Delaunay triangulation of the configuration of zeros to perform signal detection and reconstruction, \cite{bardenet2020zeros, pascal2022covariant} use tools from spatial statistics \citep{moller2003statistical}. 
This section follows the latter path, and introduces a modicum of spatial statistics before illustrating their use in zero-based signal detection. 

\subsection{Summary functions}
Besides the joint intensities, which correspond to the \emph{moments} of a Point Process distribution, several functional quantities have been proposed to either describe a Point Process, or serve as objective in an inference pipeline. 
A common class of such summary functions, well suited to \emph{stationary} Point Processes, are related to the second-order properties of the Point process.
In particular, \emph{Ripley's $K$ function} is useful to quantify the repulsiveness between the points of a stationary and isotropic Point Process.

\begin{definition}
    Let $\mathfrak{X}$ be a stationary isotropic Point Process on $\mathbb{C}$, with constant first intensity $\rho_1 > 0$ and $\mathfrak{K} $ any compact of $\mathbb{C}$.
    \emph{Ripley's $K$ function} is defined, at $r>0$, as the expected number of pairs of point in $\mathfrak{X}$ which are distant of less than $r$~\cite[Chapter 4]{moller2003statistical}, that is,
    \begin{align}
    \label{eq:def_K}
    \forall r >0, \quad K(r) = \frac{1}{4\pi \rho_1^2 \mathrm{Leb}(\mathfrak{K})} \mathbb{E} \left[\sum_{z \in \mathfrak{X}\cap\mathfrak{K}, z' \in \mathfrak{X}}^{\neq}   \boldsymbol{1}  \left(\left\lvert z - z'\right\rvert < r \right) \right]
    \end{align}
    where $\mathrm{Leb}(\mathfrak{K}) < \infty$ is the \emph{area} of $\mathfrak{K}$; the sum runs over all pairs of \textit{distinct} points $z \neq z'$ of $\mathfrak{X}$ such that $z \in \mathfrak{K}$, and $ \boldsymbol{1}(\cdot)$ is the indicator function, taking value 1 if the condition is met and zero otherwise.
        It can be shown that $K(r)$ is well defined and,  by stationarity of $\mathfrak{X}$,  independent of the compact $\mathfrak{K}$~\cite[Section~4.1.2]{moller2003statistical}.
    Moreover, Ripley's $K$ function is linked to the pair correlation function of Equation~\eqref{eq:pair-cor-fun} through
    \begin{align}
    K(r) = 2\pi \int_{0}^r u \mathsf{g}(u) \, \mathrm{d}u.
    \end{align}
    Practitioners often prefer the so-called \emph{variance-stabilized} version of $K$: 
    \begin{align}
        \forall r >0, \quad L(r) = \sqrt{\frac{K(r)}{\pi}}.
        \label{eq:def:L}
    \end{align}
    \label{def:K-L}
\end{definition}

\begin{figure}
\centering
\includegraphics[width = 7.5cm]{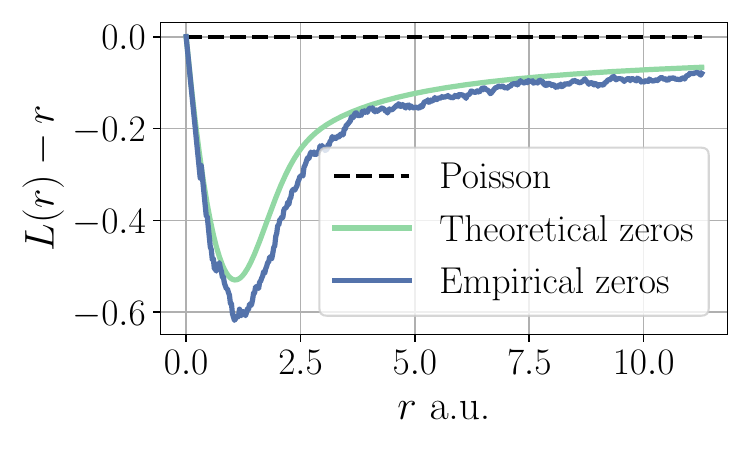}
\caption{\label{fig:Lstab}\textit{Variance stabilized}-$L$ function of the zeros of the Gaussian spectrogram of white noise. The estimation is made with the \textit{R} package \texttt{spatstat}~\cite{baddeley2015spatial}.}
\end{figure}

For the reference Poisson Point Process,  corresponding no neither attraction nor repulsion, the $L$-function is the identity, which is one reason to use $L$ rather than $K$. 
Figure~\ref{fig:Lstab} displays $r\mapsto L(r)-r$ for the Poisson Point Process, and for the zeros of the Gaussian spectrogram of white noise.
The $L$-functions of the zeros, in green for the function defined in \eqref{eq:def:L} and in blue for an estimate, show a clear lack of pairs at small scales, which is the signature of the repulsiveness between spectrogram zeros.

\begin{remark}
    From a practical point of view, Equation~\eqref{eq:def_K} means that estimating Ripley's $K$ function amounts to count the number of zeros distant from $r>0$.
    In practice, however, Point Processes are only observed on \emph{bounded} windows, so that the $K$-functional is only estimated on a bounded range $[0, r_{\max}]$, for $r_{\max}>0$ determined by the diameter of the observation window. 
    To handle border effects arising from the \emph{partial} observation of the Point Process, \cite[Section 4.3.3]{moller2003statistical} reviews sophisticated \emph{edge corrections}.
    Estimation of Figures~\ref{fig:g0} and~\ref{fig:Lstab} used Ohser and Stoyan's translation edge correction~\cite{ohser1983estimators}.
\end{remark}

Related to the hole probability of Equation~\ref{eq:hole-proba}, a second class of summary functions focuses on interpoint distances, and provides information on the size of \emph{holes} in a Point Process.

\begin{definition}
    Let $\mathfrak{X}$ be a stationary Point Process on $\mathbb{C}$ and for $r > 0$,  let $\mathsf{D}(0,r)$ denote the disk of radius $r$ centered at the origin. 
    The empty-space function of $\mathfrak{X}$ is defined as~\cite[Chapter 4]{moller2003statistical}
    \begin{align}
    \label{eq:def_F}
    F(r) = \mathbb{P}\left(\mathsf{D}(0,r) \cap \mathfrak{X}\neq \emptyset \right).
    \end{align}
    In words, the empty-space function of a stationary Point Process is defined as the distribution of the distance from a given fixed point, which by stationarity can be set arbitrarily as the origin, to the nearest point of the Point Process.\footnote{The empty space function at $r$ is linked to the hole probability \eqref{eq:hole-proba} through $F(r) = 1 - p_r$.}
    \label{def:F}
\end{definition}

\begin{remark}
    In Figure~\ref{fig:ex_spec}, the presence of a deterministic signal creates large \emph{holes} in the pattern of zeros.
    Such holes are absent in the Point Process made of zeros of the Gaussian spectrogram of white noise displayed in Figure~\ref{sfig:wnoise_spec}. 
    This advocates for the use of the empty-space function to identify measurements that correspond to a deterministic signal covered by a noise, with respect to pure noise.
\end{remark}

\subsection{A zero-based Monte Carlo envelope test for signal detection}
\label{ssec:zero_based_detect}

A fundamental question in signal processing is whether a noisy observation contains an underlying signal of interest.
Zero-based procedures using spatial statistics have proved accurate for this detection task~\cite{bardenet2020zeros,pascal2022covariant}.

Formally, in the signal-plus-noise model, signal detection consists in a statistical test enabling to reject, with a fixed confidence level $1 - \alpha$, the null hypothesis\footnote{In such tests, $\alpha$ corresponds to the probability, under $\mathrm{H}_0$, of rejecting $\mathrm{H}_0$.} 
$$\mathrm{H}_0: \snr = 0$$ 
corresponding to pure white noise, with the alternative hypothesis 
$$\mathrm{H}_1: \snr > 0$$ 
corresponding to a non-centered measurement $y = \snr \times s + \wn$, in which $s$ possibly carries interesting information. 
The smaller the $\snr$, the harder the detection.

A standard statistical procedure was followed in \cite{bardenet2020zeros},  plugging the summary functions from Section~\ref{ssec:SPat_stat} into a \emph{Monte Carlo envelope} test.
This procedure relies on the construction of an intermediate statistic $\mathcal{S} : \mathfrak{G} \rightarrow \mathbb{R}$ such that large values $\mathcal{S}y$ tend to correspond to situations where the measurements contain an underlying signal, so that $\mathrm{H}_0$ should be rejected.

\begin{definition}
    Given a summary function $J$, say $L$ or $F$ from Section~\ref{ssec:SPat_stat}, defined on a range of distances $[r_{\min}, r_{\max}]$, the associated $\mathcal{S}_{\infty}$ and $\mathcal{S}_{2}$ statistics are defined as
    \begin{align}
    \mathcal{S}_{\infty}y = \underset{r \in [r_{\min}, r_{\max}]}{\sup} \left\lvert \widehat{J}_y(r) - J_0(r) \right\rvert  ~;~ \mathcal{S}_2 y = \sqrt{\int_{r_{\min}}^{r_{\max}} \left\lvert \widehat{J}_y(r) - J_0(r)\right\rvert^2 \, \mathrm{d}r},
    \end{align}
    where $J_0$ is the corresponding summary function of the zeros of the Gaussian spectrogram of white noise, while $\widehat{J}_y$ is an estimator of this summary function, computed on the observed data $y$.
    \label{def:summary_stat}
\end{definition}
The spatial statistics $\mathcal{S}_{\infty}$ and $\mathcal{S}_{2}$ quantify the discrepancy between the observation $y$ and pure white noise.
By a symmetry argument, the following Monte Carlo test then performs detection at the confidence level $\alpha$, without further distributional assumption.

\begin{proposition}
Let $J$ a functional statistic, and $\mathcal{S}$ the associated spatial statistic. 
Let $\alpha \in ]0, 1[$, and two integers $k \leq m$ such that $\alpha = k/(m+1)$.
The following testing procedure
\begin{enumerate}[label = (\roman*)]
\item generate $m$ samples of white noise $\wn^{(1)}, \hdots, \wn^{(m)}$;
\item compute the summary statistics $\mathcal{S}^{(j)} = \mathcal{S} \wn^{(j)}$, $j \in \lbrace 1, \hdots, m\rbrace$;
\item sort them such that $\mathcal{S}^{(1)} \geq \hdots \geq \mathcal{S}^{(m)}$;
\item compute the empirical summary statistic $\mathcal{S}y$;
\item if $\mathcal{S}y \geq \mathcal{S}^{(k)}$, then reject the null hypothesis $\mathrm{H}_0$;
\end{enumerate}
provides a signal detection test with confidence level $1 - \alpha$.
\label{prop:test}
\end{proposition}

One advantage of the detection test of Proposition~\ref{prop:test} is that no user-defined \emph{threshold} is needed.  Instead, the value of $\mathcal{S}y$ above which the null hypothesis should be rejected is determined fully by the desired confidence level and by the numerical simulations of the $m$ white noise samples.
Several functional statistics $J$ can be used, among which Ripley's $K$ function, its variance-stabilized version $L$, introduced in Definition~\ref{def:K-L}, or the empty-space function $F$ of Definition~\ref{def:F}. 
The power of the resulting tests are numerically investigated in Section~\ref{ssec:numerical_exp}.

\begin{remark}
    If the functional statistic $J_0$ is not known explicitly, it can be replaced by 
    \begin{align}
        \forall r \in [r_{\min}, r_{\max}], \quad \overline{J}_0(r) = \frac{1}{m+1} \left( \sum_{j = 1}^m \widehat{J}_{\wn^{(j)}}(r) + \widehat{J}_{y}(r) \right)
    \end{align}
    without impacting the confidence level of the test provided in Proposition~\ref{prop:test}.
      Including $\widehat{J}_{y}$ in the computation of $\overline{J}_0$ crucially preserves the symmetry ensuring that all summary statistics associated with both the noise samples and the observation have the same distribution under the null hypothesis $\mathrm{H}_0$~\cite[Equation (2)]{baddeley2014tests}.
\end{remark}

Before demonstrating the test, it remains to discuss how to numerically extract zeros from a discretized approximate spectrogram.

\subsection{Numerical algorithms to find zeros}

As explained in Section~\ref{ssec:num_impl}, in practice, signals are sampled at a given frequency,  yielding discrete observations, from which a \emph{discrete spectrogram} is computed taking advantage of the discrete Fast Fourier Transform algorithm.
This discrete spectrogram is thus a Riemann-like approximation of the actual spectrogram, at a finite grid of $N$ values, sampled between $-T$ and $T$ with time step $T_s$, and thus available on a bounded uniform grid in the time-frequency plane,
\begin{align}
(t,\omega)\in T_s\mathbb{Z}\times \omega_s\mathbb{Z} \cap [-T,T]\times [0,\Omega];
\end{align}
Here, $\omega_s$ denotes the frequency step at which the spectrogram is evaluated, and the maximal frequency $\Omega=NT^{-1}/4$ satisfies the Nyquist-Shannon condition.\footnote{The Nyquist-Shannon theorem states that, if a signal $s\in L^2(\mathbb{R})$ contains no frequency larger than $B >0$, which means that $ \mathcal{F}s(\omega) = 0$ as soon as $\omega$ is larger than $2\pi B$, then $s$ can be fully reconstructed from the sequence $(s(nT_s), \, n \in \mathbb{Z})$ as soon as the sampling frequency $f_s = T_s^{-1}$ is (strictly) larger than $2B$~\cite[Theorem 5.15]{vetterli2014foundations}}

\begin{figure}
\centering
\begin{subfigure}{0.32\linewidth}
\centering
\includegraphics[width = 0.75\linewidth]{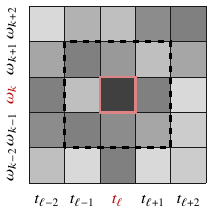}
\end{subfigure}\hspace{0.11\linewidth}
\begin{subfigure}{0.32\linewidth}
\centering
\includegraphics[width =\linewidth]{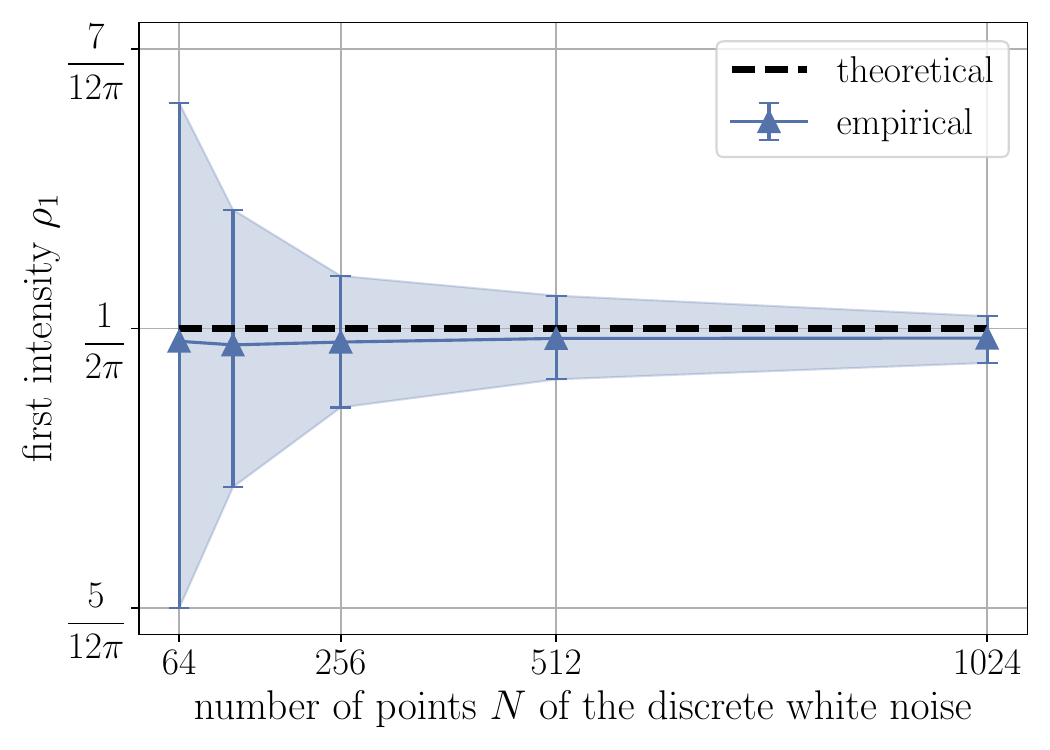}
\end{subfigure}
\caption{\label{fig:extract_zeros}\emph{Minimal Grid Neighbor} method for localization of spectrogram zeros. (left) Characterization of a local minimum adapted from~\cite{pascal2022covariant}. (right) Empirical vs. theoretical first intensity of the zeros of the Gaussian spectrogram of white noise.}
\end{figure}

In the code\footnote{https://perso.ens-lyon.fr/patrick.flandrin/zeros.html} accompanying the seminal paper~\cite{flandrin2015time}, the zeros of the spectrogram are computed using \textit{minimal grid neighbors}, as illustrated in Figure~\ref{fig:extract_zeros}.
This method first seeks \emph{local minima}, defined as points of the grid, such as $(t_k,\omega_\ell)$ in red in Figure~\ref{fig:extract_zeros}, at which the spectrogram value is lower than at the eight nearest neighbors, contained in the black dashed square in Figure~\ref{fig:extract_zeros}.
Then, the local minima which are below a fixed threshold are considered as \emph{zeros}.
This algorithm is used in most practical studies involving zeros of Short-Time Fourier Transforms~\cite{flandrin2015time,bardenet2020zeros,bardenet2021time} and~\cite[Chapters 13 and 15]{flandrin2018explorations}, zeros of wavelet transforms~\cite{abreu2018filtering,koliander2019filtering}, or zeros of generalized covariant representations~\cite{pascal2022covariant,pascal2022famille}.

\begin{remark}
    Interestingly, by a super-harmonicity argument, \cite[Corollary~1]{abreu2022local} showed that any local minimum of the spectrogram is a zero. 
    If the approximation introduced by the Discrete Fourier Transform is neglected, it is thus in principle unnecessary to further threshold the local minima.
\end{remark}

To assess the Minimal Grid Neighbor detection algorithm, the right plot of Figure~\ref{fig:extract_zeros} displays in blue the estimated first intensity of the zeros of the Gaussian spectrogram of white noise for several numbers points in the discrete signal, averaged over $200$ samples of white noise, accompanied with $95\%$ Gaussian confidence level, represented as the light blue area.
It is in fair agreement with the expected constant first intensity $\rho_1 = 1/(2\pi)$, obtained by setting $\alpha = 1/\sqrt{2}$ in Equation~\eqref{eq:rho1_pGAF}.

\begin{remark}
    Recently, ~\cite{escudero2021efficient} proposed an adaptation of the Minimal Grid Neighbors approach, namely the \textit{Adaptive Minimal Grid Neighbors} algorithm, coming with theoretical guarantees that all the zeros of the Gaussian spectrogram of a noisy signal are detected with an arbitrarily high precision, determined by the resolution of the discrete time-frequency grid. 
    The theoretical result is supported by intensive numerical experiments. 
    Interestingly,~\cite{escudero2021efficient} yields the minimal resolution needed to detect all zeros with a given probability.
\end{remark}

\subsection{Numerical experiments}
\label{ssec:numerical_exp}

This section describes numerical experiments conducted in~\cite[Section 5.2]{bardenet2020zeros} to assess the \emph{power} of the zero-based Monte Carlo detection test described in Section~\ref{ssec:zero_based_detect}, defined as the probability to reject $\mathrm{H}_0$ under the alternative hypothesis $\mathrm{H}_1$.
To make the power numerically accessible, the authors restricted the alternative to a linear chirp model \eqref{eq:chirp}.
For the sake of completeness, these experiments are reproduced \emph{verbatim}, with courtesy of all the authors.

\begin{figure}
\centering
\includegraphics[width=0.32\linewidth]{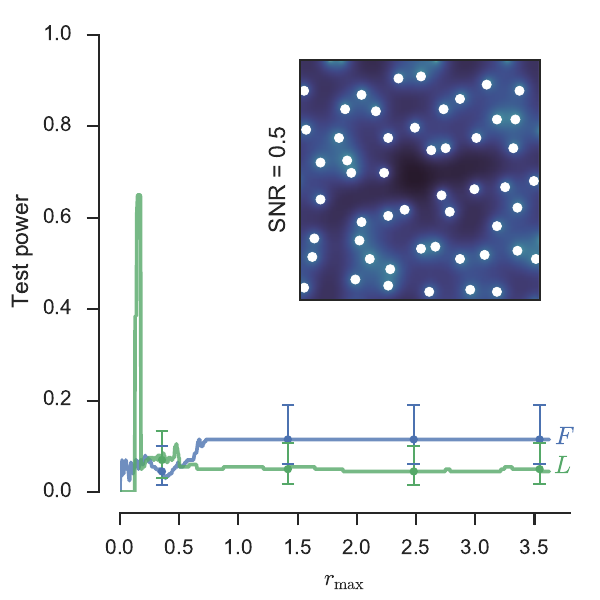} \hspace{0.11\linewidth}
\includegraphics[width=0.32\linewidth]{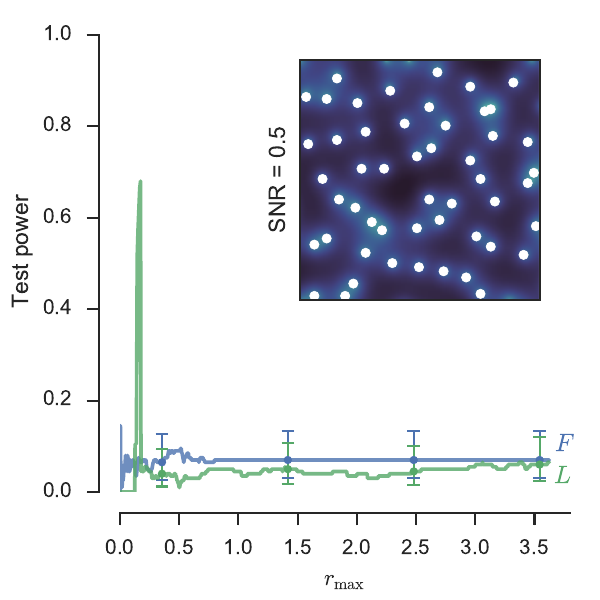}\\
\includegraphics[width=0.32\linewidth]{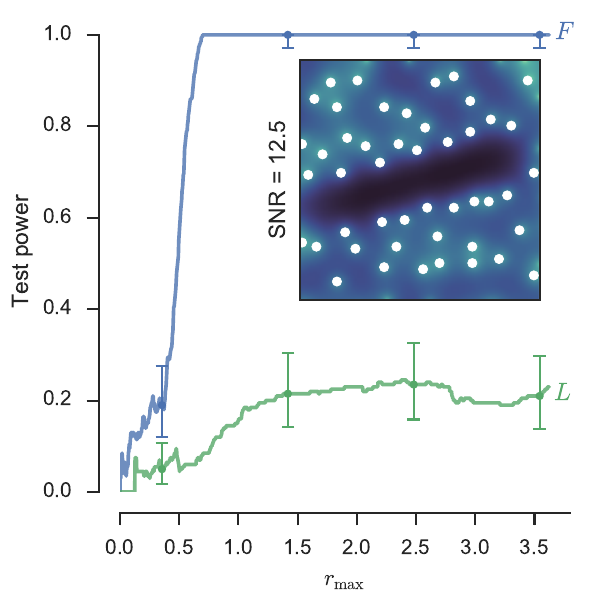} \hspace{0.11\linewidth}
\includegraphics[width=0.32\linewidth]{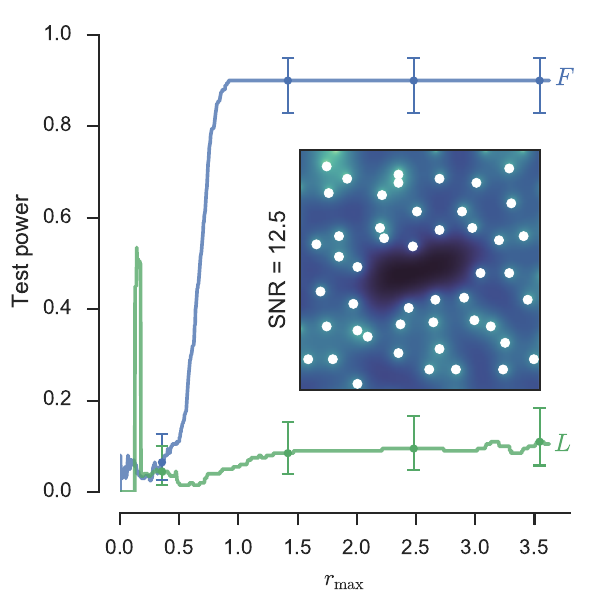}\\
\includegraphics[width=0.32\linewidth]{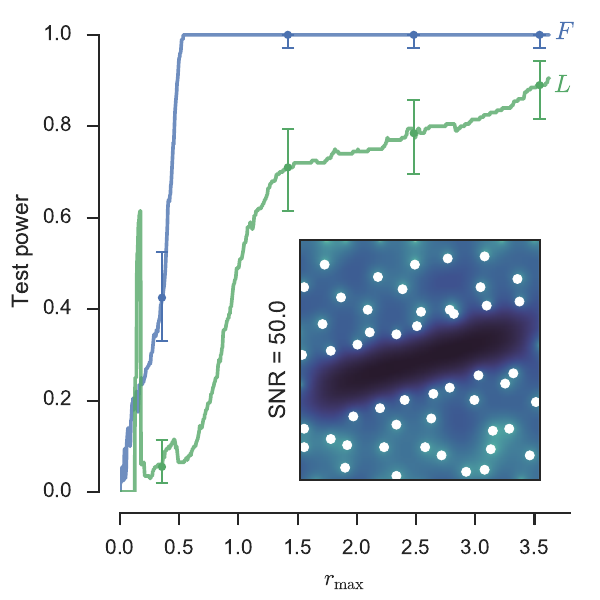}\hspace{0.11\linewidth}
\includegraphics[width=0.32\linewidth]{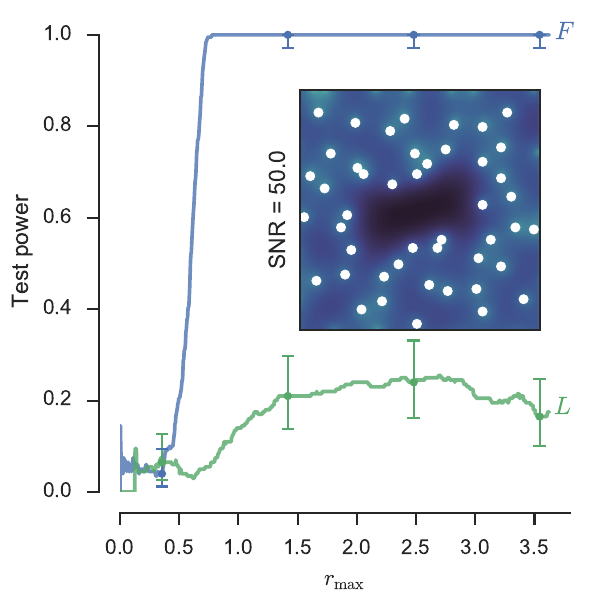}
\caption{\label{fig:testpower}\emph{Reproduced from~\cite{bardenet2020zeros} with courtesy of J. Flamant and P. Chainais.}}
\end{figure}

Synthetic noisy linear chirps following the signal-plus-noise model of Section~\ref{sssec:signalnoise} are generated.
Examples of noisy chirps are provided in Figure~\ref{fig:snr}.
Three levels of noise are considered, $\snr \in \lbrace 1, 5, 10 \rbrace$, and two support sizes, called $T$ in~\eqref{eq:chirp}, the chirp occupying either the entire observation window, left column of Figure~\ref{fig:testpower}, or half of it, right column. 
The smaller $\snr$ and $T$, the lower the expected detection power, as can be observed on the example spectrograms associated to each subplot of Figure~\ref{fig:testpower}.

The level of confidence is fixed to $\alpha = 0.05$, and the number of samples of white noise to $m=199$, imposing $k = 10$.
Following Definition~\ref{def:summary_stat}, two summary statistics using the $2$-norm, i.e. of type $\mathcal{S}_2$,  are compared, respectively based on the variance-stabilized $L$-function of Definition~\ref{def:K-L}, in green, and on the empty-space function $F$ of Definition~\ref{def:F}, in blue.
The power $\beta$ of the test is computed on 200 independent samples of each noisy chirp, and reported in Figure~\ref{fig:testpower} as a function of the upper bound $r_{\max}$ on which the summary statistic is computed.
The reported Clopper-Pearson confidence intervals for the five values of $r_{\max}$ take into account a Bonferroni correction~\cite{wasserman2004all}, corresponding to the ten multiple tests per plot.

As expected, the power increases with $\snr$ and is lower for shorter signals. 
Furthermore, it is advantageous to use the maximal range of values of $r$, so that as many points as possible are taken into account in the summary statistics. 
For large $r_{\max}$ and medium to high $\snr$, the empty-space function $F$ leads to higher power than the variance-stabilized $L$ function, and otherwise is of comparable performance. 
It thus appears as a robust choice, enabling to reach powers close to unity for $\snr$ not too small.

These experiments have shown the ability of zero-based detection procedures to detect linear chirps as long as the noise level is not prohibitive.
Moreover, this study is complemented in~\cite[Section 5.3]{bardenet2020zeros} with reconstruction experiments relying on the estimation of a mask delimiting the chirp in the spectrogram estimated jointly with the empty-space function.
Together with the zero-based denoising and component-separation strategy designed in~\cite{flandrin2015time}, these experiments draw a promising path for signal processing based on spatial statistics on the spectrogram zeros.

There exist plenty of detection and reconstruction methods in signal processing to which zero-based methods should now be compared, among which matched filtering~\cite{giannakis1990signal} and spectrogram correlators~\cite{altes1980detection} for detection, ridge extraction combined with synchrosqueezing~\cite{meignen2017demodulation} or minimization of the mean-square error on the spectrogram via a Griffin-Lim algorithm~\cite{griffin1984signal} for reconstruction.
Remarkably, a public benchmark~\cite{juanpublic} has recently been developed, enabling to compare zero-based approaches to other state-of-the-art procedures, responding to the great interest raised by the topic in the signal processing community.
Furthermore, everyone is invited to contribute to the benchmark by contributing the code of their method of choice.\footnote{https://github.com/jmiramont/benchmarks-detection-denoising/}

\section{Extensions and new research avenues}
\label{sec:extensions}

This chapter has focused on the zeros of the Short-Time-Fourier Transform.
In this opening section,  pointers to recent works that goes beyond that case are given.
The present section does not intend to exhaustively list all recent works combining harmonic analysis and spatial statistics but rather to shed light on some interesting  directions of research on zeros of signal representations.

\subsection{A time-frequency transform for each Gaussian Analytic Function}

As demonstrated in Section~\ref{ssec:SPat_stat}, the zero set of the planar Gaussian Analytic Function, defined on the complex plane, is invariant under the group of isometries of $\mathbb{C}$.
It turns out that there exist two other Gaussian Analytic Functions, whose zeros are respectively invariant under isometries of the hyperbolic plane $\mathbb{H}$ and the unit sphere $\mathbb{S}$ respectively~\cite[Section 2.3]{hough2009zeros}. 
Following the pioneering work~\cite{bardenet2020zeros} connecting the planar Gaussian Analytic Function and time-frequency analysis, a natural question is whether these two Gaussian Analytic Functions are linked to some signal processing transforms.

Such connections have been established in~\cite{abreu2018filtering,bardenet2021time,pascal2022covariant, pascal2022famille}. Notably, \cite{bardenet2021time,abreu2018filtering} have simultaneously established that the zeros of the \emph{scalogram} of white noise obtained from the Daubechies-Paul wavelet transform coincides with the zeros of the \emph{hyperbolic} Gaussian Analytic Function, though with subtly different definitions of white noise.
A preliminary connection was established in~\cite{bardenet2021time} between the spherical Gaussian Analytic Function and an ad-hoc generalized time-frequency transform, acting on vectors of measurements, thus bypassing the need to approximate integrals in the computation of Fourier transforms. 
The resulting transform was however numerically unstable. 
Later on, a novel connection was established between the spherical Gaussian Analytic Function and a numerically stable signal processing transform, expressed as a decomposition on coherent states~\cite{pascal2022covariant,pascal2022famille}\footnote{See~\cite[Chapter 9]{grochenig2001foundations} and~\cite[Chapter 2]{ali2000coherent} for comprehensive description of the notion of \emph{coherent state} and their link with \emph{covariance} properties.}.

\subsection{Toward non analytic representations}

In order for the Short-Time Fourier Transform to write as the product of a non-vanishing function and an \emph{analytic} function, it is necessary to use the \emph{circular Gaussian window}~\cite{ascensi2009model}.
Analyticity is a crucial property both in~\cite{flandrin2015time,bardenet2020zeros,bardenet2021time}, ensuring that the zeros of the spectrogram of a noisy signal constitute a well-defined Point Process and enabling to connect to the rich theory of Gaussian Analytic Functions.
However, a recent work~\cite{haimi2020zeros} proposed a more general framework enabling to study the zeros of the spectrogram of white noise for non-Gaussian windows thanks to the introduction of \textit{Gaussian Weyl-Heisenberg Functions}. To characterize the zeros of a non-Gaussian spectrogram of Gaussian white noise, the authors of \cite{haimi2020zeros} rely on the Kac-Rice formulae~\cite[Chapter 3]{azais2009level}, describing the level sets of Gaussian functions.

In particular, assuming that the analysis window is a Schwartz function, the Point Process of zeros of the spectrogram of Gaussian white noise is shown to be \textit{stationary}, though not isotropic, and the first intensity of its zeros is expressed in terms of the ambiguity function of the analysis window~\cite[Theorems 1.6 and 1.9]{haimi2020zeros}.
More precisely,  time translations and frequency modulations defined in Section~\ref{sssec:covariance}, combined with genuinely designed phase shifts, constitute the so-called \emph{Weyl-Heisenberg} group, under which the spectrogram with circular Gaussian window is covariant.
As for the Short-Time Fourier Transform with a Schwartz window, it is covariant under the action of an \emph{extended} Weyl-Heisenberg group.
These results provide a solid theoretical background to the numerical study of the zeros of the spectrogram of white noise with Hermite window performed in~\cite[Chapter 16]{flandrin2018explorations}, as discussed in~\cite[Corollary 1.10]{haimi2020zeros}.
Going beyond the restricted framework of the Short-Time Fourier Transform with circular Gaussian window is of utmost importance, since a significant proportion of applied signal processing works relies on non-Gaussian windows~\cite{karnik2022thomson}.  

\bibliographystyle{plain}
\bibliography{my_biblio}

\end{document}